\documentclass[12pt]{article} 

\pdfoutput=1
\usepackage[top=80pt,bottom=85pt,left=85pt,right=85pt]{geometry}
\usepackage[all]{xy}
\usepackage{amssymb}
\usepackage{amsmath}
\usepackage{mathtools}
\usepackage{setspace}
\usepackage{sectsty}
\usepackage{accents}
\usepackage{comment}
\usepackage[english]{babel}

\usepackage[utf8]{inputenc}
\usepackage{uniinput}
\usepackage{changepage,cite}
\usepackage[plain]{fancyref}

\usepackage{makecell} % line break within a cell
\usepackage{longtable} % long table spanning multiple pages
\usepackage[normalem]{ulem} % \sout for strikethrough text
\usepackage{transparent} % pdf_tex needs this
\usepackage{bm} % \bm{} for bold math symbols

\usepackage[usenames]{xcolor}
\usepackage{graphicx,subcaption}
\usepackage{setspace}
\usepackage{float}
\usepackage{booktabs}
\usepackage[vcentermath]{youngtab}

\usepackage[debug,pageanchor=false]{hyperref}
\definecolor{link}{rgb}{.8,.15,.1}
\definecolor{pigment}{rgb}{0.36, 0.54, 0.66}
\definecolor{pigment2}{rgb}{0.19, 0.55, 0.91}
\definecolor{pigment3}{rgb}{0.2, 0.2, 0.6}
\definecolor{light-gray}{gray}{0.75}
\hypersetup{colorlinks=true,linkcolor=link,citecolor=link,urlcolor=link,linktocpage}
\usepackage{cleveref}

\usepackage{array,multirow,booktabs,longtable}
\usepackage{mathrsfs}
\usepackage{tikz-cd} 
\usetikzlibrary{backgrounds, arrows,calc,shapes,decorations.pathreplacing, decorations.markings, automata,positioning,matrix}

\newcommand\vertarrowbox[3][6ex]{%
  \begin{array}[t]{@{} c c @{}} #2 \\
  \left\downarrow\vcenter{\hrule height #1}\right.\kern-\nulldelimiterspace & #3
  \end{array}%
}

\tikzset{
        cvertex/.style={circle,draw=black,inner sep=1pt,outer sep=3pt},
        vertex/.style={circle,fill=black,inner sep=1pt,outer sep=3pt},
        star/.style={circle,fill=yellow,inner sep=0.75pt,outer sep=0.75pt},
        tvertex/.style={inner sep=1pt,font=\scriptsize},
        gap/.style={inner sep=0.5pt,fill=white}}

\tikzstyle{mybox} = [draw=black, fill=blue!10, very thick,
    rectangle, rounded corners, inner sep=10pt, inner ysep=20pt]
\tikzstyle{boxtitle} =[fill=blue!50, text=white,rectangle,rounded corners]
\allowdisplaybreaks

\def\node#1#2{\overset{#1}{\underset{#2}{\circ}}}

\usetikzlibrary{arrows}
\tikzstyle{every picture}+=[remember picture]
\tikzstyle{na} = [baseline=-.5ex]
\tikzstyle{mine}= [arrows={angle 90}-{angle 90},thick]

\def\Llleftarrow{%
\lower2pt\hbox{\begingroup
\tikz
\draw[shorten >=0pt,shorten <=0pt] (0,3pt) -- ++(-1em,0) (0,1pt) -- ++(-1em-1pt,0) (0,-1pt) -- ++(-1em-1pt,0) (0,-3pt) -- ++(-1em,0) (-1em+1pt,5pt) to[out=-105,in=45] (-1em-2pt,0) to[out=-45,in=105] (-1em+1pt,-5pt);
\endgroup}
}

\renewcommand{\Re}{\mathrm{Re}}
\renewcommand{\Im}{\mathrm{Im}}

\DeclareMathOperator{\depth}{depth}

\DeclareMathOperator{\rank}{rank}

\newcommand{\todo}[1]{}
\renewcommand{\todo}[1]{{\color{red} TODO: {#1}}}
\newcommand{\red}[1]{}
\renewcommand{\red}[1]{{\color{red} {#1}}}
\newcommand{\blue}[1]{}
\renewcommand{\blue}[1]{{\color{blue} {#1}}}

\newcommand{\su}[1]{}
\renewcommand{\su}[1]{{\mathfrak{su}({#1})}}
\newcommand{\uu}[1]{}
\renewcommand{\uu}[1]{{\mathfrak{u}({#1})}}
\newcommand{\so}[1]{}
\renewcommand{\so}[1]{{\mathfrak{so}({#1})}}
\newcommand{\usp}[1]{}
\renewcommand{\usp}[1]{{\mathfrak{usp}({#1})}}

\makeatletter
\renewcommand\xleftrightarrow[2][]{%
  \ext@arrow 9999{\longleftrightarrowfill@}{#1}{#2}}
\newcommand\longleftrightarrowfill@{%
  \arrowfill@\leftarrow\relbar\rightarrow}
\makeatother

\makeatletter
\@addtoreset{equation}{section}
\makeatother
\usepackage[en-US,showdow,showzone]{datetime2}
\DTMlangsetup[en-US]{abbr=true,showyear=false,zone=atlantic}

\graphicspath{{figures/}}

\begin{document}

% titlepage 

\begin{titlepage}

\begin{center}

\phantom{bu}

\vskip .3in \noindent

{\Huge \textbf{The Higgs Branch}}

\medskip

{\Huge \textbf{of Heterotic ALE Instantons}}

\bigskip

%{\large \textsc{part iv}}

\vskip .6in \noindent

{\Large Michele Del Zotto$^{\dagger,\ddagger}$, Marco Fazzi$^{\ddagger,\sharp}$}

\medskip

{\Large and Suvendu Giri$^{\ddagger,\ast,\star}$}

\bigskip

%{\color{blue}Version of \DTMNow~(w.r.t. GMT)}

\bigskip
{\small 

$^\dagger$ Department of Mathematics,  Uppsala University,  SE-75120 Uppsala, Sweden \\
\vspace{.25cm}
$^\ddagger$ Department of Physics and Astronomy,  Uppsala University,  SE-75120 Uppsala, Sweden \\
\vspace{.25cm}
$^\sharp$ NORDITA, Hannes Alfv\'ens v\"ag 12, SE-10691 Stockholm, Sweden\\
\vspace{.25cm}
$^\ast$ Department of Physics, Princeton University, Princeton, New Jersey 08544, USA\\
\vspace{.25cm}
$^\star$ Princeton Gravity Initiative, Princeton University, Princeton, New Jersey 08544, USA
}
 
\vskip .3cm
{\small \tt \href{mailto:michele.delzotto@math.uu.se}{michele.delzotto@math.uu.se}  \hspace{.5cm} \href{mailto:marco.fazzi@physics.uu.se}{marco.fazzi@physics.uu.se} \hspace{.5cm} \href{mailto:suvendu.giri@physics.uu.se}{suvendu.giri@princeton.edu}}

\vskip .6cm
     	{\Large {\bf Abstract }}
\vskip .1in

\end{center}

\noindent We begin a study of the Higgs branch of six-dimensional $(1,0)$ little string theories governing the worldvolumes of heterotic ALE instantons. We give a description of this space by constructing the corresponding magnetic quiver. The latter is a three-dimensional $\mathcal{N}=4$ quiver gauge theory that flows in the infrared to a fixed point whose quantum corrected Coulomb branches is the Higgs branch of the six-dimensional theory of interest. We present results for both types of heterotic strings, and mostly for $\mathbb C^2/\mathbb Z_k$ ALE spaces. Our analysis is valid both in the absence and in the presence of small instantons. Along the way, we also describe small $SO(32)$ instanton transitions in terms of the corresponding magnetic quiver, which parallels a similar treatment of the small $E_8$ instanton transitions in the context of the $E_8\times E_8$ heterotic string.
\vfill

\begin{flushright}
UUITP-11/23\\
NORDITA 2023-110
\end{flushright}
\eject

\end{titlepage}

% end titlepage

\tableofcontents
%\newpage

%%%%%%%%%%%%%%%%%%%%%%%%%%%%%%%%%%%%%%%

%%%%%%%%%%%%%%%%%%%%%%%%%%%%%%%
\section{Introduction}
%%%%%%%%%%%%%%%%%%%%%%%%%%%%%%%

Little string theories (LSTs) are six-dimensional (6D) nonlocal quantum field theories (QFTs) enjoying a form of T-duality.\footnote{\ More precisely, they are examples of ``quasilocal'' QFTs \cite{Kapustin:1999ci}. For instance, several different operators may be interpreted as a valid energy-momentum tensor.} Examples of such systems have originally been obtained by taking the $g_\text{s} \to 0$ limit (while $M_\text{s}=1/\sqrt{\alpha'}$ is held fixed) in the worldvolume theory of NS5-branes inside 10D string theories \cite{Seiberg:1997zk}.\footnote{\ Bulk modes of the 10D string decouple, whereas those on the worldvolume remain interacting. For a review circa 2000 on LSTs with sixteen supercharges see the classic reference \cite{Aharony:1999ks}. The name was coined in \cite{Losev:1997hx}.} Further investigations, in the context of classifications of six-dimensional theories, unveiled several other LSTs that can be geometrically engineered exploiting F-theory \cite{Vafa:1996xn} -- see e.g. \cite{Bhardwaj:2015xxa,Bhardwaj:2015oru,Bhardwaj:2018jgp,Bhardwaj:2019hhd}. Describing the LST moduli spaces unearthed several intriguing features \cite{Intriligator:1999cn} and an interplay with 3D $\mathcal N=4$ mirror symmetries \cite{Intriligator:1996ex,deBoer:1996ck,deBoer:1996mp} and related string duality chains \cite{Porrati:1996xi}. This interplay, together with recent improvements in our understanding of T-duality of LSTs via their two-group structure \cite{DelZotto:2020sop}, are among the core motivations for our study.

Of interest to us will be the LSTs governing heterotic ALE instantons. These are obtained from the so-called $(e)$ theory, the 6D $(1,0)$ LST (with eight Poincaré supercharges) coming from $M$ parallel NS5-branes of the $E_8 \times E_8$ heterotic string, acting as ``small'' instantons for the heterotic gauge group. (Namely, these instantons are pointlike: the curvature of the gauge bundle is concentrated at a point, parameterizing the location of the NS5s in the transverse $\mathbb{R}^4$ in 10D.) The $(e)$ theory contains only tensor multiplets, and is believed to flow to a nontrivial interacting fixed point (the so called rank-$M$ E-string) in the infrared (IR), upon decoupling the little string modes. The theories governing heterotic ALE instantons are close cousins of the $(e)$ theory, and are obtained by placing the parent heterotic string on a $\mathbb{C}^2/\Gamma_G$ orbifold transverse to say $M$ NS5s (with $\Gamma_G$ denoting one of the finite subgroups of $SU(2)$, associated to $G$ via the usual McKay correspondence) \cite{Intriligator:1997dh,Aspinwall:1997ye}.\footnote{\ This work hinges upon earlier results \cite{Aspinwall:1996vc,Intriligator:1997kq,Blum:1997fw,Blum:1997mm}, which are mostly concerned with $M$ NS5s of the $Spin(32)/\mathbb{Z}_2$ heterotic string on $\mathbb{C}^2/\Gamma_G$, i.e. with the $(o)$ LST on the orbifold, called $(o')$.} This latter LST is sometimes known as $(e')$ \cite{Gremm:1999hm}, and we will adopt this notation in the following. To fully specify the $(e')$ theory one should also provide the data of a flat connection at infinity, which in the $E_8 \times E_8$ case is encoded in two group homomorphisms $\mu_\text{L,R}:\Gamma_G \to E_8$. %Oftentimes the choice of the trivial embedding is given; in this case the dependence on $\mu_\text{L,R}$ will be omitted. 
In this paper we are interested in the moduli space of the $(e')$ LSTs in presence of nontrivial flat connections at infinity.

The above setup gives rise to intricate 6D models with (dynamical) tensor multiplets (say $n_\text{T}$ of them), vector multiplets, and matter hypermultiplets in various representations of the (product) gauge group, of rank $r_\text{V}$. For ALE singularities of type $\mathbb C^2/\mathbb Z_k$ and $\mathbb C^2/\mathbb D_k$ these models can be understood via a dual description in Type I' (adding O6-planes for the orbifolds of D type) -- see \cite{Hanany:1997gh,Brunner:1997gf,Hanany:1999sj} for a detailed description. In this paper we focus on type A, i.e. $\mathbb C^2/\mathbb Z_k$. In this case the tensor (or Coulomb) branch of the 6D vacuum moduli space --the branch where tensor multiplet scalars take vacuum expectation values (VEVs)-- is the Coxeter box of $USp(2M)$, i.e. topologically $(S^1)^{\otimes M}/\text{Weyl}(USp(2M))$ if there are $n_\text{T}=M$ dynamical tensors. This space is compact and has size $M_\text{s}^2$ \cite{Intriligator:1999cn}. Upon compactification on a $T^3$, for a trivial choice of flat connection at infinity leaving the heterotic gauge symmetry unbroken, there is an exact (quantum corrected) Coulomb branch (CB)\footnote{\ For a modern perspective on CBs in 3D see e.g. \cite{Bullimore:2015lsa} and references therein.} of quaternionic dimension
\begin{equation}
\label{eq:firstdim}
\dim_\mathbb{H} \text{CB}_{T^3} = \tfrac{1}{4}\dim_\mathbb{R} \text{CB}_{T^3}= r_\text{V}+n_\text{T}= r_\text{V}+M = h^\vee_G M - \dim_\mathbb{R} G=kM-(k^2-1)\ ,
\end{equation}
where $h^\vee_G$ is the dual Coxeter number of $G=SU(k)$. 

It was proposed in \cite{Intriligator:1999cn} that this space is the moduli space of $M$ instantons for the gauge group $G$ on a compact K3 surface of volume $M_\text{s}^2$. The appearance of the K3 can be understood via duality with M-theory: the $E_8\times E_8$ heteortic string on $T^3$ is believed to be dual to M-theory on a ($T^3$-fibered) K3, therefore we obtain a dual M-theory background $\mathbb{C}^2/\Gamma_G\times \text{K3}$. The singularity is being probed by $M$ transverse M2s, corresponding to the $M$ heterotic NS5s wrapped on the $T^3$ fiber of the K3 surface.\footnote{\ For the details of the relevant geometry we refer our readers to the slides of the talk \emph{Half K3 surfaces, or K3, $G_2$, $E_8$, M, and all that} by David Morrison,  given at Strings 2002. Currently the slides are available at the \href{https://member.ipmu.jp/yuji.tachikawa/stringsmirrors/2002/morrison.pdf}{unofficial Strings mirror website} maintained by Yuji Tachikawa.} Since M2-branes are pointlike instantons for the 7D gauge theory of type $G$ corresponding to M-theory on $\mathbb C^2/\Gamma_G$, this explains the fact that the CB of this 3D theory is simply the moduli space of $M$ instantons for the gauge group $G$ on a compact K3. The resulting moduli space is a compact hyperk\"ahler space with $c_1=0$ (its metric being the unique Ricci-flat one). These spaces have several interesting singularities that can be characterized exploiting corresponding 3D IR fixed points. In particular, taking the limit $M_\text{s} \to \infty$ produces a 3D field theory with CB given by the moduli space of $M$  instantons for the gauge group $G$ on a noncompact singular patch of the K3. For reasons that will become clear later, we will call the 3D QFT which flows to such a fixed point an \textit{electric quiver} for the 6D theory. As will be argued in the main body of the paper, for $G=SU(k)$ this quiver QFT reads:
\begin{equation}
\label{eq:first3D}
    1 - 2 - 3 - \cdots - (k-1) - \underbrace{\overset{\underset{\scriptstyle \vert}{\displaystyle 1}}{k} - k - \cdots - k - \overset{\underset{\scriptstyle \vert}{\displaystyle 1}}{k}}_{M-2k+1} - (k-1) - \cdots - 3 -2 -1\ ,
\end{equation}
with $M-2k+1 \geq 1$, i.e. $M\geq 2k$. (Henceforth, and unless stated otherwise, $k$ denotes an $\mathcal{N}=4$ $U(k)$ vector multiplet, an edge a bifundamental hypermultiplet, and $\boxed{p}-k$ denotes $p$ fundamentals of the gauge group $U(k)$.) The dimension of this CB is easily computed by summing all gauge ranks and subtracting one. (All gauge groups are unitary, so the product gauge group can be broken to a maximal torus. Moreover an overall $U(1)$ decouples from the dynamics -- see \cite[Sec. 6.3]{Hanany:1996ie}.) That is,
\begin{equation}
\dim_\mathbb{H}\text{CB}_\text{3D}\eqref{eq:first3D} = k(M-k)+1=h^\vee_{SU(k)}M-\dim_\mathbb{R}SU(k)\ ,    
\end{equation}
as expected from \eqref{eq:firstdim}. At the singularity of this space a \emph{Higgs branch} (HB) emanates, with $\dim_\mathbb{H}\text{HB}_\text{3D}\eqref{eq:first3D} = k-1$, and at the intersection of the two branches lives the 3D interacting superconformal field theory (SCFT) with $\mathcal{N}=4$ supersymmetry capturing the corresponding singularity of the moduli space.\footnote{\ The existence of such a nontrivial fixed point is guaranteed by the fact that each node in the quiver is balanced (i.e. $2N_\text{c}=N_\text{f}$) or overbalanced (i.e. $2N_\text{c} < N_\text{f}$) -- this is obtained with the understanding that neighboring gauge groups act as flavors for the gauge group with rank $N_\text{c}$ -- and the quiver is therefore ``good'' in the sense of \cite{Gaiotto:2008ak}.}

This HB is the space of interest for us. In fact, the 3D electric quiver \eqref{eq:first3D} comes from the torus compactification of a 6D generalized quiver (containing massless vector multiplets and tensor multiplets)\footnote{\ Both 6D $(1,0)$ vectors and tensors reduce to vectors in 3D.} which we will encounter in \eqref{eq:little2}, and reads:
\begin{equation}
    \label{eq:first6D}
    {\scriptstyle
    \boxed{\scriptstyle 1} - SU(2) - SU(3) - \cdots - SU(k-1) - \underbrace{\overset{\underset{\scriptstyle \vert}{\scriptstyle \boxed{\scriptstyle 1}}}{\scriptstyle SU(k)} \scriptstyle - SU(k) - \cdots - SU(k) - \overset{\underset{\scriptstyle \vert}{\scriptstyle \boxed{\scriptstyle 1}}}{SU(k)}}_{M-2k+1} - SU(k-1) - \cdots - SU(3) -SU(2) -\boxed{\scriptstyle 1}}\ ,
\end{equation}
with $\dim_{\mathbb{H}}\text{HB}_\text{6D}\eqref{eq:first6D}=M+k-1$. The $M$ extra moduli (w.r.t. the HB dimension in 3D, i.e. $k-1$) come from the 6D special unitary groups (as opposed to unitary in 3D).\footnote{\ The $U(1)$ center of $SU(n)$ is massive in 6D, and decouples from the low-energy dynamics. See \cite[Eq. (2.6)]{Hanany:1997gh}.} They correspond to the locations of $M$ identical small instantons on the ALE space, whereas the other $k − 1$ to the resolution parameters of the $\mathbb{C}^2/\mathbb{Z}_k$ orbifold \cite{Intriligator:1997dh}. These two numbers sum up to give the dimension of the hypermultiplet (i.e. Higgs) branch of the associated 10D heterotic moduli space \cite{Sen:1997js,Witten:1999fq}, which coincides with the HB of the $(e')$ LST for an ALE singularity of type $\mathbb C^2/\mathbb Z_k$ (and a trivial flat connection at infinity leaving the gauge group unbroken).

The main focus of this paper we will be to generalize the above construction to the cases where the  $(e')$ LST is enriched with choices of nontrivial flat connections at infinity breaking the $E_8\times E_8$ gauge group to the commutant of the embedding $(\mu_\text{L},\mu_\text{R}): \Gamma_G \to E_8 \times E_8$. Rather than focusing on the corresponding tensor (or Coulomb) branch, we will be interested in the HB. The latter is the branch where scalars in the matter hypermultiplets take VEVs, and thus corresponds to the hypermultiplet moduli space of the parent $E_8 \times E_8$ heterotic string on the orbifold, as discussed in our companion paper \cite{DelZotto:2023myd}. Applying the same logic as above, and because the HB is invariant under torus compactification (assuming \emph{no} Wilson lines are turned on breaking further the flavor symmetry in the toroidal reduction), we want to study the HB of the electric quiver. Thanks to mirror symmetry \cite{Intriligator:1996ex}, this is equivalent to the CB of the 3D mirror, that is a \emph{different} QFT. 

For instance, applying the mirror map to \eqref{eq:first3D} we obtain \cite{Hanany:1996ie}
\begin{equation}
\label{eq:1kCBtrick-pre}
    \overset{\overset{\displaystyle \boxed{M}}{\vert}}{SU(k)} \ ,
\end{equation}
with 
\begin{align}
    &\dim_\mathbb{H}\text{CB}_\text{3D}\eqref{eq:1kCBtrick-pre} =  \dim_\mathbb{H}\text{HB}_\text{3D}\eqref{eq:first3D} = k-1\ , \\
    &\dim_\mathbb{H}\text{HB}_\text{3D}\eqref{eq:1kCBtrick-pre} = \dim_\mathbb{H}\text{CB}_\text{3D}\eqref{eq:first3D} = kM-(k^2-1)\ .
\end{align}
This (single-node) quiver is a generalization to the case with $M$ flavors of the pure 3D $\mathcal{N}=4$ $G$ gauge theory conjectured in \cite{Witten:1999fq} to capture the hypermultiplet moduli space of the heterotic string on ALE via its CB (here $G=SU(k)$ and the ALE space is of type A, i.e. $\mathbb{C}^2/\mathbb{Z}_k$). Moreover \eqref{eq:1kCBtrick-pre} is closely related to the 3D \emph{magnetic quiver} introduced in \cite{Cabrera:2019izd} (and reviewed in detail below), which in this case reads:
\begin{equation}
\label{eq:1kCBtrick}
  \overset{\overset{\displaystyle \overbrace{1 \cdots 1}^{M}}{ \rotatebox[origin=c]{190}{$\setminus$}\ \rotatebox[origin=c]{-190}{$/$} }}{k}\ .
\end{equation}
The $SU(k)$ node in \eqref{eq:1kCBtrick-pre} is replaced by $U(k)$, while $M$ flavors are replaced by a ``bouquet'' of $M$ gauge $U(1)$'s, the opposite of an operation termed ``hyperk\"ahler implosion'' in \cite{Dancer:2012sv,Dancer:2020wll} which preserves the hyperk\"ahler structure of the moduli space and the action of (a maximal torus of) the flavor symmetry group. In physics terms, implosion corresponds to ungauging a (or more, as in this case) $U(1)$ by gauging the topological $U(1)_J$ symmetry associated to it \cite{Hanany:1996ie,Kapustin:1999ha,Witten:2003ya}. The origin of the ``explosion'' needed to go from \eqref{eq:1kCBtrick-pre} to \eqref{eq:1kCBtrick} can be traced to fact that in 6D an $S_M$ symmetry (exchanging the $M$ identical NS5s) is gauged \cite{Hanany:2018vph}.\footnote{\ This was also confirmed holographically for 6D $(1,0)$ T-brane theories in \cite{Bergman:2020bvi}.}

The CB dimension of \eqref{eq:1kCBtrick} is
\begin{equation}
\dim_\mathbb{H}\text{CB}_\text{3D}\eqref{eq:1kCBtrick}= \dim_{\mathbb{H}}\text{HB}_\text{6D}\eqref{eq:first6D}=M+k-1 \ ,  
\end{equation}
as an overall $U(1)$ decouples from the IR dynamics (similarly to the examples of \cite{Cabrera:2019izd}).
As we have already said, the dimension of the associated heterotic hypermultiplet moduli space (or 6D HB) is given by the number of resolution parameters of the $\mathbb{C}^2/\mathbb{Z}_k$ orbifold (i.e. $k − 1$) plus the locations of $M$ identical small instantons on the ALE space \cite{Intriligator:1997dh}. Conveniently, and because of mirror symmetry, this moduli space is captured by the CB of the 3D magnetic quiver.

In this work we explicitly construct the magnetic quivers for more general $(e')$ LSTs, with choices of nontrivial flat connections at infinity for ALE spaces with $\mathbb C^2/\mathbb Z_k$ singularities. The various possibilities are classified by breaking patterns of the $E_8 \times E_8$ gauge group of the 10D heterotic parent on the ALE orbifold $\mathbb{C}^2/\mathbb Z_k$. To understand the origin of both electric and magnetic quiver, it will be most instructive to realize the $E_8 \times E_8$ heterotic string as a Type I' setup via duality to the Ho\v{r}ava--Witten M-theory background.

\medskip

This paper is intended as a continuation of a double series of papers by the three authors, whose first installments are \cite{Fazzi:2022hal,Fazzi:2022yca,Fazzi:2023ulb,DelZotto:2022ohj,DelZotto:2022xrh,DelZotto:2023ahf,delzotto-liu-oehlman}. It is organized as follows. In section \ref{sec:het} we give a lightning review of the construction of the 6D tensor branches (or electric quivers) for the relevant LSTs of interest. In section \ref{sec:quivs} we give an algorithmic construction for the dual magnetic quivers. In section \ref{sec:checks} we discuss several consistency checks of our proposal. The main one comes from realizing that the $(e')$ theories with nontrivial flat connections at infinity (dubbed $\mathcal K_N(\mu_\text{L},\mu_\text{R};\mathfrak{a}_{k-1})$ in \cite{DelZotto:2022ohj,DelZotto:2022xrh,DelZotto:2023ahf}) are realized by fusing together two orbi-instanton theories \cite{DelZotto:2014hpa,Heckman:2018jxk}. The latter is an operation in 6D that generalizes a diagonal gauging of two identical global symmetry groups for two theories in four dimesions \cite{Heckman:2018pqx,DelZotto:2018tcj}. As a result we expect our magnetic quiver CB should have the features of a hyperk{\"a}hler quotient of the two HBs of the orbi-instanton theories involved in the glueing \cite{Hitchin:1986ea,Moore:2011ee}. Since the HBs of orbi-instanton theories can be computed in many different ways, this provides several interesting consistency checks of our proposal. We conclude our discussion in section \ref{sec:3Ddual} with some preliminary remarks about the behavior of the HBs upon T-duality.\footnote{\ We stress that it is not the HBs of the 6D LSTs that have to match across T-dualities, rather the HBs of the 5D theories obtained upon circle reduction. Since often these involve turning on nontrivial flavor symmetry Wilson lines, the 6D HBs will get corrected.} In section \ref{sec:conc} we present our conclusions.

%%%%%%%%%%%%%%%%%%%%%%%%%%%%%%%
\section{\texorpdfstring{The $E_8\times E_8$ heterotic and $(e')$ little string theories}{The E8xE8 heterotic and (e') little string theories}}
\label{sec:het}
%%%%%%%%%%%%%%%%%%%%%%%%%%%%%%%

In 9D the heterotic strings are dual to orientifolds of Type II. The $Spin(32)/\mathbb{Z}_2$ heterotic string is S-dual to Type I on a circle, i.e. the O9$^-$ orientifold of Type IIB with 16 physical D9-branes.\footnote{\ The gauge group is a quotient of $Spin(32)$ by $\mathbb{Z}_2$ which is not $SO(32)$, as there are no particles in the vector representation of the $D_{16}$ algebra \cite{Witten:1997bs}.} Type I on $S^1$ is in turn T-dual to Type I', the orientifold of Type IIA with two O8$^-$-planes at the endpoints of $S^1/\mathbb{Z}_2$ with 16 D8's along this interval. The latter setup can be lifted to a Ho\v{r}ava--Witten M-theory background on $S^1\times S^1/\mathbb{Z}_2 $ with two M9-walls at the endpoints of the interval; compactification along the former circle brings us back to Type I', whereas on the latter to the 9D $E_8 \times E_8$ heterotic string, the two being related (i.e. dual) by a so-called 9-11 flip in M-theory \cite{Porrati:1996xi}. All in all, when the gauge group is broken to $SO(16)\times SO(16)$ the two 9D heterotic strings are related by a form of T-duality sending the radius of one circle to the inverse of the other (see e.g. \cite{Ginsparg:1986bx,Mohaupt:1992jf}).

Consider now the $E_8\times E_8$ heterotic string compactified on a K3; its low-energy dynamics (once gravity is decoupled) is captured by a $(1,0)$ LST \cite{Seiberg:1997zk}, conventionally called $(e)$. %, a peculiar 6D \emph{nonlocal} quantum field theory (QFT) which enjoys a form of T-duality (inherited from its 10D parent) and a characteristic mass scale $M_\text{string}$, at which the description as a local QFT must break down. 
Rather conveniently to us, the heterotic string on a K3 can also be dualized to F-theory on an elliptically fibered Calabi--Yau complex threefold (CY), and this setup, in presence of small instantons (i.e. heterotic NS5-branes), was analyzed in detail long ago \cite{Aspinwall:1997ye}. The K3 is viewed as an elliptically fibered twofold, with a compact $\mathbb{P}^1$ base. Moreover, to decouple gravity, we let the volumes of the K3 (and CY) go to infinity. (See e.g. \cite{Aspinwall:1999xs} for the precise limit.) Namely, once the orbifold is added we are \emph{only} interested in the physics near the singularity $\mathbb{C}^2/\Gamma_G$ of the K3. 

\subsection{\texorpdfstring{6D electric quivers for the $(e')$ LSTs}{6D electric quivers for the (e') LSTs}}

The aforementioned T-duality is also present in the 6D version of the heterotic string, when one compactifies not on a circle but on a K3 as we just did. In fact, before taking the orbifold, one can study the dynamics of $N$ NS5-branes of the heterotic string. These play the role of small instantons \cite{Witten:1995gx}, and their 6D dynamics is captured by an LST whose generalized Lagrangian may be compactly written as
\begin{equation}
    \text{($o$):}\quad [SO(32)] \overset{\mathfrak{usp}(2N)}{0}[SU(2)]
\end{equation}
in the $Spin(32)/\mathbb{Z}_2$ case,\footnote{\ The 0 curve (which can never appear in the construction of 6D SCFTs) decorated by a $\mathfrak{usp}$ gauge algebra is necessarily a $\mathbb{P}^1$ with $\mathbb{C}\times \mathbb{P}^1$ normal bundle \cite{Bhardwaj:2015oru}.} whereas as
\begin{equation}
\label{eq:little1}
   \text{$(e)$:}\quad  [E_8]  \underbrace{1\,2\,2\cdots 2\,2\,1}_{N+1}\, [E_8]
\end{equation}
in the $E_8 \times E_8$ case.\footnote{\ This was already observed in \cite{Hanany:1997gh}.} In the above ``electric quivers'' we have used standard F-theory notation \cite{Aspinwall:1997ye,DelZotto:2014hpa} (see also the review \cite{Heckman:2018jxk}),\footnote{\ Briefly, $\overset{\mathfrak{g}}{n}$ denotes an algebraic curve ($\mathbb{P}^1$) in the base of F-theory with negative self-intersection $n$ and hosting a gauge algebra $\mathfrak{g}$; $[F]$ denotes a noncompact flavor curve, i.e. a base divisor hosting an algebra $\mathfrak{f}$. Adjacency of two curves means transversal intersection, unless otherwise stated.} and this is because both heterotic string setups can be mapped via S and T-dualities (as explained above) to a configuration of compact curves ($\mathbb{P}^1$'s) of negative self-intersection 0, 1 or 2 in 6D Type IIB with varying axiodilaton (with $\mathbb{C}^2$ internal space and seven-branes wrapped on the noncompact curves indicated by $[H]$, providing an $H$ flavor group). Not all compact curves may be simultaneously shrunk to a point; the size of the curve which remains finite sets the mass scale $M_\text{s}$ of the LST \cite{Bhardwaj:2015oru}. Let us focus on the $(e)$ theory. We can add the $\mathbb{C}^2/\Gamma_G$ orbifold to the heterotic string with $N$ small instantons (i.e. the four internal dimensions span a singular K3 surface), and turn \eqref{eq:little1} into the $(e')$ theory, with the following F-theory configuration of curves:\footnote{\ The quiver for $(o')$ can be found in \cite{Intriligator:1997kq}.}
\begin{equation}
\label{eq:littlee'}
 \text{$(e')$:}\quad [E_8] \,\underbrace{\overset{\mathfrak{g}}{1} \,  \overset{\mathfrak{g}}{2} \, \overset{\mathfrak{g}}{2} \,\cdots \overset{\mathfrak{g}}{2} \, \overset{\mathfrak{g}}{2} \, \overset{\mathfrak{g}}{1}}_{N+1}\, [E_8]\ .
\end{equation}%
This is \emph{not} the end of the story however, as the presence of the orbifold generically requires (in order to have a well-defined Weierstrass model in F-theory) further blowups in the base. %\footnote{\ Sometimes we must also add noncompact curves supporting a nontrivial fibration and intersecting transversally some of the compact ones, which correspond in Type IIB to flavor seven-branes with infinite worldvolume, i.e. flavors. This is necessary to insure gauge anomaly cancellation in the ensuing 6D electric quiver. The bases and fibration structures of all such F-theory models can be classified \cite{Bhardwaj:2015oru,DelZotto:2022ohj,DelZotto:2022xrh,DelZotto:2023ahf}.}  
These extra blow-up modes are interpreted in terms of 6D conformal matter after \cite{DelZotto:2014hpa}. For more general $(e')$ theories, characterized by an embedding (injective homomorphism) $\mu_\text{L,R}: \Gamma_G \to E_8$ (one per $E_8$ factor, left and right, of the $E_8\times E_8$ string) one obtains more general F-theory geometries, dictated by the commutant of the embedding in $E_8 \times E_8$. The latter have been determined in \cite{DelZotto:2022ohj,DelZotto:2022xrh} building upon \cite{Aspinwall:1997ye,DelZotto:2014hpa,Frey:2018vpw}, and have the structure
\begin{equation}
\xymatrix{    \Omega_1(F(\mu_\text{L}),G) \ar@{-}[r]^{G}& \mathcal T_{N-2}(G,G) \ar@{-}[r]^{G\,}&\Omega_1(F(\mu_\text{R}),G)}
\end{equation}
where $\Omega_1(F(\mu),G)$ is the theory of one orbi-instanton (i.e. one M5) with global symmetry $F(\mu) \times G$ corresponding to the embedding $\mu: \Gamma_G \to E_8$, $\mathcal T_{N-2}(G,G)$ is the $G$-type conformal matter corresponding to $N-2$ M5s probing a $G$-type singularity \cite{DelZotto:2014hpa}, and $ \xymatrix{\ar@{-}[r]^{G\,}&}$ denotes a fusion operation \cite{Heckman:2018pqx,DelZotto:2018tcj} on the corresponding 6D SCFTs replacing a global symmetry $G\times G$ with a gauge node $\overset{\mathfrak g}{n}$. See \cite{DelZotto:2023ahf} for a review of the resulting systems. An equivalent presentation of the above result is as follows
\begin{equation}
\xymatrix{    \Omega_{N_\text{L}}(F(\mu_\text{L}),G) \ar@{-}[r]^{G}& \Omega_{N_\text{R}}(F(\mu_\text{R}),G)}
\end{equation}
where we represent the system as the fusion of two higher orbi-instanton teories, with $N_\text{L} + N_\text{R} = N$ the total number of M5s in the dual Ho\v{r}ava-Witten setup. 

Let us specialize to the case $G=SU(k)$. (We will say a few words on the other cases in the outlook section at the end of the paper.) To fully specify the instanton configuration in the 6D heterotic string, on top of the instanton number $N$ we should also specify a nontrivial flat connection $F=0$ for the gauge group at the spatial infinity $S^3/\Gamma_G$ of the orbifold (since $\pi_1 (S^3/\Gamma_G) \neq 0$). This is given by a representation $\rho_\infty : \Gamma_G \to E_8$, i.e. the embedding $\mu_\text{L,R}$ we just introduced which encodes the F-theory configuration. For $G=SU(k)$ these embeddings can be conveniently classified in terms of so-called Kac labels \cite{kac1990infinite} (also known as Kac diagrams in the mathematics literature), i.e. integer partitions of the order $k$ of the orbifold in terms of the Coxeter labels $1,\ldots,6,4',3',2'$ of the \emph{affine} $E_8$ Dynkin:
\begin{equation}
\label{eq:kac}
    k=\left(\sum_{i=1}^6 i n_i\right) + 4 n_{4'} + 3 n_{3'} +2 n_{2'}\ ,
\end{equation}
which will be denoted $k=[1^{n_1},\ldots,6^{n_6},4^{n_{4'}},3^{n_{3'}},2^{n_{2'}}]$ (and we will also say that the $n_i,n_{i'}$ --some of which may be zero-- are the multiplicities of the parts of the Kac label). 

\begin{figure}
	\centering
\begin{tikzpicture}[scale=1,baseline]
	% NS5
	\node at (0,0) {};
	\draw[fill=black] (0.5,0) circle (0.075cm);
	\draw[fill=black] (1.25,0) circle (0.075cm);
	\draw[fill=black] (2,0) circle (0.075cm);
	\draw[fill=black] (2.75,0) circle (0.075cm);
	\draw[fill=black] (3.5,0) circle (0.075cm);
	\draw[fill=black] (4.25,0) circle (0.075cm);
 	\draw[fill=black] (5,0) circle (0.075cm);
	\draw[fill=black] (5.75,0) circle (0.075cm);
	\draw[fill=black] (6.5,0) circle (0.075cm);
        \draw[fill=black] (7.25,0) circle (0.075cm);
        \draw[fill=black] (8,0) circle (0.075cm);
        \draw[fill=black] (8.75,0) circle (0.075cm);
	
	%D6
	
	\draw[solid,black,thick] (0.5,0)--(1.25,0) node[black,midway,yshift=0.2cm] {\footnotesize $1$};
	\draw[solid,black,thick] (1.25,0)--(2,0) node[black,midway,yshift=0.2cm] {\footnotesize $2$};
	  \path (2,0)--(2.75,0) node[black,midway] {\footnotesize $\cdots$};
	\draw[solid,black,thick] (2.75,0)--(3.5,0) node[black,midway,yshift=0.2cm] {\footnotesize $k$};
	\draw[solid,black,thick] (3.5,0)--(4.25,0) node[black,midway,yshift=0.2cm] {\footnotesize $k$};
	\path (4.25,0)--(5,0) node[black,midway] {\footnotesize $\cdots$};
        \draw[solid,black,thick] (5,0)--(5.75,0) node[black,midway,yshift=0.2cm] {\footnotesize $k$};
        \draw[solid,black,thick] (5.75,0)--(6.5,0) node[black,midway,yshift=0.2cm] {\footnotesize $k$};
        \path (6.5,0)--(7.25,0) node[black,midway] {\footnotesize $\cdots$};
        \draw[solid,black,thick] (7.25,0)--(8,0) node[black,midway,yshift=0.2cm] {\footnotesize $2$};
	\draw[solid,black,thick] (8,0)--(8.75,0) node[black,midway,yshift=0.2cm] {\footnotesize $1$};

	%D8-O8
	\draw[dashed,black,very thick] (0,-.5)--(0,.5) node[black,midway, xshift =0cm, yshift=-1.5cm] {} node[black,midway, xshift =0cm, yshift=1.5cm] {};
	\draw[solid,black,very thick] (0.25,-.5)--(0.25,.5) node[black,midway, xshift =0cm, yshift=+.75cm] {\footnotesize $7$};
	\draw[solid,black,very thick] (3.5-0.2,-.5)--(3.5-0.2,.5) node[black,midway, xshift =0cm, yshift=+.75cm] {\footnotesize $1$};
        \draw[solid,black,very thick] (5.75+0.2,-.5)--(5.75+0.2,.5) node[black,midway, xshift =0cm, yshift=+.75cm] {\footnotesize $1$};
        \draw[solid,black,very thick] (9,-.5)--(9,.5) node[black,midway, xshift =0cm, yshift=+.75cm] {\footnotesize $7$};
        \draw[dashed,black,very thick] (9.25,-.5)--(9.25,.5) node[black,midway, xshift =0cm, yshift=-1.5cm] {} node[black,midway, xshift =0cm, yshift=1.5cm] {};
        
\end{tikzpicture}
\caption{Type I' engineering of \eqref{eq:little2}. A vertical dashed line represents an O8$^-$-plane, vertical solid lines represent D8's, horizontal solid lines represent D6's (with their number in the stack on top of the line), circles represent NS5s. The total D8 charge vanishes (as it should, in a compact space) because of the two negatively charged orientifolds.}
\label{fig:IIAlittle1}
\end{figure}
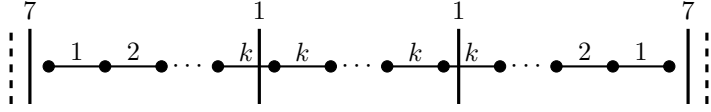

Each embedding preserves a subalgebra of $E_8$ determined via a simple algorithm:\footnote{\ In this paper we do not pay attention to the global structure of the flavor group, so that it can be identified with its Lie algebra.} one simply ``deletes'' all nodes with nonzero multiplicity $n_i,n_{i'}$ in this partition, and reads off the Dynkin of the leftover algebra, which may be a sum of nonabelian algebras, plus a bunch of $\mathfrak{u}(1)$'s to make the total rank eight. E.g. the trivial flat connection (embedding), which exists for any $k$, is given by the label $k=[1^k]$ and preserves the full $E_8$. In this case a further $k$ blowups are required in the middle of each of the two pairs $[E_8]\overset{\mathfrak{su}(k)}{1}$ in \eqref{eq:littlee'} (introducing each time a new 1 curve, decorated by $\mathfrak{su}(k-i)$, $i=1,\ldots,k$, and turning the ``old'' 1 into a 2), and the full electric quiver reads \cite{Aspinwall:1997ye}:
\begin{equation}
\label{eq:little2}
[E_8] \, \underbrace{\overset{\emptyset}{1}\, \overset{\mathfrak{su}(1)}{2}\, \overset{\mathfrak{su}(2)}{2} \cdots \overset{\mathfrak{su}(k-1)}{2}}_k \underbrace{\overset{\mathfrak{su}(k)}{\underset{[N_\text{f}=1]}{2}} \overset{\mathfrak{su}(k)}{2}\cdots \overset{\mathfrak{su}(k)}{2} \overset{\mathfrak{su}(k)}{\underset{[N_\text{f}=1]}{2}}}_{N+1} \underbrace{\overset{\mathfrak{su}(k-1)}{2} \cdots \overset{\mathfrak{su}(2)}{2}\, \overset{\mathfrak{su}(1)}{2} \, \overset{\emptyset}{1}}_{k} \,[E_8] \ .
\end{equation}
Equivalently, we may engineer this F-theory configuration in Type I'. We first go to M-theory on an interval \cite{Horava:1995qa,Horava:1996ma}. Each $E_8$ gauge group of the heterotic string (represented by a noncompact $E_8$ seven-brane in Type IIB) is engineered in M-theory by an M9-wall; each of the original $N$ instantons (NS5-branes) corresponds to an M5; the orbifold lifts to an equivalent orbifold probed by the M5's. We can now reduce the system to Type I': the M9 becomes an O8$^-$-plane plus eight D8's, each M5 reduces to an NS5, and the orbifold to $k$ D6's suspended between the NS5s. See figure \ref{fig:IIAlittle1}. Importantly, because of the $2k$ extra blowups ($k$ per ``tail'' in \eqref{eq:little2}) we have $2k$ \emph{new} NS5s in Type I', giving rise to ``fractional instantons''. In M-theory, they can be explained by considering that the M9-wall actually fractionates in presence of the orbifold.

\subsection{Fractional instantons and 6D SCFTs}

In the fully blown-up electric quiver \eqref{eq:little2} (which represents the generic point on the tensor branch of the LST), and using the heterotic/F-theory/Type I' dictionary (see again figure \ref{fig:IIAlittle1}), we see that we have a total of
\begin{equation}
\label{eq:MN}
    M \equiv (N+1 + k + k) -1 = N + k + k
\end{equation}
heterotic NS5s (small instantons). $N$ of them correspond to the ``full'' M5's originally present in the Ho\v{r}ava--Witten setup \eqref{eq:little1}, whereas the other $2k$ correspond to new \emph{fractional} instantons: the M9 in presence of the orbifold fractionates \cite{DelZotto:2014hpa,Hanany:1997gh}, and the number of fractions depends on the chosen Kac label. (In F-theory, these $k$ fractions correspond to $k$ blowups in the base.%\footnote{\ It is reasonable to expect that, in M-theory, this number comes from a torsional $C_3$ (perhaps \`a la \cite{deBoer:2001wca,Tachikawa:2015wka,Bhardwaj:2018jgp}) once the M9 is on top of the orbifold. For instance, it is known \cite{Witten:1996md,Diaconescu:2000wy} that $\int_U \frac{G_4}{2\pi} = \frac{1}{2}\int_U \lambda \mod \zz$ on any 4-cycle $U$ (such as $\cc^2/\Gamma_G$) inside the M-theory spacetime $Y$. Here $G_4=dC_3$ is the M-theory flux and $2\lambda=p_1(Y) = \frac{1}{8\pi^2} \Tr R \wedge R =w_4 \mod 2$, with $R$ the curvature two-form (i.e. the curvature of the spin connection $\omega$). The class ${\lambda}/{2}$ is always integral but may not be even.  Then if $a,b$ are the characteristic classes of the $E_8$ bundles of the heterotic string, to ensure gauge anomaly cancellation we must have $a+b=\lambda$. This is derived via circle compactification (i.e. $Y$ is an $S^1$-bundle). This picture may be modified further to accommodate the orbifold. We also know the Bianchi identity for $G_4$ gets a contribution from each of the M9's:
%\begin{equation}
%    dG_4= c(\kappa_{11}) \delta(x^{10})(\Tr F\wedge F - \Tr R \wedge R)
%\end{equation} 
%(which would read $dG_4=0$ in absence of sources), with $c(\kappa_{11})$ a constant in terms of the 11D gravitational coupling, and $F$ the $E_8$ gauge field strength localized at $x^{10}=0$ (and analogously for the other M9 at $x^{10}=L$). The integral of this equation along the interval should reflect the fractionalization of each of the M9's, yielding exactly $N_{\mu_\textsc{l}}+N_{\mu_\textsc{r}}$ fractions.}
) E.g. for $\mu_\text{L,R}=[1^k]$ we have $k$ new fractions. Let us call this number $N_{\mu_\textsc{l,r}}$ for a general choice of $\mu_\text{L,R}$.

We have two tails from $\mathfrak{su}(1)=\emptyset$ to $\mathfrak{su}(k)$ (this latter gauge algebra with one flavor), and $N+1$ ``central'' $\mathfrak{su}(k)$'s (i.e. a plateau in the gauge ranks). All 2 curves but one may be shrunk to a point; its size sets the LST scale $M_\text{s}$. In each of the two ``halves'' of the Type I' setup (left and right), $N_{\mu_\textsc{l,r}}=k$ also corresponds to the largest linking number $l_\text{L,R}$: for each of the 8 D8's this number is defined as the number of D6's ending on it from the right minus from the left plus the number of NS5s to the immediate left of it \cite{Cabrera:2019izd} (where for concreteness we have assumed the O8 sits on the left of each half, when considered individually). The linking numbers are read off after having brought all D8's close to the O8 via a series of simple Hanany--Witten moves (as done e.g. below in figure \ref{fig:orbi-iia}). Therefore in general we will have
\begin{equation}
    M = N + N_{\mu_\textsc{l}}+N_{\mu_\textsc{r}} = (N_\text{L}+N_{\mu_\textsc{l}})+(N_\text{R}+N_{\mu_\textsc{r}}) = M_\text{L}+M_\text{R}
\end{equation}
heterotic small instantons, or NS5-branes in the $(e')$ LST. Let us explain the meaning of this formula.

In \eqref{eq:little2} we recognize the same quiver as in \eqref{eq:first3D}, with $M-2k+1=N+1$ in light of \eqref{eq:MN}. This is not a coincidence, as the latter is the $T^3$ compactification of the former, as mentioned in the introduction. Now consider the two halves of the Type I' setup. We may split the the number $N+1$ of 2 curves into $N_\text{L}+N_\text{R}+1$ arbitrarily (i.e. the pleateaux of the two halves need not be of the same length). Each of the two halves provides the Type IIA engineering of a 6D $(1,0)$ SCFT (rather than an LST) on the tensor branch, which is known as A-type orbi-instanton \cite{DelZotto:2014hpa,Fazzi:2022hal}. The instantonic NS5s contribute tensor multiplets; at strong coupling the $N_\text{L,R}+N_{\mu_\textsc{l,r}}$ (i.e. full plus fractional) NS5s are on top of each other and get absorbed into the O8-D8 wall. The NS5s can then move freely along this wall, thereby liberating a 6D HB. We should keep track of this effect in the QFT: at strong string coupling there is a phase transition whereby each tensor multiplet turns into twenty-nine hypermultiplets \cite{Ganor:1996mu}. In 6D this transition appears as we hit the origin of the tensor branch, so that the orbi-instanton HB dimension\footnote{\ At finite coupling, i.e. classically, the quaternionic dimension of the HB in any spacetime dimension with eight supercharges can easily be computed as the total number of hypermultiplets minus that of vector multiplets.} at ``infinite (gauge) coupling'' \cite{Cremonesi:2015lsa} (i.e. in the SCFT) is 
\begin{equation}
\dim_\mathbb{H} \text{HB}_{\text{6D},\mu_\text{L,R}}^\infty = \dim_\mathbb{H} \text{HB}_{\text{6D},\mu_\text{L,R}} + 29(N_\text{L,R}+N_{\mu_\textsc{l,r}})
\end{equation}
if there are $N_\text{L,R}+N_{\mu_\textsc{l,r}}$ (dynamical) tensor multiplet scalars whose VEVs can be simultaneously tuned to zero. This should also correspond to a ``jump'' in the dimension of the 3D HB and CB (compactifying on a $T^3$ and taking the mirror, respectively). In the simplest case of $\mu_\text{L}=\mu_\text{R}=[1^k]$ where $N_{\mu_\textsc{l,r}}=k$ (and $N=N_\text{L}+N_\text{R})$, and gluing the two halves into an LST setup, we predict that
\begin{equation}
\label{eq:dimensions}
    \dim_\mathbb{H}\text{HB}_\text{6D}\eqref{eq:first6D} = \dim_\mathbb{H}\text{CB}_\text{3D}\text{\eqref{eq:1kCBtrick}} = M+k-1 \underset{M=N+2k}{=} N+3k-1 = \dim_\mathbb{H}\text{HB}_\text{6D}\eqref{eq:little2}
\end{equation}
should ``jump'' to
\begin{equation}
\label{eq:diminf}
\dim_\mathbb{H}\text{HB}_\text{6D}^\infty = \dim_\mathbb{H}\text{CB}_\text{3D}^\infty = 29M+M+k-1 \underset{M=N+2k}{=} 30N+61k-1 = \dim_\mathbb{H}\text{HB}_\text{6D}^{M_\text{s}}\ .
\end{equation}
Here $\text{CB}_\text{3D}^\infty$ stands for the CB of a \emph{new} 3D theory capturing $\text{HB}_\text{6D}^{M_\text{s}}$. By the latter we mean the HB of the LST at energies of order $M_\text{s}$ or higher. (We will sometimes say that the LST is at infinite coupling, in the sense just explained.) As we said earlier, this is the size of the 2 curve that remains compact in the F-theory picture, while all other 2's are shrunk to a point (i.e. all NS5s are absorbed into the O8's). Equivalently, $M_\text{s}^2=1/g_\text{YM}^2$ is the finite gauge coupling of the LST, and the distance between two consecutive NS5s (all other distances being zero) -- see again figure \ref{fig:IIAlittle1}.

Going back to the description of the two halves as orbi-instantons, i.e. 
\begin{equation}
\label{eq:orbi-trivial}
    [E_8] \underbrace{\overset{\emptyset}{1} \,\overset{\mathfrak{su}(1)}{2}\, \overset{\mathfrak{su}(2)}{2} \cdots \overset{\mathfrak{su}(k-1)}{2}}_{N_{\mu_\textsc{l,r}}=k} \underbrace{\overset{\mathfrak{su}(k)}{\underset{[N_\text{f}=1]}{2}}\overset{\mathfrak{su}(k)}{2}\cdots \overset{\mathfrak{su}(k)}{2} [SU(k)]}_{N_\text{L,R}+1}
\end{equation}
in F-theory (see e.g. \cite{Fazzi:2022hal} for more details), we see that gluing two orbi-instantons to create an LST means gauging together the flavor $[SU(k)]$ at the end of their respective plateaux of length $N_\text{L,R}+1$ (see \eqref{eq:little2}). This is the new $\mathfrak{su}(k)$ at finite coupling $1/g_\text{YM}^2$ that sets the LST mass scale. The above electric quiver is the low-energy description (i.e. quiver gauge theory plus tensors) of the UV SCFT that resides at infinite coupling (i.e. at the origin of the tensor branch). Luckily, we already have a description of the latter's HB$^\infty_\text{6D}$ as the CB of a magnetic quiver, CB$^\infty_\text{3D}$, which will make its appearance in section \ref{sec:quivs}.

\subsection{General electric quivers}

More generally, for $G=SU(k)$ the orbi-instanton has an electric quiver given by
\begin{equation}\label{eq:genelequiv}
    [F_\text{L,R}] \underbrace{\overset{\mathfrak{g}_\text{L,R}}{1} \, \overset{\mathfrak{su}(m_1)}{2}\, \overset{\mathfrak{su}(m_2)}{2} \cdots \overset{\mathfrak{su}(m_{N_{\mu_\textsc{l,r}}-1})}{2}}_{\max\left(N_{\mu_\textsc{l,r}},1\right)} \underbrace{\overset{\mathfrak{su}(k)}{\underset{[N_\text{f}=k-m_{N_{\mu_\textsc{l,r}}-1}]}{2}}\overset{\mathfrak{su}(k)}{2}\cdots \overset{\mathfrak{su}(k)}{2} [SU(k)]}_{N_\text{L,R}+1}\ ,
\end{equation}
where $[F_\text{L,R}]$ is (the nonabelian part of) a maximal subalgebra of $E_8$,\footnote{\ For the abelian factors needed to make the total rank 8, see \cite{Fazzi:2022hal}.} and
where $\mathfrak{g}_\text{L,R}$ is one among $\{\emptyset,\mathfrak{usp}(m_0),\mathfrak{su}(m_0)\}$. In the last case we also have one (half) hypermultiplet in the two-index (three-index) antisymmetric representation of $\mathfrak{su}(m_0)$ for all $m_0\neq 6$ ($m_0=6$). All ranks are determined by the chosen $\mu_\text{L,R}$. The algorithm to determine from the Kac label the full electric quiver (i.e. including matter representations, which we have mostly omitted, except for the fundamentals at the beginning of the plateau) can be found in \cite{Mekareeya:2017jgc}. 

In particular, the number $N_{\mu_\textsc{l,r}}$ is given by \cite{Fazzi:2023ulb}
\begin{equation}\label{eq:Nmu}
    N_{\mu_\textsc{l,r}}=\sum_{i=1}^6 n_i^\textsc{l,r}+p_\text{L,R}\ , \quad p_\text{L,R}=\min\left(\left\lfloor \frac{n_{3'}^\textsc{l,r}+n_{4'}^\textsc{l,r}}{2} \right\rfloor, \left\lfloor \frac{n_{2'}^\textsc{l,r}+n_{3'}^\textsc{l,r}+2n_{4'}^\textsc{l,r}}{3} \right\rfloor\right)\ ,
\end{equation}
and is identical to the total number of unprimed parts in a Kac label when it does not contain any primes.\footnote{\ Notice that our $p_\text{L,R}$ also appears in \cite{Cabrera:2019izd} with the same name, and in \cite{Mekareeya:2017jgc} denotes the difference between $N_S$ and $N_6$ in their five-case classification of electric quivers.} For some primes-only labels it may still happen that $p_\text{L,R}=0$ (e.g. for $[4']$);\footnote{\ $k=7$ is the first case where we can have a nonzero $p$.} then $N_{\mu_\textsc{l,r}}=0$ but $\max\left(N_{\mu_\textsc{l,r}},1\right)=1$, and the only surviving curve is the leftmost $\overset{\mathfrak{g}_\text{L,R}}{1}$, which is a remnant of $\overset{\mathfrak{su}(k)}{1}$ in \eqref{eq:littlee'}. In other words, the F-theory configuration does not require any extra blowups in this case.

In light of the above, the general $(\mu_\text{L},\mu_\text{R})$ LST will have an electric quiver given by
\begin{equation}\label{eq:genelequivLST}
{\scriptstyle [F_\text{L}] \,\underbrace{\overset{\scriptstyle \mathfrak{g}_\text{L}}{\scriptstyle 1} \, \overset{\scriptstyle \mathfrak{su}(m_1)}{\scriptstyle 2}\, \overset{\scriptstyle \mathfrak{su}(m_2)}{\scriptstyle 2} \scriptstyle \cdots \overset{\scriptstyle\mathfrak{su}(m_{\scriptscriptstyle N_{\mu_\textsc{l}}-1})}{\scriptstyle 2}}_{\scriptstyle \max\left(N_{\mu_\textsc{l}},1\right)} \underbrace{\overset{\scriptstyle \mathfrak{su}(k)}{\underset{\scriptscriptstyle [N_\text{f}=k-m_{\scriptscriptstyle N_{\mu_\textsc{l}}-1}]}{\scriptstyle 2}}  \overset{\scriptstyle \mathfrak{su}(k)}{\scriptstyle 2}\scriptstyle\cdots \overset{\scriptstyle\mathfrak{su}(k)}{\scriptstyle 2} \overset{\scriptstyle\mathfrak{su}(k)}{\underset{\scriptscriptstyle[N_\text{f}=k-\ell_{\scriptscriptstyle N_{\mu_\textsc{r}}-1}]}{\scriptstyle2}}}_{N_\text{L}+N_\text{R}+1} \underbrace{\overset{\scriptstyle \mathfrak{su}(\ell_{N_{\mu_\textsc{r}}-1})}{\scriptstyle 2}\scriptstyle \cdots \overset{\scriptstyle \mathfrak{su}(\ell_2)}{\scriptstyle2} \, \overset{\scriptstyle\mathfrak{su}(\ell_1)}{\scriptstyle2} \, \overset{\scriptstyle\mathfrak{g}_\text{R}}{\scriptstyle1}}_{\scriptstyle \max\left(N_{\mu_\textsc{r}},1\right)} \,[\scriptstyle F_\text{R}]}
    \ ,
\end{equation}
having identified (i.e. gauged a diagonal subgroup of) the two $[SU(k)]$ factors in \eqref{eq:genelequiv}. This generalizes \eqref{eq:little2}; all possibilities have been classified in \cite{Bhardwaj:2015xxa}.

The ``minimal'' choice with $N=0$ deserves some attention. In this case the LST has only fractional instantons, exactly $M=M_\text{L}+M_\text{R}= N_{\mu_\textsc{l}}+N_{\mu_\textsc{r}}$ of them, which are created by the fractionalization of the two M9's against the orbifold. E.g. for $k=2$ we have the electric quivers \cite{DelZotto:2022ohj}
\begin{align}
    &([1^2],[1^2]),\ N=0,\ N_{\mu_\textsc{l}}=N_{\mu_\textsc{r}}=l_\text{L,R}=2: && [E_8] \, \overset{\emptyset}{1}\, \overset{\mathfrak{su}(1)}{2} \overset{\mathfrak{su}(2)}{\underset{[N_\text{f}=2]}{2}} \overset{\mathfrak{su}(1)}{2}\, \overset{\emptyset}{1}\,[E_8] \ ,\\
    &([2],[2]),\ N=0,\ N_{\mu_\textsc{l}}=N_{\mu_\textsc{r}}=l_\text{L,R}=1: && [E_7] \, \overset{\emptyset}{1}\, \overset{\mathfrak{su}(2)}{\underset{[N_\text{f}=4]}{2}} \, \overset{\emptyset}{1}\,[E_7] \ ,
\end{align}
both with a plateau of only one $\mathfrak{su}(2)$. However notice that
\begin{equation}\label{eq:usp}
    ([2'],[2']),\ N=-1,\ N_{\mu_\textsc{l}}=N_{\mu_\textsc{r}}=l_\text{L,R}=0: \quad [SO(16)] \, \overset{\mathfrak{usp}(2)}{1}\,  \overset{\mathfrak{usp}(2)}{1} \,[SO(16)]
\end{equation}
is a gauge anomaly-free electric quiver as is (corresponding to zero full or fractional instantons). In the notation of \eqref{eq:genelequivLST}, this is equivalent to formally continuing $N$ to $-1$. This is because the $+1$ in the $N_\text{L}+N_\text{R}+1$-long plateau of the \emph{generic} LST \eqref{eq:genelequivLST} comes from fusing the two $[SU(k)]$'s from left and right orbi-instantons (i.e. gauging a diagonal subgroup). However sometimes we may be able to build anomaly-free LSTs even without a central plateau, just as in \eqref{eq:usp}.

%%%%%%%%%%%%%%%%%%%%%%%%%%%%%%%%%%%%%%%%%%%%%%%%%%%%%%%%%%%
\section{3D magnetic quivers}
\label{sec:quivs}
%%%%%%%%%%%%%%%%%%%%%%%%%%%%%%%%%%%%%%%%%%%%%%%%%%%%%%%%%%%

We are now ready to formulate our proposal for the HB of the LST. As we just explained, the latter is obtained by gluing two A-type orbi-instantons. The electric quiver of each is engineered by an NS5-D6-D8-O8$^-$ configuration (half of the Type I' setup), and is obtained by reading off the massless (electric) degrees of freedom obtained by stretching F1's between D6's when the latter are suspended between NS5s. The HB of the orbi-instanton at a generic point on the tensor branch (i.e. when the SCFT is approximated by a quiver as in \eqref{eq:orbi-trivial}) is captured by the CB of a 3D $\mathcal{N}=4$ quiver gauge theory colloquially known as magnetic quiver \cite{Cabrera:2019izd}. In this case the massless degrees of freedom are provided by D4's stretched between D6's, NS5s, or D6-NS5s in a phase where the D6's are suspended between the D8's and the NS5s are ``lifted off'' of the D6's. 

The magnetic quiver is star-shaped,\footnote{\ It made its first appearance in \cite{Benini:2010uu}, even though this name was not adopted at the time.} and is obtained by gluing a $T(SU(k))$ tail \cite{Gaiotto:2008ak}, 
\begin{equation}
    1-2-\cdots-(k-1)-\boxed{k}\ ,
\end{equation}
to another quiver of affine $E_8$ Dynkin shape (which we will call $E_8^{(1)}$ following \cite{kac1990infinite}) along the $\boxed{k}$ node. Moreover, there is a ``bouquet'' of $1$'s attached to $k$, representing the NS5s suspended over the D6's. The shape of the generic magnetic quiver is thus \cite{Cabrera:2019izd}:
\begin{equation}\label{eq:genericmagquiv}
 {\displaystyle 
 1 - 2 - \cdots - (k-1) - \hspace{-.8cm}\overset{\overset{\displaystyle \overbrace{\displaystyle 1 \cdots 1}^{N_\text{L,R}+\sum_{i=1}^6 n_i}}{ \rotatebox[origin=c]{190}{$\setminus$}\ \rotatebox[origin=c]{-190}{$/$} }}{k} \hspace{-.8cm} - r_1 -r_2 -r_3 - r_4 -r_5-\overset{\overset{\displaystyle r_{3'}}{\vert}}{r_6}-r_{4'}-r_{2'}
 }
\end{equation}
or, equivalently,
\begin{equation}\label{eq:genericmagquivmod}
 {\scriptstyle 
 1 - 2 - \cdots - (k-1) - \hspace{-1cm}\overset{\overset{\scriptstyle \overbrace{\scriptstyle 1 \ \cdots \ 1}^{M_\text{L,R}=N_\text{L,R}+N_{\mu_\textsc{l,r}}}}{ \rotatebox[origin=c]{180}{$\setminus$}\ \rotatebox[origin=c]{-180}{$/$} }}{k} \hspace{-1cm} - (r_1-p) -(r_2-2p) -(r_3-3p) - (r_4-4p) -(r_5-5p)-\overset{\overset{\scriptstyle (r_{3'}-3p)}{\vert}}{(r_6-6p)}-(r_{4'}-4p)-(r_{2'}-2p)
 }\ ,
\end{equation}
remembering the definition in \eqref{eq:Nmu}. (In the above formula we have omitted the $\text{L,R}$ subscripts (on $n_i,r_i, r_{i'},p$) to avoid clutter.) Let us also call
\begin{equation}\label{eq:twoMs}
    \widetilde{M}_\text{L,R}=M_\text{L,R}-p_\text{L,R}=N_\text{L,R}+\sum_{i=1}^6 n_i^\textsc{l,r}\ .
\end{equation}
Of course, $\widetilde{M}_\text{L,R}=M_\text{L,R}$ for Kac labels for which $p=0$. 

The ranks $r_i, r_{i'}$ of the $U$ gauge groups along the $E_8^{(1)}$ tail (some of which may be zero) are determined by the specific Kac label chosen to determine the embedding $\mu_\text{L,R}: \mathbb{Z}_k \to E_8$. Concretely, for both left and right orbi-instanton:
\begin{equation}\label{eq:rj}
    r_j = (1-\delta_{j6})\sum_{i=1}^{6-j} i n_{i+j} + 2n_{2'}+3n_{3'}+4n_{4'} =k-\sum_{i=1}^6 i n_i + (1-\delta_{j6})\sum_{i=1}^{6-j} i n_{i+j}
\end{equation}
for $j=1,\ldots,6$, and
\begin{subequations}\label{eq:rjprime}
\begin{align}
& r_{2'}= n_{3'}+n_{4'}\ , \\
& r_{3'} = n_{2'}+n_{3'}+2n_{4'}\ , \\
& r_{4'} = n_{2'}+2n_{3'}+2n_{4'}\ .
\end{align}
\end{subequations}
Going to the origin of the tensor branch requires computing the infinite-coupling HB of the orbi-instanton. As we explained above, this is achieved via $M_\text{L,R}$ small $E_8$ instanton transitions which simply ``add'' $M_\text{L,R}$ times an $E_8^{(1)}$ Dynkin to the right tail of the magnetic quiver.\footnote{\ This quiver addition may be thought of as the reverse of ``quiver subtraction'' \cite{Cabrera:2018ann,Gledhill:2021cbe,Bourget:2021siw}.} Using the simpler version of the latter in \eqref{eq:genericmagquiv}, we get:
\begin{equation}\label{eq:genericmagquivinf}
 {\scriptscriptstyle 
 1 - 2 - \cdots - k - (r_1+\widetilde M_\text{L,R}) -(r_2+2\widetilde M_\text{L,R}) -(r_3+3\widetilde M_\text{L,R}) - (r_4+4\widetilde M_\text{L,R}) -(r_5+5\widetilde M_\text{L,R})-\overset{\overset{\scriptscriptstyle r_{3'}+3\widetilde M_\text{L,R}}{\vert}}{(r_6+6\widetilde M_\text{L,R})}-(r_{4'}+4\widetilde M_\text{L,R})-(r_{2'}+2\widetilde M_\text{L,R})}\ .
\end{equation}
Computing the CB dimension of the above quiver (and substituting \eqref{eq:rj}-\eqref{eq:rjprime}-\eqref{eq:twoMs}) yields
\begin{equation}\label{eq:dimmodspE8}
    \dim_\mathbb{H}\text{CB}_\text{3D}^\infty \eqref{eq:genericmagquivinf} = 30 (N_\text{L,R}+k) +\frac{k}{2}(k+1) -\left\langle \bm{w}_\text{L,R},\bm{\rho} \right\rangle-1\ ,\footnotemark
\end{equation}
\footnotetext{\ The dimension of the moduli space of $E_8$ instantons on the deformation/resolution of $\mathbb{C}^2/\mathbb{Z}_k$ reads instead $\dim_\mathbb{H}  = 30 (N_\text{L,R}+k)  -\left\langle \bm{w}_\text{L,R},\bm{\rho} \right\rangle$ \cite{Mekareeya:2017jgc}.} just as predicted in \cite{Mekareeya:2017jgc},
where $\left\langle \bm{w}_\text{L,R},\bm{\rho} \right\rangle$ is the so-called height pairing in $E_8$:
\begin{equation}\label{eq:height}
    \left\langle \bm{w}_\text{L,R},\bm{\rho} \right\rangle  =  29 n_2+57 n_3+84 n_4+110 n_5+135 n_6+46 n_{2'}+68 n_{3'}+91 n_{4'}\ .
\end{equation}
As a final note, one may expect the electric and magnetic quivers to be related. This is indeed the case, as one may obtain the latter by taking three T-dualities and an S-duality in the NS5-D6-D8-O8$^-$ setup (along directions spanned by all branes) ``probed'' by F1-strings which engineers the former. At the QFT level, the magnetic quiver is the mirror dual to the $T^3$ compactification of the electric one, as mentioned multiple times by now.

\subsection{\texorpdfstring{$\mu=[1^k]$ SCFT}{mu=[1k] SCFT}}

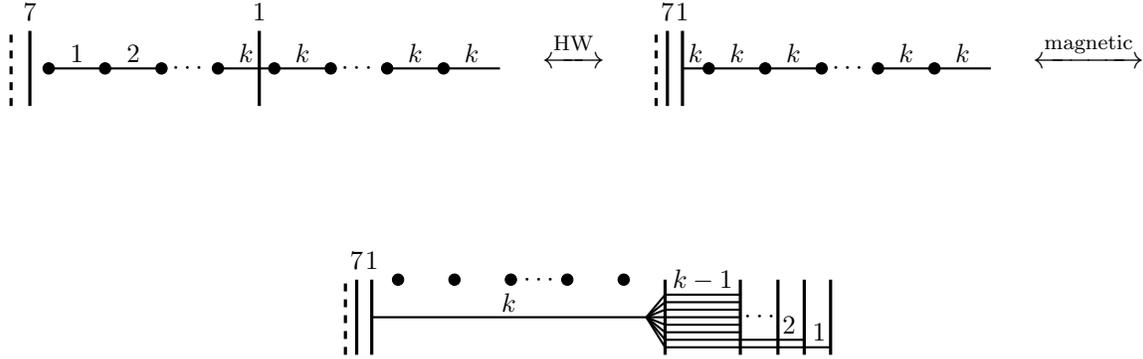
\begin{figure}
	\centering
\begin{tikzpicture}[scale=1,baseline]
	% NS5
	\node at (0,0) {};
	\draw[fill=black] (0.5,0) circle (0.075cm);
	\draw[fill=black] (1.25,0) circle (0.075cm);
	\draw[fill=black] (2,0) circle (0.075cm);
	\draw[fill=black] (2.75,0) circle (0.075cm);
	\draw[fill=black] (3.5,0) circle (0.075cm);
	\draw[fill=black] (4.25,0) circle (0.075cm);
	\draw[fill=black] (5,0) circle (0.075cm);
	\draw[fill=black] (5.75,0) circle (0.075cm);
	%\draw[fill=black] (8.5,0) circle (0.075cm);
	
	%D6
	
	\draw[solid,black,thick] (0.5,0)--(1.25,0) node[black,midway,yshift=0.2cm] {\footnotesize $1$};
	\draw[solid,black,thick] (1.25,0)--(2,0) node[black,midway,yshift=0.2cm] {\footnotesize $2$};
        \path (2,0)--(2.75,0) node[black,midway] {\footnotesize $\cdots$};
	\draw[solid,black,thick] (2.75,0)--(3.5,0) node[black,midway,yshift=0.2cm] {\footnotesize $k$};
	\draw[solid,black,thick] (3.5,0)--(4.25,0) node[black,midway,yshift=0.2cm] {\footnotesize $k$};
        \path (4.25,0)--(5,0) node[black,midway] {\footnotesize $\cdots$};
        \draw[solid,black,thick] (5,0)--(5.75,0) node[black,midway,yshift=0.2cm] {\footnotesize $k$};
        \draw[solid,black,thick] (5.75,0)--(6.5,0) node[black,midway,yshift=0.2cm] {\footnotesize $k$};
 
	%D8-O8
	\draw[dashed,black,very thick] (0,-.5)--(0,.5) node[black,midway, xshift =0cm, yshift=-1.5cm] {} node[black,midway, xshift =0cm, yshift=1.5cm] {};
	\draw[solid,black,very thick] (0.25,-.5)--(0.25,.5) node[black,midway, xshift =0cm, yshift=+.75cm] {\footnotesize $7$};
	\draw[solid,black,very thick] (3.5-0.2,-.5)--(3.5-0.2,.5) node[black,midway, xshift =0cm, yshift=+.75cm] {\footnotesize $1$};
\end{tikzpicture}
%	\hspace*{2cm}
$\quad \xleftrightarrow{\text{HW}}\quad$
	%\vspace*{.5cm}
 \begin{tikzpicture}[scale=1,baseline]
	% NS5
	\node at (0,0) {};
	\draw[fill=black] (0.5+.2,0) circle (0.075cm);
	\draw[fill=black] (1.25+.2,0) circle (0.075cm);
	\draw[fill=black] (2+.2,0) circle (0.075cm);
	\draw[fill=black] (2.75+.2,0) circle (0.075cm);
	\draw[fill=black] (3.5+.2,0) circle (0.075cm);
	%\draw[fill=black] (5.5,0) circle (0.075cm);
	%\draw[fill=black] (6.5,0) circle (0.075cm);
	%\draw[fill=black] (7.5,0) circle (0.075cm);
	%\draw[fill=black] (8.5,0) circle (0.075cm);
	
	%D6
	
	\draw[solid,black,thick] (0.35,0)--(0.5+.2,0) node[black,midway,yshift=0.2cm] {\footnotesize $k$};
	\draw[solid,black,thick] (0.5+.2,0)--(1.25+.2,0) node[black,midway,yshift=0.2cm] {\footnotesize $k$};
	\draw[solid,black,thick] (1.25+.2,0)--(2+.2,0) node[black,midway,yshift=0.2cm] {\footnotesize $k$};
        \path (2+.2,0)--(2.75+.2,0) node[black,midway] {\footnotesize $\cdots$};	
        %\draw[dashed,black,thick] (2+.2,0)--(2.75+.2,0);% node[black,midway,yshift=0.2cm] {\footnotesize $k$};
	\draw[solid,black,thick] (2.75+.2,0)--(3.5+.2,0) node[black,midway,yshift=0.2cm] {\footnotesize $k$};
        \draw[solid,black,thick] (3.5+.2,0)--(4.25+.2,0) node[black,midway,yshift=0.2cm] {\footnotesize $k$};
	
	%D8-O8
	\draw[dashed,black,very thick] (0,-.5)--(0,.5) node[black,midway, xshift =0cm, yshift=-1.5cm] {} node[black,midway, xshift =0cm, yshift=1.5cm] {};
	\draw[solid,black,very thick] (0.15,-.5)--(0.15,.5) node[black,midway, xshift =0cm, yshift=+.75cm] {\footnotesize $7$};
	\draw[solid,black,very thick] (0.35,-.5)--(0.35,.5) node[black,midway, xshift =0cm, yshift=+.75cm] {\footnotesize $1$};
\end{tikzpicture}
%	\hspace*{2cm}
$\quad \xleftrightarrow{\text{magnetic}}\quad$
	%\vspace*{.5cm}
 \begin{tikzpicture}[scale=1,baseline]
	% NS5
	\node at (0,0) {};
	\draw[fill=black] (0.5+.2,0.5) circle (0.075cm);
	\draw[fill=black] (1.25+.2,0.5) circle (0.075cm);
	\draw[fill=black] (2+.2,0.5) circle (0.075cm);
	\draw[fill=black] (2.75+.2,0.5) circle (0.075cm);
	\draw[fill=black] (3.5+.2,0.5) circle (0.075cm);
	
	%D6
	
	\draw[solid,black,thick] (0.35,0)--(4,0) node[black,midway,yshift=0.2cm] {\footnotesize $k$};
	%\draw[solid,black,thick] (0.75,0)--(1.25,0) node[black,midway,yshift=0.2cm] {\footnotesize $k$};
	%\draw[solid,black,thick] (1.25,0)--(2,0) node[black,midway,yshift=0.2cm] {\footnotesize $k$};
	\path (2+0.225,0.5)--(2.75+0.225,0.5) node[black,midway] {\footnotesize $\cdots$};
	%\draw[solid,black,thick] (2.75,0)--(3,0) node[black,midway,yshift=0.2cm] {\footnotesize $k$};
        %\draw[solid,black,thick] (3,0)--(3.75,0) node[black,midway,yshift=0.2cm] {\footnotesize $k$};
        
        \draw[solid,black,thick] (4.25,-0.3)--(6.1,-0.3) node[black,xshift=-0.2cm,yshift=0.2cm] {\footnotesize $2$};
        \draw[solid,black,thick] (4.25,-0.4)--(6.45,-0.4) node[black,xshift=-0.15cm,yshift=0.2cm] {\footnotesize $1$};
        \draw[solid,black,thick] (4.25,-0.2)--(5.25,-0.2); %node[black,xshift=-0.125cm,yshift=0.2cm] {\footnotesize $1$};
        \draw[solid,black,thick] (4.25,-0.1)--(5.25,-0.1); %node[black,xshift=-0.125cm,yshift=0.2cm] {\footnotesize $1$};
        \draw[solid,black,thick] (4.25,-0)--(5.25,-0); %node[black,xshift=-0.125cm,yshift=0.2cm] {\footnotesize $1$};
        \draw[solid,black,thick] (4.25,0.1)--(5.25,0.1); %node[black,xshift=-0.125cm,yshift=0.2cm] {\footnotesize $1$};
        \draw[solid,black,thick] (4.25,0.2)--(5.25,0.2); %node[black,midway,yshift=0.2cm] {\footnotesize $k-1$};
        \draw[solid,black,thick] (4.25,0.3)--(5.25,0.3) node[black,midway,yshift=0.2cm] {\footnotesize $k-1$};
        \path (4.25-0.05,0)--(5.75-0.05,0) node[black,xshift=-.165cm] {\footnotesize $\cdots$};
        
	%D8-O8
	\draw[dashed,black,very thick] (0,-.5)--(0,.5) node[black,midway, xshift =0cm, yshift=-1.5cm] {} node[black,midway, xshift =0cm, yshift=1.5cm] {};
	\draw[solid,black,very thick] (0.15,-.5)--(0.15,.5) node[black,midway, xshift =0cm, yshift=+.75cm] {\footnotesize $7$};
	\draw[solid,black,very thick] (0.35,-.5)--(0.35,.5) node[black,midway, xshift =0cm, yshift=+.75cm] {\footnotesize $1$};
        \draw[solid,black,very thick] (4.25,-.5)--(4.25,.5) node[black,midway, xshift =0cm, yshift=+.75cm] {};
        \draw[solid,black,very thick] (5.25,-.5)--(5.25,.5) node[black,midway, xshift =0cm, yshift=+.75cm] {};
        \draw[solid,black,very thick] (5.75,-.5)--(5.75,.5) node[black,midway, xshift =0cm, yshift=+.75cm] {};
        \draw[solid,black,very thick] (6.1,-.5)--(6.1,.5) node[black,midway, xshift =0cm, yshift=+.75cm] {};
        \draw[solid,black,very thick] (6.45,-.5)--(6.45,.5) node[black,midway, xshift =0cm, yshift=+.75cm] {};

        \draw[solid,black,thick] (4,0)--(4.25,-0.4);
        \draw[solid,black,thick] (4,0)--(4.25,-0.3);
        \draw[solid,black,thick] (4,0)--(4.25,-0.2);
        \draw[solid,black,thick] (4,0)--(4.25,-0.1);
        \draw[solid,black,thick] (4,0)--(4.25,0);
        \draw[solid,black,thick] (4,0)--(4.25,0.1);
        \draw[solid,black,thick] (4,0)--(4.25,0.2);
        \draw[solid,black,thick] (4,0)--(4.25,0.3);
\end{tikzpicture}

\caption{\textbf{\emph{Top:}} The Type IIA engineering of \eqref{eq:orbi-trivial}. \textbf{\emph{Middle}:} an equivalent configuration up to Hanany--Witten moves, i.e. a D8 in position $m$ is equivalent to a D8 in position 0 (close to the O8$^-$) with $m$ D6's ending on it. \textbf{\emph{Bottom:}} the magnetic phase, obtained by lifting all NS5s off of the D6's, and brining in $k$ D8's from the right infinity (i.e. having each semi-infinite D6 end on a separate D8).}
	\label{fig:orbi-iia}
\end{figure}

Consider for simplicity the orbi-instanton with electric quiver in \eqref{eq:orbi-trivial}. It is specified by Kac label $[1^k]$ and admits the Type IIA engineering of figure \ref{fig:orbi-iia}.
From the bottom frame we can directly read off the magnetic quiver in the phase where all NS5s are still separated:
\begin{equation}
\label{eq:orbi-simple-mag}
     1-2-\cdots- (k-1)-\hspace{-1.35cm} \overset{\overset{\displaystyle \overbrace{1 \cdots 1}^{\widetilde M_\text{L,R}=\widetilde{M}_\text{L,R}=N_\text{L,R}+k}}{ \rotatebox[origin=c]{190}{$\setminus$}\ \rotatebox[origin=c]{-190}{$/$} }}{k} \hspace{-1cm} .
\end{equation}
That is, all $r_i,r_{i'}$ in \eqref{eq:genericmagquiv} are zero in this case. The symmetry on the CB can be read off as follows \cite{Gaiotto:2008ak,Ferlito:2017xdq,Gledhill:2021cbe,Cabrera:2018uvz}. Separate the nodes between balanced, i.e. those for which $2N_\text{c}=N_\text{f}$, and unbalanced (those which are not balanced -- they could be either overbalanced, $2N_\text{c}< N_\text{f}$, or underbalanced, $2N_\text{c}>N_\text{f}$). The subset of the balanced nodes gives the Dynkin of the nonabelian part of the symmetry $G_J^\text{IR}$ on the CB at the IR fixed point. The number of unbalanced nodes minus one gives the number of $U(1)$'s in the abelian part of the symmetry. (There may be enhancements in the IR, so $G_J^\text{IR}$ is only the minimum symmetry we must have. Such an enhancement can be checked by computing the spectrum of 3D monopole operators \cite{Bashkirov:2010kz,Bashkirov:2010hj} or the superconformal index.) The quaternionic dimension of the CB is given by the total rank of the gauge group of the magnetic quiver minus one. 

For instance, for \eqref{eq:orbi-simple-mag} we have $G_J^\text{IR}=SU(k)\times U(1)^{N_\text{L,R}+k}$, since the $1-2-\cdots-(k-1)$ portion of the $T(SU(k))$ tail is balanced (while $U(k)$ as well as the collection of $U(1)$'s is generically overbalanced), and $\dim_\mathbb{H} \text{CB}_\text{3D} \eqref{eq:orbi-simple-mag}= k(k+1)/2+N_\text{L,R}+k-1$ (which is obviously integer for any $k$). When $N_\text{L,R}=0$ we have $G_J^\text{IR}=SU(k)\times U(1)^k$ and $\dim_\mathbb{H} \text{CB}_\text{3D}\eqref{eq:orbi-simple-mag}\vert_{N_\text{L,R}=0} = k(k+1)/2+k-1$. The infinite-coupling HB is found where all NS5s are coincident and brought on top of the O8; upon performing $k$ small $E_8$ instanton transitions, \eqref{eq:orbi-simple-mag} turns into\footnote{\ It has been conjectured \cite{Hanany:2018vph} that the CB of the magnetic quiver at infinite-coupling (as a hyperk\"ahler space) is obtained via a discrete gauging by $S_{M_\text{L,R}}$ of the finite-coupling CB. (This also reflects into an equivalent statement on the 6D HBs.) In the case of conformal matter of type $(A,A)$ (i.e. just bifundamentals) \cite{DelZotto:2014hpa}, this can also be confirmed via a gravity calculation \cite{Bergman:2020bvi}.}
\begin{equation}
\label{eq:1kinfcoup}
%\scriptstyle
    1-2-\cdots-(k-1)-k-k-2k-3k-4k-5k-\overset{\overset{\displaystyle 3k}{\vert}}{6k}-4k-2k\ .
\end{equation}
All nodes but the rightmost $U(k)$ in the left tail and the extending (i.e. leftmost) $U(k)$ node of $E_8^{(1)}$ are balanced, hence $G_J^\text{IR}=SU(k)\times E_8 \times U(1)$ (which coincides with the flavor symmetry in \eqref{eq:orbi-trivial} and is generically smaller than that at finite coupling),\footnote{\ See \cite{Apruzzi:2020eqi} for the M/F-theory origin of this $U(1)$ in the 6D SCFT. For the special case $k=2$, the $U(1)$ is known to enhance to $SU(2)$. The reduction of the symmetry on the CB passing from finite to infinite coupling has been linked to the discrete gauging of $S_k$ in \cite{Hanany:2018vph}.} and 
\begin{equation}
\dim_\mathbb{H} \text{CB}_\text{3D}^\infty\eqref{eq:1kinfcoup} = k (k + 1)/2 + 30 k -1 = \dim_\mathbb{H} \text{CB}_\text{3D}\eqref{eq:orbi-simple-mag}\vert_{N_\text{L,R}=0} +29k\ ,
\end{equation}
as expected.

\subsection{\texorpdfstring{$(\mu_\text{L},\mu_\text{R})=([1^k],[1^k])$ LST}{(muL,muR)=([1k,1k]) LST}}

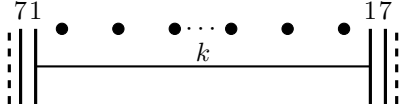
\begin{figure}[tpb]
	\centering
 \begin{tikzpicture}[scale=1,baseline]
	% NS5
	\node at (0,0) {};
	\draw[fill=black] (0.5+.2,0.5) circle (0.075cm);
	\draw[fill=black] (1.25+.2,0.5) circle (0.075cm);
	\draw[fill=black] (2+.2,0.5) circle (0.075cm);
	\draw[fill=black] (2.75+.2,0.5) circle (0.075cm);
	\draw[fill=black] (3.5+.2,0.5) circle (0.075cm);
        \draw[fill=black] (4.25+.2,0.5) circle (0.075cm);
    
	%D6
	
	\draw[solid,black,thick] (0.35,0)--(4.8,0) node[black,midway,yshift=0.2cm] {\footnotesize $k$};
	%\draw[solid,black,thick] (0.75,0)--(1.25,0) node[black,midway,yshift=0.2cm] {\footnotesize $k$};
	%\draw[solid,black,thick] (1.25,0)--(2,0) node[black,midway,yshift=0.2cm] {\footnotesize $k$};
	\path (2+0.2,0.5)--(2.75+0.2,0.5) node[black,midway] {\footnotesize $\cdots$};
	%\draw[solid,black,thick] (2.75,0)--(3,0) node[black,midway,yshift=0.2cm] {\footnotesize $k$};
        %\draw[solid,black,thick] (3,0)--(3.75,0) node[black,midway,yshift=0.2cm] {\footnotesize $k$};

	%D8-O8
	\draw[dashed,black,very thick] (0,-.5)--(0,.5) node[black,midway, xshift =0cm, yshift=-1.5cm] {} node[black,midway, xshift =0cm, yshift=1.5cm] {};
	\draw[solid,black,very thick] (0.15,-.5)--(0.15,.5) node[black,midway, xshift =0cm, yshift=+.75cm] {\footnotesize $7$};
	\draw[solid,black,very thick] (0.35,-.5)--(0.35,.5) node[black,midway, xshift =0cm, yshift=+.75cm] {\footnotesize $1$};
        \draw[solid,black,very thick] (4.8,-.5)--(4.8,.5) node[black,midway, xshift =0cm, yshift=+.75cm] {\footnotesize $1$};
        \draw[solid,black,very thick] (5,-.5)--(5,.5) node[black,midway, xshift =0cm, yshift=+.75cm] {\footnotesize $7$};
        \draw[dashed,black,very thick] (5.15,-.5)--(5.15,.5) node[black,midway, xshift =0cm, yshift=-1.5cm] {} node[black,midway, xshift =0cm, yshift=1.5cm] {};
\end{tikzpicture}

\caption{The magnetic phase of the configuration in figure \ref{fig:IIAlittle1}.}
	\label{fig:little1}
\end{figure}

We are now ready to derive the magnetic quiver for the simplest $(e')$ LST of type A (i.e. for $G=SU(k)$). We simply glue two orbi-instantons of type A specified by $\mu_\text{L,R}=[1^k]$ along their common $[SU(k)]$, as done in \eqref{eq:little2}. We now see the usefulness of figure \ref{fig:IIAlittle1}: we can easily read off the magnetic phase (see figure \ref{fig:little1}) and write down the magnetic quiver. 
When all NS5s are separated it is simply given by 
\begin{equation}
\label{eq:maglittle1}
    \overset{\overset{\displaystyle \overbrace{1 \cdots 1}^{\widetilde M_\text{L}+\widetilde M_\text{R}}}{ \rotatebox[origin=c]{190}{$\setminus$}\ \rotatebox[origin=c]{-190}{$/$} }}{k}\ ,
\end{equation}
with $\widetilde M_\text{L,R}=M_\text{L,R}=N_\text{L,R}+N_{\mu_\textsc{l,r}}=N_\text{L,R}+k$ and thus $M_\text{L}+M_\text{R}= N+2k$. The $U(k)$ node in 3D comes from the $T^3$ compactification of the 6D vector multiplet obtained by gauging the common $[SU(k)]$ of left and right orbi-instanton. Now every node is generically overbalanced, so $G_J^\text{IR}=U(1)^{M_\text{L}+M_\text{R}}$ and $\dim_\mathbb{H} \text{CB}_\text{3D}\eqref{eq:maglittle1}=M_\text{L}+M_\text{R}+k-1=N+3k-1$. This is nothing but \eqref{eq:1kCBtrick} with $M=M_\text{L}+M_\text{R}$, so the LST in the phase of separated NS5s has a HB captured by the CB of $U(k)$ with $M$ flavors, which as shown in \eqref{eq:dimensions} has dimension $N+3k-1$.

We can now explore the ``infinite-coupling'' limit of the LST (i.e. we probe the theory at energies of order $M_\text{s}$ or higher) for the choice $(\mu_\text{L},\mu_\text{R})=([1^k],[1^k])$ and determine the associated $\text{HB}_\text{6D}^{M_\text{s}}$ as the $\text{CB}_\text{3D}^\infty$ of a \emph{new} magnetic quiver, which is the first result of this paper. We simply need to perform $M_\text{L}+M_\text{R}$ instanton transitions, i.e. bringing the left $M_\text{L}$ NS5s on top of each other and onto the left O8$^-$-plane and repeating the same procedure for the right stack of $M_\text{R}$ NS5s. Doing so, we obtain the magnetic quiver
\begin{equation}
\label{eq:long1k1klst}
\scriptstyle  2M_\text{L}-4M_\text{L}-\overset{\overset{\scriptstyle 3M_\text{L}}{\vert}}{6M_\text{L}}-5M_\text{L}-4M_\text{L}-3M_\text{L}-2M_\text{L}-M_\text{L}-k-M_\text{R}-2M_\text{R}-3M_\text{R}-4M_\text{R}-5M_\text{R}-\overset{\overset{\scriptstyle 3M_\text{R}}{\vert}}{6M_\text{R}}-4M_\text{R}-2M_\text{R}\ ,
\end{equation}
which we will compactly write as 
\begin{equation}
\label{eq:short1k1klst}
M_\text{L}{E_8^{(1)}}^\vee \hspace{-0.1cm} - k - M_\text{R}E_8^{(1)}\ ,    
\end{equation}
where once again $E_8^{(1)}$ stands for the quiver of \emph{affine} $E_8$ Dynkin shape, with the ranks of the $U$ groups appearing therein being equal to the Coxeter labels, and ${E_8^{(1)}}^\vee$ is the Dynkin mirrored around the vertical axis, i.e. with the bifurcated tail on the left. Generically, all nodes in \eqref{eq:long1k1klst} but $U(k),U(M_\text{L}),U(M_\text{R})$ are balanced, producing $G_J^\text{IR} = E_8 \times E_8 \times U(1)^2$ as expected from the $E_8 \times E_8$ heterotic string on an A-type singularity with trivial flat connections at infinity (i.e. for $\mu_\text{L}=\mu_\text{R}=[1^k]$).\footnote{\ Here the two $U(1)$'s can be seen as arising from the rotation symmetry of probe M5's inside each of the two M9's in presence of the $\mathbb{C}^2/\mathbb{Z}_k$ orbifold, which preserves a $U(1) \subset SU(2)\times SU(2) = SO(4)$ at the level of Lie algebras.} Moreover $\dim_\mathbb{H} \text{CB}_\text{3D}^\infty \text{\eqref{eq:short1k1klst}} = 30(M_\text{L}+M_\text{R})+k-1=30(N+2k)+k-1$. This nicely matches our prediction in \eqref{eq:diminf}. We will see another application of this formula in \eqref{eq:dimLST1k1k}.

\subsection{General rule}
\label{sub:rule}

Suppose we now glue (by gauging the common $[SU(k)]$) two orbi-instantons of type A defined by two different embeddings $\mu_\text{L,R} : \mathbb{Z}_k \to E_8$ (i.e. two different Kac labels, producing two different sets of $\{r_i,r_{i'}\}_\text{L,R}$ as in \eqref{eq:genericmagquiv}), and different lengths $N_\text{L,R}+1$ of the respective plateaux. The electric quiver is the one in \eqref{eq:genelequiv}. Repeating the above game (i.e. writing down the Type I' configuration realizing the electric quiver, and moving to the magnetic phase), it is easy to convince oneself that the general rule for the magnetic quiver of the LST at finite coupling of the constituent orbi-instantons is given by 
\begin{equation}
    {\displaystyle r_{2'}^\text{L}-r_{4'}^\text{L}-\overset{\overset{\displaystyle r_{3'}^\text{L}}{\vert}}{r_6^\text{L}} - r_5^\text{L} -r_4^\text{L} -r_3^\text{L} - r_2^\text{L} -r_1^\text{L}- \hspace{-0.6cm}\overset{\overset{\scriptstyle \overbrace{\displaystyle 1 \cdots 1}^{M=\widetilde M_\textsc{l}+\widetilde M_\textsc{r}}}{ \rotatebox[origin=c]{190}{$\scriptstyle \setminus$}\ \rotatebox[origin=c]{-190}{$\scriptstyle /$} }}{k} \hspace{-0.6cm} - r_1^\text{R} -r_2^\text{R} -r_3^\text{R} - r_4^\text{R} -r_5^\text{R}-\overset{\overset{\displaystyle r_{3'}^\text{R}}{\vert}}{r_6^\text{R}}-r_{4'}^\text{R}-r_{2'}^\text{R}}\ , \label{eq:generalfin}
\end{equation}
and by
\begin{equation}
        \Bigg(\widetilde M_\text{L}{E_8^{(1)}}^\vee + {\scriptstyle r_{2'}^\text{L} - r_{4'}^\text{L} - \overset{\underset{\scriptscriptstyle \vert}{\scriptstyle r_{3'}^\text{L}}}{r_6^\text{L}}-r_5^\text{L}-r_4^\text{L}-r_3^\text{L}-r_2^\text{L}-r_1^\text{L} }
               %\begin{smallmatrix} & &r_{3'}^\text{L} & & &  & & \\ r_{2'}^\text{L} & r_{4'}^\text{L} & r_6^\text{L} & r_5^\text{L} & r_4^\text{L} & r_3^\text{L} & r_{2}^\text{L} & r_{1}^\text{L}\end{smallmatrix}
               \Bigg) \hspace{-0.1cm} - k - \hspace{-0.1cm} \Bigg({\scriptstyle r_1^\text{R}-r_2^\text{R}-r_3^\text{R}-r_4^\text{R}-r_5^\text{R}-\overset{\underset{\scriptscriptstyle \vert}{\scriptstyle r_{3'}^\text{R}}}{r_6^\text{R}}-r_{4'}^\text{R}-r_{2'}^\text{R}}
               %\begin{smallmatrix} & & & & & r_{3'}^\text{R} & & \\ r_1^\text{R} & r_2^\text{R} & r_3^\text{R} & r_4^\text{R} & r_5^\text{R} & r_6^\text{R} & r_{4'}^\text{R} & r_{2'}^\text{R}\end{smallmatrix}
               +\widetilde M_\text{R}E_8^{(1)} 
               \Bigg)
         \label{eq:generalinf}
\end{equation}
at infinite coupling for all gauge algebras but one (setting the scale $M_\text{s}^2=1/g_\text{YM}^2$ and corresponding to the $U(k)$ node in the 3D quiver), i.e. the ``infinite-coupling'' phase of the LST we are interested in. (The sums in the parentheses in \eqref{eq:generalinf} are performed node-by-node.) This is our second result. Notice also that, by construction, the collection $r_1^\text{L,R},\ldots,r_6^\text{L,R}$ is non-decreasing, i.e. $r_1^\text{L,R}\geq \ldots \geq r_6^\text{L,R}$, and $r_{2'}^\text{L,R},r_{3'}^\text{L,R},r_{4'}^\text{L,R} < r_6^\text{L,R}$.

In light of \eqref{eq:rj}-\eqref{eq:rjprime}, the CB dimension of the above quiver can be shown to be equal to %\red{CHECK MISMATCH OF $p$ when 3' and 4' present}
\begin{align}\label{eq:dim3DLST}
    \dim_\mathbb{H}\text{CB}_\text{3D}^\infty \eqref{eq:generalinf} = &
    \left(30 (N_\text{L}+k) +\frac{k}{2}(k+1) -\left\langle \bm{w}_\text{L},\bm{\rho} \right\rangle-1\right) + \nonumber \\
    &+\left(30 (N_\text{R}+k) +\frac{k}{2}(k+1) -\left\langle \bm{w}_\text{R},\bm{\rho} \right\rangle-1\right) - (k^2-1)\ .
\end{align}
The meaning of the $- (k^2-1)$ term will be clarified at the beginning of section \ref{sec:checks}.

A final observation is in order here. If one thinks of an LST as being obtained by fusion of two orbi-instanton constituents (in the sense of \cite{Heckman:2018pqx}), then there is no ambiguity in how many transitions we should perform ``on the left'' and how many ``on the right''. These two numbers are dictated by the $M_\text{L}$ and $M_\text{R}$ numbers (respectively) of the orbi-instantons we started off with. However, when taken at face value (i.e. ``forgetting'' about its orbi-instanton origin), for an LST with $M$ instantons there are multiple shape-inequivalent choices for the infinite-coupling magnetic quiver \eqref{eq:generalinf}, corresponding to the number of different partitions of $M$ into two integers $(\widetilde M_\text{L}+p_\text{L},\widetilde M_\text{R}+p_\text{R})$. These shape-inequivalent 3D quivers will have the same CB dimension, but generically different $G_J^\text{IR}$ when read off from the UV Lagrangian. Our guiding principle should be to reproduce the \emph{same} symmetry preserved by the choice of labels $(\mu_\text{L},\mu_\text{R})$ in the LST. As a heuristic explanation, consider the following. It is easy to convince oneself that most choices of $(\widetilde M_\text{L},\widetilde M_\text{R})$ in \eqref{eq:generalinf} produce magnetic quivers which preserve the ``right'' $G_J^\text{IR}$, i.e. identical to the algebra preserved by the left and right Kac labels. However some choices (e.g. $(0,M)$ or, equivalently, $(M,0)$ for $([1^k],[1^k])$ for any $k$) do not, and hence have to be excluded. We do not know of a deeper explanation of this fact at the moment. It may signal that most, but not all, infinite-coupling magnetic quivers represent IR-dual UV Lagrangians.

Finally, in the above quiver $N_{\mu_\textsc{l,r}}$ (see the discussion below \eqref{eq:genelequiv}) coincides with the largest linking number $l_\text{L,R}$ (out of the nine $l_1^\text{L,R},\ldots,l_8^\text{L,R},l_9^\text{L,R}$) in the Type IIA engineering of the left, respectively right orbi-instanton, i.e. in the two halves of the Type I' setup \cite{Cabrera:2019izd}. The same paper also provides a convenient map between linking numbers $l_1,\ldots,l_8,l_9$ and multiplicities $n_i,n_{i'}$ of the parts in a Kac label, so that both the set $\{r_i,r_{i'}\}_\text{L,R}$ and $N_{\mu_\textsc{l,r}}$ are fully determined by the (left or right) Kac label. Therefore the number $M_\text{L,R}=N_\text{L,R}+N_{\mu_\textsc{l,r}}$ will generically differ for different choices of $\mu_\text{L,R}$ (with $N_\text{L,R}$ being arbitrary while $N_{\mu_\textsc{l,r}}$ determined by the chosen label).
 
\subsection{A simple case study: \texorpdfstring{$k=2$}{k=2}}
\label{sub:k=2study}

For low enough $k$ it is easy to list all possibilities (i.e. all LSTs) and their magnetic quivers, both at finite and infinite gauge coupling. Take $k=2$. The possible Kac labels are $\mu_
\text{L,R}=[1^2],[2],[2']$. For all these labels $p=0$, so there is no need to distinguish $M_\text{L,R}$ from $\widetilde M_\text{L,R}$. The associated orbi-instantons with length-$(N_\text{L,R}+1)$ plateau  have the following electric quiver and Type IIA engineering:\footnote{\ After having performed all necessary Hanany--Witten moves to bring all D8's close to the O8.}
\begin{equation}
\label{eq:1^2}
    [1^2]:\ [E_8] \, \overset{\emptyset}{1} \, \overset{\mathfrak{su}(1)}{2} \underbrace{\overset{\mathfrak{su}(2)}{\underset{[N_\text{f}=1]}{2}} \overset{\mathfrak{su}(2)}{2} \cdots \overset{\mathfrak{su}(2)}{2} [SU(2)]}_{N_\text{L,R}+1}\ , \quad
   \begin{tikzpicture}[baseline,scale=1]
        \node at (0,0) {};
        \draw[fill=black] (0.5+.2,0) circle (0.075cm);
	\draw[fill=black] (1.25+.2,0) circle (0.075cm);
	\draw[fill=black] (2+.2,0) circle (0.075cm);
        \draw[fill=black] (2.75+.2,0) circle (0.075cm);
        	
	%D6
	
	\draw[solid,black,thick] (0.35,0)--(0.5+.2,0) node[black,midway,yshift=0.2cm] {\footnotesize $2$};
	\draw[solid,black,thick] (0.5+.2,0)--(1.25+.2,0) node[black,midway,yshift=0.2cm] {\footnotesize $2$};
	\draw[solid,black,thick] (1.25+.2,0)--(2+.2,0) node[black,midway,yshift=0.2cm] {\footnotesize $2$};
        \path (2+.2,0)--(2.75+.2,0) node[black,xshift=-.25cm,yshift=-.025cm] {\footnotesize $\cdots$};
        \draw[solid,black,thick] (2.75+.2,0)--(3.5+.2,0) node[black,midway,yshift=0.2cm] {\footnotesize $2$};
	
	%D8-O8
	\draw[dashed,black,very thick] (0,-.5)--(0,.5) node[black,midway, xshift =0cm, yshift=-1.5cm] {} node[black,midway, xshift =0cm, yshift=1.5cm] {};
	\draw[solid,black,very thick] (0.15,-.5)--(0.15,.5) node[black,midway, xshift =0cm, yshift=+.75cm] {\footnotesize $7$};
	\draw[solid,black,very thick] (0.35,-.5)--(0.35,.5) node[black,midway, xshift =0cm, yshift=+.75cm] {\footnotesize $1$};
   \end{tikzpicture}
\end{equation}
with $r_i=r_{i'}=0$ for all $i$, $M_\text{L,R}=N_\text{L,R}+N_{\mu_\textsc{l,r}}=N_\text{L,R}+2$ (in the notation of \eqref{eq:orbi-trivial}) and largest linking number $l_\text{L,R}=2$;
\begin{equation}
\label{eq:2}
    [2]:\ [E_7] \, \overset{\emptyset}{1} \, \underbrace{\overset{\mathfrak{su}(2)}{\underset{[N_\text{f}=2]}{2}} \overset{\mathfrak{su}(2)}{2} \cdots \overset{\mathfrak{su}(2)}{2} [SU(2)]}_{N_\text{L,R}+1}\ , \quad
   \begin{tikzpicture}[baseline,scale=1]
        \node at (0,0) {};
        \draw[fill=black] (1.25,0) circle (0.075cm);
	\draw[fill=black] (2,0) circle (0.075cm);
        \draw[fill=black] (2.75,0) circle (0.075cm);
	
	%D6

	\draw[solid,black,thick] (0.35,-0.2)--(0.85,-0.2) node[black,xshift=-0.08cm,yshift=0.2cm] {\footnotesize $1$};
	\draw[solid,black,thick] (0.55,0.2)--(0.85,0.2) node[black,xshift=-0.18cm,yshift=0.2cm] {\footnotesize $1$};
	\draw[solid,black,thick] (0.85,0.2)--(1.25,0) node[black,midway,yshift=0.2cm] {};
        \draw[solid,black,thick] (0.85,-0.2)--(1.25,0) node[black,midway,yshift=0.2cm] {};

	\draw[solid,black,thick] (1.25,0)--(2,0) node[black,midway,yshift=0.2cm] {\footnotesize $2$};
        \path (2,0)--(2.75,0) node[black,xshift=-.25cm,yshift=-.025cm] {\footnotesize $\cdots$};
        \draw[solid,black,thick] (2.75,0)--(3.5,0) node[black,midway,yshift=0.2cm] {\footnotesize $2$};
	
	%D8-O8
	\draw[dashed,black,very thick] (0,-.5)--(0,.5) node[black,midway, xshift =0cm, yshift=-1.5cm] {} node[black,midway, xshift =0cm, yshift=1.5cm] {};
	\draw[solid,black,very thick] (0.15,-.5)--(0.15,.5) node[black,midway, xshift =0cm, yshift=+.75cm] {\footnotesize $6$};
 
	\draw[solid,black,very thick] (0.35,-.5)--(0.35,.5) node[black,midway, xshift =0cm, yshift=+.75cm] {\footnotesize $1$};
        \draw[solid,black,very thick] (0.55,-.5)--(0.55,.5) node[black,midway, xshift =0cm, yshift=+.75cm] {\footnotesize $1$};
   \end{tikzpicture}
\end{equation}
with $r_i=r_{i'}=0$ for all $i$ but $r_1=1$, $M_\text{L,R}=N_\text{L,R}+N_{\mu_\textsc{l,r}}=N_\text{L,R}+1$ and largest linking number $l_\text{L,R}=1$;
\begin{equation}
\label{eq:2'}
    [2']:\ [SO(16)] \,\overset{\mathfrak{usp}(2)}{1} \,\underbrace{\overset{\mathfrak{su}(2)}{2} \cdots \overset{\mathfrak{su}(2)}{2}  [SU(2)]}_{N_\text{L,R}+1}\ , \quad
   \begin{tikzpicture}[baseline,scale=1]
        \node at (0,0) {};
        \draw[fill=black] (1,0) circle (0.075cm);
	\draw[fill=black] (1.75,0) circle (0.075cm);
        \draw[fill=black] (2.5,0) circle (0.075cm);
	
	%D6
	
	\draw[solid,black,thick] (0,0)--(1,0) node[black,midway,yshift=0.2cm] {\footnotesize $2$};
        \draw[solid,black,thick] (1,0)--(1.75,0) node[black,midway,yshift=0.2cm] {\footnotesize $2$};
        \path (1.75,0)--(2.5,0) node[black,xshift=-.25cm,yshift=-.025cm] {\footnotesize $\cdots$};
        \draw[solid,black,thick] (2.5,0)--(3.25,0) node[black,midway,yshift=0.2cm] {\footnotesize $2$};
	
	%D8-O8
	\draw[dashed,black,very thick] (0,-.5)--(0,.5) node[black,midway, xshift =0cm, yshift=-1.5cm] {} node[black,midway, xshift =0cm, yshift=1.5cm] {};
	\draw[solid,black,very thick] (0.25,-.5)--(0.25,.5) node[black,midway, xshift =0cm, yshift=+.75cm] {\footnotesize $8$};
   \end{tikzpicture}
\end{equation}
with $r_1=\ldots=r_6=2, r_{4'}=r_{3'}=1, r_{2'}=0$, $M_\text{L,R}=N_\text{L,R}+N_{\mu_\textsc{l,r}}=N_\text{L,R}+0$ and largest linking number is $l_\text{L,R}=0$, but $\max\left(N_{\mu_\textsc{l,r}},1\right)=1$.

There are six inequivalent $(e')$ LSTs we can build out of these Kac labels; namely: 
\begin{equation}
    (\mu_\text{L},\mu_\text{R}) = ([1^2],[1^2]), \ ([2],[2]), \ ([2'],[2']),\ ([1^2],[2]),\ ([1^2],[2']), \ ([2],[2'])\ .
\end{equation}
To build them, we simply glue any two among \eqref{eq:1^2}-\eqref{eq:2'} along $[SU(2)]$. We have already analyzed the case $([1^2],[1^2])$ for any $k$ in \eqref{eq:long1k1klst}-\eqref{eq:short1k1klst}; for $k=2$ it has infinite-coupling magnetic quiver $(N_\text{L}+2){E_8^{(1)}}^\vee - 2 -(N_\text{R}+2)E_8^{(1)}$. The other cases read:
\begin{equation}
\label{eq:1^2-2}
([1^2],[2]):  \overset{\overset{\displaystyle \overbrace{1 \cdots 1}^{M_\text{L}+M_\text{R}}}{ \rotatebox[origin=c]{190}{$\setminus$}\ \rotatebox[origin=c]{-190}{$/$} }}{2} \hspace{-.5cm} - 1 \ ,
\quad
\begin{tikzpicture}[baseline,scale=1]
	% NS5
	\node at (0,0) {};
	%\draw[fill=black,label=right:$\times 5$] (0.5+.2,0.5) circle (0.075cm);
	\node[draw,fill=black,circle,label=right:{\footnotesize $M$},minimum size=3.5pt,inner sep=0pt, outer sep=0pt] (CircleNode) at (1,0.5) {};
	%D6
 
	\draw[solid,black,thick] (0.35,0)--(1.5,0) node[black,midway,yshift=-0.2cm] {\footnotesize $2$};
	\draw[solid,black,thick] (1.5,0)--(1.75,0.2); %node[black,midway,yshift=0.2cm] {\footnotesize $k$};
        \draw[solid,black,thick] (1.5,0)--(1.75,-0.2); %node[black,midway,yshift=0.2cm] {\footnotesize $k$};
        \draw[solid,black,thick] (1.75,0.2)--(2,0.2) node[black,midway,yshift=0.2cm] {};
        \draw[solid,black,thick] (1.75,-0.2)--(2.2,-0.2) node[black,xshift=-0.09cm,yshift=0.2cm] {};

	%D8-O8
	\draw[dashed,black,very thick] (0,-.5)--(0,.5) node[black,midway, xshift =0cm, yshift=-1.5cm] {} node[black,midway, xshift =0cm, yshift=1.5cm] {};
	\draw[solid,black,very thick] (0.15,-.5)--(0.15,.5) node[black,midway, xshift =0cm, yshift=+.75cm] {\footnotesize $7$};
	\draw[solid,black,very thick] (0.35,-.5)--(0.35,.5) node[black,midway, xshift =0cm, yshift=+.75cm] {\footnotesize $1$};

        \draw[dashed,black,very thick] (2.55,-.5)--(2.55,.5) node[black,midway, xshift =0cm, yshift=-1.5cm] {} node[black,midway, xshift =0cm, yshift=1.5cm] {};
	\draw[solid,black,very thick] (2,-.5)--(2,.5) node[black,midway, xshift =0cm, yshift=+.75cm] {\footnotesize $1$};
 	\draw[solid,black,very thick] (2.2,-.5)--(2.2,.5) node[black,midway, xshift =0cm, yshift=+.75cm] {\footnotesize $1$};
        \draw[solid,black,very thick] (2.4,-.5)--(2.4,.5) node[black,midway, xshift =0cm, yshift=+.75cm] {\footnotesize $6$};
\end{tikzpicture}
\end{equation}
with $M_\text{L}{E_8^{(1)}}^\vee - 2 - \Big(\raisebox{-5pt}{$\scriptstyle 1-0-0-0-0-\overset{\underset{\scriptscriptstyle \vert}{\scriptstyle 0}}{0}-0-0$} \hspace{0.05cm} + M_\text{R}E_8^{(1)}\Big)$ as infinite-coupling limit;
\begin{equation}
\label{eq:1^2-2p}
([1^2],[2']):  \overset{\overset{\displaystyle \overbrace{1 \cdots 1}^{M_\text{L}+M_\text{R}}}{ \rotatebox[origin=c]{190}{$\setminus$}\ \rotatebox[origin=c]{-190}{$/$} }}{2}\hspace{-.5cm} - 2 -2-2-2-2-\overset{\overset{\displaystyle 1}{\vert}}{2}-1\ , 
\quad
\begin{tikzpicture}[baseline,scale=1]
	% NS5
	\node at (0,0) {};
	%\draw[fill=black,label=right:$\times 5$] (0.5+.2,0.5) circle (0.075cm);
	\node[draw,fill=black,circle,label=right:{\footnotesize $M$},minimum size=3.5pt,inner sep=0pt, outer sep=0pt] (CircleNode) at (0.6,0.5) {};
	%D6
	
	\draw[solid,black,thick] (0.35,0)--(1.5,0) node[black,midway,yshift=-0.2cm] {\footnotesize $2$};

        \draw[solid,black,thick] (1.5,0.2)--(1.65,0.2) node[black,midway,yshift=-0.2cm] {};
        \draw[solid,black,thick] (1.5,-0.2)--(1.65,-0.2) node[black,midway,yshift=-0.2cm] {};
        \draw[solid,black,thick] (1.65,0)--(1.8,0) node[black,midway,yshift=-0.2cm] {};
        \draw[solid,black,thick] (1.65,-0.4)--(1.8,-0.4) node[black,midway,yshift=-0.2cm] {};
        \draw[solid,black,thick] (1.8,0.2)--(1.95,0.2) node[black,midway,yshift=-0.2cm] {};
        \draw[solid,black,thick] (1.8,-0.2)--(1.95,-0.2) node[black,midway,yshift=-0.2cm] {};
        \draw[solid,black,thick] (1.95,0)--(2.1,0) node[black,midway,yshift=-0.2cm] {};
        \draw[solid,black,thick] (1.95,-0.4)--(2.1,-0.4) node[black,midway,yshift=-0.2cm] {};
        \draw[solid,black,thick] (2.1,0.2)--(2.25,0.2) node[black,midway,yshift=-0.2cm] {};
        \draw[solid,black,thick] (2.1,-0.2)--(2.25,-0.2) node[black,midway,yshift=-0.2cm] {};
        \draw[solid,black,thick] (2.25,0)--(2.4,0) node[black,midway,yshift=-0.2cm] {};
        \draw[solid,black,thick] (2.25,-0.4)--(2.4,-0.4) node[black,midway,yshift=-0.2cm] {};
        \draw[solid,black,thick] (2.4,0.2)--(2.55,0.2) node[black,midway,yshift=-0.2cm] {};
        \draw[solid,black,thick] (2.4,-0.2)--(2.55,-0.2) node[black,midway,yshift=-0.2cm] {};
        \draw[solid,black,thick] (2.55,0.2)--(2.7,0.1) node[black,midway,yshift=-0.2cm] {};
        \draw[solid,black,thick] (2.7,0.1)--(2.55,0) node[black,midway,yshift=-0.2cm] {};

	%D8-O8
	\draw[dashed,black,very thick] (0,-.5)--(0,.5) node[black,midway, xshift =0cm, yshift=-1.5cm] {} node[black,midway, xshift =0cm, yshift=1.5cm] {};
	\draw[solid,black,very thick] (0.15,-.5)--(0.15,.5) node[black,midway, xshift =0cm, yshift=+.75cm] {\footnotesize $7$};
	\draw[solid,black,very thick] (0.35,-.5)--(0.35,.5) node[black,midway, xshift =0cm, yshift=+.75cm] {\footnotesize $1$};

        \draw[solid,black,very thick] (1.5,-.5)--(1.5,.5) node[black,midway, xshift =0cm, yshift=+.75cm] {\footnotesize $1$};
        \draw[solid,black,very thick] (1.65,-.5)--(1.65,.5) node[black,midway, xshift =0cm, yshift=+.75cm] {\footnotesize $1$};
        \draw[solid,black,very thick] (1.8,-.5)--(1.8,.5) node[black,midway, xshift =0cm, yshift=+.75cm] {\footnotesize $1$};
        \draw[solid,black,very thick] (1.95,-.5)--(1.95,.5) node[black,midway, xshift =0cm, yshift=+.75cm] {\footnotesize $1$};
        \draw[solid,black,very thick] (2.1,-.5)--(2.1,.5) node[black,midway, xshift =0cm, yshift=+.75cm] {\footnotesize $1$};
        \draw[solid,black,very thick] (2.25,-.5)--(2.25,.5) node[black,midway, xshift =0cm, yshift=+.75cm] {\footnotesize $1$};
        \draw[solid,black,very thick] (2.4,-.5)--(2.4,.5) node[black,midway, xshift =0cm, yshift=+.75cm] {\footnotesize $1$};
        \draw[solid,black,very thick] (2.55,-.5)--(2.55,.5) node[black,midway, xshift =0cm, yshift=+.75cm] {\footnotesize $1$};

        \draw[dashed,black,very thick] (2.7,-.5)--(2.7,.5) node[black,midway, xshift =0cm, yshift=-1.5cm] {} node[black,midway, xshift =0cm, yshift=1.5cm] {};
\end{tikzpicture}
\end{equation}
with $M_\text{L}{E_8^{(1)}}^\vee - 2 - \Big(\raisebox{-5pt}{$\scriptstyle 2-2-2-2-2-\overset{\underset{\scriptscriptstyle \vert}{\scriptstyle 1}}{2}-1-0$} \hspace{0.05cm} + M_\text{R}E_8^{(1)}\Big)$ as infinite-coupling limit;
\begin{equation}
\label{eq:2-2}
([2],[2]):  1-\hspace{-.35cm }\overset{\overset{\displaystyle \overbrace{1 \cdots 1}^{M_\text{L}+M_\text{R}}}{ \rotatebox[origin=c]{190}{$\setminus$}\ \rotatebox[origin=c]{-190}{$/$} }}{2} \hspace{-.35cm} - 1 \ , \quad
\begin{tikzpicture}[baseline,scale=1]
	% NS5
	\node at (0,0) {};
	%\draw[fill=black,label=right:$\times 5$] (0.5+.2,0.5) circle (0.075cm);
	\node[draw,fill=black,circle,label=right:{\footnotesize $M$},minimum size=3.5pt,inner sep=0pt, outer sep=0pt] (CircleNode) at (1,0.5) {};
	%D6
	
        \draw[solid,black,thick] (0.55,0.2)--(0.8,0.2); 
        \draw[solid,black,thick] (0.35,-0.2)--(0.8,-0.2); 
        \draw[solid,black,thick] (1.5,0)--(1.75,-0.2);
        \draw[solid,black,thick] (1.05,0)--(0.8,0.2); 
        \draw[solid,black,thick] (1.05,0)--(0.8,-0.2);	
        \draw[solid,black,thick] (1.05,0)--(1.5,0) node[black,midway,yshift=-0.2cm] {\footnotesize $2$};
	\draw[solid,black,thick] (1.5,0)--(1.75,0.2); 
        \draw[solid,black,thick] (1.5,0)--(1.75,-0.2);
        \draw[solid,black,thick] (1.75,0.2)--(2,0.2) node[black,midway,yshift=0.2cm] {};
        \draw[solid,black,thick] (1.75,-0.2)--(2.2,-0.2) node[black,xshift=-0.09cm,yshift=0.2cm] {};
    
	%D8-O8
	\draw[dashed,black,very thick] (0,-.5)--(0,.5) node[black,midway, xshift =0cm, yshift=-1.5cm] {} node[black,midway, xshift =0cm, yshift=1.5cm] {};
	\draw[solid,black,very thick] (0.15,-.5)--(0.15,.5) node[black,midway, xshift =0cm, yshift=+.75cm] {\footnotesize $6$};
	\draw[solid,black,very thick] (0.35,-.5)--(0.35,.5) node[black,midway, xshift =0cm, yshift=+.75cm] {\footnotesize $1$};
 \draw[solid,black,very thick] (0.55,-.5)--(0.55,.5) node[black,midway, xshift =0cm, yshift=+.75cm] {\footnotesize $1$};

	\draw[solid,black,very thick] (2,-.5)--(2,.5) node[black,midway, xshift =0cm, yshift=+.75cm] {\footnotesize $1$};
 	\draw[solid,black,very thick] (2.2,-.5)--(2.2,.5) node[black,midway, xshift =0cm, yshift=+.75cm] {\footnotesize $1$};
        \draw[solid,black,very thick] (2.4,-.5)--(2.4,.5) node[black,midway, xshift =0cm, yshift=+.75cm] {\footnotesize $6$};
        \draw[dashed,black,very thick] (2.55,-.5)--(2.55,.5) node[black,midway, xshift =0cm, yshift=-1.5cm] {} node[black,midway, xshift =0cm, yshift=1.5cm] {};
\end{tikzpicture}
\end{equation}
with $\Big(M_\text{L}{E_8^{(1)}}^\vee + \hspace{0.05cm} \raisebox{-5pt}{$\scriptstyle 0-0-\overset{\underset{\scriptscriptstyle \vert}{\scriptstyle 0}}{0}-0-0-0-0-1$}\Big) - 2 - \Big(\raisebox{-5pt}{$\scriptstyle 1-0-0-0-0-\overset{\underset{\scriptscriptstyle \vert}{\scriptstyle 0}}{0}-0-0$} \hspace{0.05cm} + M_\text{R}E_8^{(1)}\Big)$ as infinite-coupling limit;
\begin{equation}
    \label{eq:2'-2'}
([2'],[2']):  {\scriptstyle 1-\overset{\overset{\scriptstyle 1}{\vert}}{2}-2-2-2-2-2-\hspace{-.4cm}\overset{\overset{\scriptstyle \overbrace{\scriptstyle 1 \cdots 1}^{M_\text{L}+M_\text{R}}}{ \rotatebox[origin=c]{170}{$\scriptstyle \setminus$}\ \rotatebox[origin=c]{-170}{$\scriptstyle /$} }}{2}\hspace{-.4cm} - 2 -2-2-2-2-\overset{\overset{\scriptstyle 1}{\vert}}{2}-1}\ , \quad 
%\overset{\overset{\scriptstyle 1 \, 1}{ \rotatebox[origin=c]{170}{$\scriptstyle \setminus$}  \rotatebox[origin=c]{-170}{$\scriptstyle /$} }}{2}\hspace{-.05cm} - 2 -2-2-2-2-\overset{\overset{\scriptstyle 1}{\vert}}{2}-1}\ , \quad 
%\end{equation}
%\begin{equation}
\begin{tikzpicture}[baseline,scale=1]
    \node at (0,0) {};
	%\draw[fill=black,label=right:$\times 5$] (0.5+.2,0.5) circle (0.075cm);
	\node[draw,fill=black,circle,label=right:{\footnotesize $M$},minimum size=3.5pt,inner sep=0pt, outer sep=0pt] (CircleNode) at (1.45,0.5) {};
	%D6
	
	\draw[solid,black,thick] (1.2,0)--(1.5+0.85,0) node[black,midway,yshift=-0.2cm] {\footnotesize $2$};

        \draw[solid,black,thick] (1.5+0.85,0.2)--(1.65+0.85,0.2) node[black,midway,yshift=-0.2cm] {};
        \draw[solid,black,thick] (1.5+0.85,-0.2)--(1.65+0.85,-0.2) node[black,midway,yshift=-0.2cm] {};
        \draw[solid,black,thick] (1.65+0.85,0)--(1.8+0.85,0) node[black,midway,yshift=-0.2cm] {};
        \draw[solid,black,thick] (1.65+0.85,-0.4)--(1.8+0.85,-0.4) node[black,midway,yshift=-0.2cm] {};
        \draw[solid,black,thick] (1.8+0.85,0.2)--(1.95+0.85,0.2) node[black,midway,yshift=-0.2cm] {};
        \draw[solid,black,thick] (1.8+0.85,-0.2)--(1.95+0.85,-0.2) node[black,midway,yshift=-0.2cm] {};
        \draw[solid,black,thick] (1.95+0.85,0)--(2.1+0.85,0) node[black,midway,yshift=-0.2cm] {};
        \draw[solid,black,thick] (1.95+0.85,-0.4)--(2.1+0.85,-0.4) node[black,midway,yshift=-0.2cm] {};
        \draw[solid,black,thick] (2.1+0.85,0.2)--(2.25+0.85,0.2) node[black,midway,yshift=-0.2cm] {};
        \draw[solid,black,thick] (2.1+0.85,-0.2)--(2.25+0.85,-0.2) node[black,midway,yshift=-0.2cm] {};
        \draw[solid,black,thick] (2.25+0.85,0)--(2.4+0.85,0) node[black,midway,yshift=-0.2cm] {};
        \draw[solid,black,thick] (2.25+0.85,-0.4)--(2.4+0.85,-0.4) node[black,midway,yshift=-0.2cm] {};
        \draw[solid,black,thick] (2.4+0.85,0.2)--(2.55+0.85,0.2) node[black,midway,yshift=-0.2cm] {};
        \draw[solid,black,thick] (2.4+0.85,-0.2)--(2.55+0.85,-0.2) node[black,midway,yshift=-0.2cm] {};
        \draw[solid,black,thick] (2.55+0.85,0.2)--(2.7+0.85,0.1) node[black,midway,yshift=-0.2cm] {};
        \draw[solid,black,thick] (2.7+0.85,0.1)--(2.55+0.85,0) node[black,midway,yshift=-0.2cm] {};

        \draw[solid,black,thick] (1.2,0.2)--(1.05,0.2) node[black,midway,yshift=-0.2cm] {};
        \draw[solid,black,thick] (1.2,-0.2)--(1.05,-0.2) node[black,midway,yshift=-0.2cm] {};
        \draw[solid,black,thick] (1.05,0)--(0.9,0) node[black,midway,yshift=-0.2cm] {};
        \draw[solid,black,thick] (1.05,-0.4)--(0.9,-0.4) node[black,midway,yshift=-0.2cm] {};
        \draw[solid,black,thick] (0.9,0.2)--(0.75,0.2) node[black,midway,yshift=-0.2cm] {};
        \draw[solid,black,thick] (0.9,-0.2)--(0.75,-0.2) node[black,midway,yshift=-0.2cm] {};
        \draw[solid,black,thick] (0.75,0)--(0.6,0) node[black,midway,yshift=-0.2cm] {};
        \draw[solid,black,thick] (0.75,-0.4)--(0.6,-0.4) node[black,midway,yshift=-0.2cm] {};
        \draw[solid,black,thick] (0.6,0.2)--(0.45,0.2) node[black,midway,yshift=-0.2cm] {};
        \draw[solid,black,thick] (0.6,-0.2)--(0.45,-0.2) node[black,midway,yshift=-0.2cm] {};
        \draw[solid,black,thick] (0.45,0)--(0.3,0) node[black,midway,yshift=-0.2cm] {};
        \draw[solid,black,thick] (0.45,-0.4)--(0.3,-0.4) node[black,midway,yshift=-0.2cm] {};
        \draw[solid,black,thick] (0.3,0.2)--(0.15,0.2) node[black,midway,yshift=-0.2cm] {};
        \draw[solid,black,thick] (0.3,-0.2)--(0.15,-0.2) node[black,midway,yshift=-0.2cm] {};
        \draw[solid,black,thick] (0.15,0.2)--(0,0.1) node[black,midway,yshift=-0.2cm] {};
        \draw[solid,black,thick] (0,0.1)--(0.15,0) node[black,midway,yshift=-0.2cm] {};

	%D8-O8
	\draw[dashed,black,very thick] (0,-.5)--(0,.5) node[black,midway, xshift =0cm, yshift=-1.5cm] {} node[black,midway, xshift =0cm, yshift=1.5cm] {};
	\draw[solid,black,very thick] (0.15,-.5)--(0.15,.5) node[black,midway, xshift =0cm, yshift=+.75cm] {\footnotesize $1$};
        \draw[solid,black,very thick] (0.3,-.5)--(0.3,.5) node[black,midway, xshift =0cm, yshift=+.75cm] {\footnotesize $1$};
        \draw[solid,black,very thick] (0.45,-.5)--(0.45,.5) node[black,midway, xshift =0cm, yshift=+.75cm] {\footnotesize $1$};
        \draw[solid,black,very thick] (0.6,-.5)--(0.6,.5) node[black,midway, xshift =0cm, yshift=+.75cm] {\footnotesize $1$};
        \draw[solid,black,very thick] (0.75,-.5)--(0.75,.5) node[black,midway, xshift =0cm, yshift=+.75cm] {\footnotesize $1$};
        \draw[solid,black,very thick] (0.9,-.5)--(0.9,.5) node[black,midway, xshift =0cm, yshift=+.75cm] {\footnotesize $1$};
        \draw[solid,black,very thick] (1.05,-.5)--(1.05,.5) node[black,midway, xshift =0cm, yshift=+.75cm] {\footnotesize $1$};
        \draw[solid,black,very thick] (1.2,-.5)--(1.2,.5) node[black,midway, xshift =0cm, yshift=+.75cm] {\footnotesize $1$};

        \draw[solid,black,very thick] (1.5+0.85,-.5)--(1.5+0.85,.5) node[black,midway, xshift =0cm, yshift=+.75cm] {\footnotesize $1$};
        \draw[solid,black,very thick] (1.65+0.85,-.5)--(1.65+0.85,.5) node[black,midway, xshift =0cm, yshift=+.75cm] {\footnotesize $1$};
        \draw[solid,black,very thick] (1.8+0.85,-.5)--(1.8+0.85,.5) node[black,midway, xshift =0cm, yshift=+.75cm] {\footnotesize $1$};
        \draw[solid,black,very thick] (1.95+0.85,-.5)--(1.95+0.85,.5) node[black,midway, xshift =0cm, yshift=+.75cm] {\footnotesize $1$};
        \draw[solid,black,very thick] (2.1+0.85,-.5)--(2.1+0.85,.5) node[black,midway, xshift =0cm, yshift=+.75cm] {\footnotesize $1$};
        \draw[solid,black,very thick] (2.25+0.85,-.5)--(2.25+0.85,.5) node[black,midway, xshift =0cm, yshift=+.75cm] {\footnotesize $1$};
        \draw[solid,black,very thick] (2.4+0.85,-.5)--(2.4+0.85,.5) node[black,midway, xshift =0cm, yshift=+.75cm] {\footnotesize $1$};
        \draw[solid,black,very thick] (2.55+0.85,-.5)--(2.55+0.85,.5) node[black,midway, xshift =0cm, yshift=+.75cm] {\footnotesize $1$};

        \draw[dashed,black,very thick] (2.7+0.85,-.5)--(2.7+0.85,.5) node[black,midway, xshift =0cm, yshift=-1.5cm] {} node[black,midway, xshift =0cm, yshift=1.5cm] {};
\end{tikzpicture}
\end{equation}
with $\Big(M_\text{L}{E_8^{(1)}}^\vee + \hspace{0.05cm} \raisebox{-5pt}{$\scriptstyle 0-1-\overset{\underset{\scriptscriptstyle \vert}{\scriptstyle 1}}{2}-2-2-2-2-2$}\Big) - 2 - \Big(\raisebox{-5pt}{$\scriptstyle 2-2-2-2-2-\overset{\underset{\scriptscriptstyle \vert}{\scriptstyle 1}}{2}-1-0$} \hspace{0.05cm} + M_\text{R}E_8^{(1)}\Big)$ as infinite-coupling limit;
    \begin{equation}
    \label{eq:2-2'}
([2]_\text{L},[2']_\text{R}):  1-\hspace{-.35cm }\overset{\overset{\displaystyle \overbrace{1 \cdots 1}^{M_\text{L}+M_\text{R}}}{ \rotatebox[origin=c]{190}{$\setminus$}\ \rotatebox[origin=c]{-190}{$/$} }}{2} \hspace{-.35cm} - 2 -2-2-2-2-\overset{\overset{\displaystyle 1}{\vert}}{2}-1\ , \quad
%\end{equation}
%\begin{equation}
    \begin{tikzpicture}[baseline,scale=1]
	% NS5
	\node at (0,0) {};
	%\draw[fill=black,label=right:$\times 5$] (0.5+.2,0.5) circle (0.075cm);
	\node[draw,fill=black,circle,label=right:{\footnotesize $M$},minimum size=3.5pt,inner sep=0pt, outer sep=0pt] (CircleNode) at (1,0.5) {};
	%D6
	
        \draw[solid,black,thick] (0.55,0.2)--(0.8,0.2); 
        \draw[solid,black,thick] (0.35,-0.2)--(0.8,-0.2); 
        \draw[solid,black,thick] (1.5,0)--(1.75,-0.2);
        \draw[solid,black,thick] (1.05,0)--(0.8,0.2); 
        \draw[solid,black,thick] (1.05,0)--(0.8,-0.2);	
        \draw[solid,black,thick] (1.05,0)--(1.5,0) node[black,midway,yshift=-0.2cm] {\footnotesize $2$};
	\draw[solid,black,thick] (1.5,0)--(1.75,0.2); 
        \draw[solid,black,thick] (1.5,0)--(1.75,-0.2);
        \draw[solid,black,thick] (1.75,0.2)--(2,0.2) node[black,midway,yshift=0.2cm] {};
        \draw[solid,black,thick] (1.75,-0.2)--(2,-0.2) node[black,xshift=-0.09cm,yshift=0.2cm] {};

	%D8-O8
	\draw[dashed,black,very thick] (0,-.5)--(0,.5) node[black,midway, xshift =0cm, yshift=-1.5cm] {} node[black,midway, xshift =0cm, yshift=1.5cm] {};
	\draw[solid,black,very thick] (0.15,-.5)--(0.15,.5) node[black,midway, xshift =0cm, yshift=+.75cm] {\footnotesize $6$};
	\draw[solid,black,very thick] (0.35,-.5)--(0.35,.5) node[black,midway, xshift =0cm, yshift=+.75cm] {\footnotesize $1$};
 \draw[solid,black,very thick] (0.55,-.5)--(0.55,.5) node[black,midway, xshift =0cm, yshift=+.75cm] {\footnotesize $1$};

        \draw[solid,black,thick] (1.5+0.5,-0)--(1.65+0.5,0) node[black,midway,yshift=-0.2cm] {};
        \draw[solid,black,thick] (1.5+0.5,-0.4)--(1.65+0.5,-0.4) node[black,midway,yshift=-0.2cm] {};
        \draw[solid,black,thick] (1.65+0.5,0.2)--(1.8+0.5,0.2) node[black,midway,yshift=-0.2cm] {};
        \draw[solid,black,thick] (1.65+0.5,-0.2)--(1.8+0.5,-0.2) node[black,midway,yshift=-0.2cm] {};
        \draw[solid,black,thick] (1.8+0.5,0)--(1.95+0.5,0) node[black,midway,yshift=-0.2cm] {};
        \draw[solid,black,thick] (1.8+0.5,-0.4)--(1.95+0.5,-0.4) node[black,midway,yshift=-0.2cm] {};
        \draw[solid,black,thick] (1.95+0.5,0.2)--(2.1+0.5,0.2) node[black,midway,yshift=-0.2cm] {};
        \draw[solid,black,thick] (1.95+0.5,-0.2)--(2.1+0.5,-0.2) node[black,midway,yshift=-0.2cm] {};
        \draw[solid,black,thick] (2.1+0.5,0)--(2.25+0.5,0) node[black,midway,yshift=-0.2cm] {};
        \draw[solid,black,thick] (2.1+0.5,-0.4)--(2.25+0.5,-0.4) node[black,midway,yshift=-0.2cm] {};
        \draw[solid,black,thick] (2.25+0.5,0.2)--(2.4+0.5,0.2) node[black,midway,yshift=-0.2cm] {};
        \draw[solid,black,thick] (2.25+0.5,-0.2)--(2.4+0.5,-0.2) node[black,midway,yshift=-0.2cm] {};
        \draw[solid,black,thick] (2.4+0.5,0)--(2.55+0.5,0) node[black,midway,yshift=-0.2cm] {};
        \draw[solid,black,thick] (2.4+0.5,-0.4)--(2.55+0.5,-0.4) node[black,midway,yshift=-0.2cm] {};
        \draw[solid,black,thick] (2.55+0.5,0)--(2.7+0.5,-0.1) node[black,midway,yshift=-0.2cm] {};
        \draw[solid,black,thick] (2.7+0.5,-0.1)--(2.55+0.5,-0.2) node[black,midway,yshift=-0.2cm] {};

	%D8-O8

        \draw[solid,black,very thick] (1.5+0.5,-.5)--(1.5+0.5,.5) node[black,midway, xshift =0cm, yshift=+.75cm] {\footnotesize $1$};
        \draw[solid,black,very thick] (1.65+0.5,-.5)--(1.65+0.5,.5) node[black,midway, xshift =0cm, yshift=+.75cm] {\footnotesize $1$};
        \draw[solid,black,very thick] (1.8+0.5,-.5)--(1.8+0.5,.5) node[black,midway, xshift =0cm, yshift=+.75cm] {\footnotesize $1$};
        \draw[solid,black,very thick] (1.95+0.5,-.5)--(1.95+0.5,.5) node[black,midway, xshift =0cm, yshift=+.75cm] {\footnotesize $1$};
        \draw[solid,black,very thick] (2.1+0.5,-.5)--(2.1+0.5,.5) node[black,midway, xshift =0cm, yshift=+.75cm] {\footnotesize $1$};
        \draw[solid,black,very thick] (2.25+0.5,-.5)--(2.25+0.5,.5) node[black,midway, xshift =0cm, yshift=+.75cm] {\footnotesize $1$};
        \draw[solid,black,very thick] (2.4+0.5,-.5)--(2.4+0.5,.5) node[black,midway, xshift =0cm, yshift=+.75cm] {\footnotesize $1$};
        \draw[solid,black,very thick] (2.55+0.5,-.5)--(2.55+0.5,.5) node[black,midway, xshift =0cm, yshift=+.75cm] {\footnotesize $1$};

        \draw[dashed,black,very thick] (2.7+0.5,-.5)--(2.7+0.5,.5) node[black,midway, xshift =0cm, yshift=-1.5cm] {} node[black,midway, xshift =0cm, yshift=1.5cm] {};
\end{tikzpicture}
\end{equation}
with 
%\begin{equation}
$\Big(M_\text{L}{E_8^{(1)}}^\vee + \hspace{0.05cm} \raisebox{-5pt}{$\scriptstyle 0-0-\overset{\underset{\scriptscriptstyle \vert}{\scriptstyle 0}}{0}-0-0-0-0-1$}\Big) - 2 - \Big(\raisebox{-5pt}{$\scriptstyle 2-2-2-2-2-\overset{\underset{\scriptscriptstyle \vert}{\scriptstyle 1}}{2}-1-0$} \hspace{0.05cm} + M_\text{R}E_8^{(1)}\Big)$
%\end{equation}
as infinite-coupling limit.

%%%%%%%%%%%%%%%%%%%%%%%%%%%%%%%%%
\section{Checks}
\label{sec:checks}
%%%%%%%%%%%%%%%%%%%%%%%%%%%%%%%%%

The first nontrivial check that we perform to confirm that the proposed 3D quiver gives a good description of the HB we are after is that the quaternionic dimension $\dim_\mathbb{H} \text{CB}_\text{3D}^\infty \text{\eqref{eq:generalinf}}$ of the former matches that of the HB of the LST at $M_\text{s}$.%, or equivalently that of the hypermultiplet moduli space of the heterotic string on an ALE space with full and fractional small instantons (a total of $M=M_{\text{L}}+M_{\text{R}}$ of them).

We will show this explicitly only for $k=2$. All other $k$'s work in the same way. Let us begin by counting, for each orbi-instanton \eqref{eq:1^2}-\eqref{eq:2'}: the dimension of the associated 3D CB both at finite and infinite coupling (after performing $N_\text{L,R}$ small instanton transitions), the dimension of the 6D HB both at finite and infinite coupling (i.e. in the low-energy quiver and in the SCFT, respectively), the dimension of the 4D HB for the class-S model \cite{Gaiotto:2009we} obtained by $T^2$ compactification of the 6D orbi-instanton.

As already stated above, the dimension of the 3D CB is given by the total gauge rank minus one, whereas the dimension of the 6D HB at finite coupling is given by counting the total number of hypermultiplets and subtracting the total number of vector multiplets, $n_\text{H}-n_\text{V}$. At infinite coupling, i.e. at the origin of the tensor branch, we should turn each tensor multiplet into twenty-nine hypers.\footnote{\ This is because each orbi-instanton half of the LST is an ``(obviously) very Higgsable'' theory in the language of \cite{Ohmori:2015pua,Ohmori:2015pia}, and thus at the origin of the tensor branch each dynamical tensor turns into twenty-nine hypers, as proven in \cite{Mekareeya:2017sqh}.} The dimension of the HB of the LST ``at infinite coupling'' (i.e. with only one 2 curve of finite K\"ahler size, $M_\text{s}^2$) is obtained simply by summing the dimensions of the infinite-coupling HB of its two orbi-instanton constitutents (i.e. $(n_\text{H}+29n_\text{T})-n_\text{V}$) and subtracting the dimension of the central $\mathfrak{su}(k)$ gauge algebra, as it provides $k^2-1$ new vectors. This explains the appeareance of the term $-(k^2-1)$ term in \eqref{eq:dim3DLST}.

\subsection{Gravitational anomaly matching}

As a further independent check, it was already shown in \cite{Intriligator:1997dh} that, for the case $(\mu_\text{L},\mu_\text{R})=([1^k],[1^k])$, 't Hooft gravitational anomaly matching imposes 
\begin{equation}
\dim_\mathbb{H}\text{HB}_\text{6D}^{M_\text{s}} (\text{LST}_{([1^k],[1^k])}) = (n_\text{H}+29n_\text{T})-n_\text{V}=30M+r\ ,    
\end{equation}
with $M$ the total number of small instantons of $E_8\times E_8$ and $r$ the number of resolution (i.e. K\"ahler) parameters of the $\mathbb{C}^2/\mathbb{Z}_k$ orbifold. Namely:
\begin{equation}
\label{eq:gravan1k1k}
    \dim_\mathbb{H}\text{HB}_\text{6D}^{M_\text{s}} (\text{LST}_{([1^k],[1^k])})  = 30M+r=30(N+2k)+k-1=30(N_\text{L}+N_\text{R})+61k-1\ .
\end{equation}
This can easily be generalized to any other choice $(\mu_\text{L},\mu_\text{R})$; we simply need to compute the gravitational anomaly of the LST, since
\begin{equation}
    I_8^\text{LST} \supset \dim_\mathbb{H}\text{HB}_\text{6D}^{M_\text{s}}(\text{LST}_{(\mu_\text{L},\mu_\text{R})})\, \frac{7p_1(T)^2-4p_2(T)}{5760}\ ,
\end{equation}
where $I_8^\text{LST}$ is the eight-form anomaly polynomial. The contribution of each of the two orbi-instanton constituents has already been computed in \cite{Mekareeya:2017jgc}; putting everything together we obtain
\begin{align}
    I_8^\text{LST} \supset &\Bigg[ \frac{7 (k (k+61)+60 N_\text{L}-2 (\left\langle \bm{w}_\text{L},\bm{\rho} \right\rangle+1))}{11520} \ + \nonumber \\ &+\frac{7 (k (k+61)+60 N_\text{R}-2 (\left\langle \bm{w}_\text{R},\bm{\rho} \right\rangle+1))}{11520} +\frac{-7 \left(k^2-1\right)}{5760}\Bigg]p_1(T)^2
\end{align}
and equivalently for $p_2(T)$. In the third term in parenthesis we have added the contribution of $k^2-1$ vectors (coming from the new decorated $\overset{\mathfrak{su}(k)}{2}$ curve), each contributing $-\tfrac{7}{5760}\,p_1(T)^2$.\footnote{\ The tensor multiplet associated to this new 2 curve does \emph{not} contribute to the total coefficient of $p_1(T)^2$ or $p_2(T)$ in $I_8^\text{LST}$: since the curve cannot be shrunk to zero size, the associated tensor scalar is nondynamical.}
All in all we obtain
\begin{equation}\label{eq:gravdim}
    \dim_\mathbb{H}\text{HB}_\text{6D}^{M_\text{s}}(\text{LST}_{(\mu_\text{L},\mu_\text{R})}) = 30(N_\text{L}+N_\text{R})+61k-\left\langle \bm{w}_\text{L},\bm{\rho} \right\rangle-\left\langle \bm{w}_\text{R},\bm{\rho} \right\rangle-1\ ,
\end{equation}
which satisfactorily matches with \eqref{eq:dim3DLST} (and reduces to \eqref{eq:gravan1k1k} for $(\mu_\text{L},\mu_\text{R})=([1^k],[1^k])$, since $\left\langle \bm{w}_\text{L,R},\bm{\rho} \right\rangle=0$ in that case -- see again \eqref{eq:height}).

\subsection{4D class-S fixtures}

The 4D theory obtained by $T^2$ compactification of the orbi-instanton \cite{Mekareeya:2017jgc} is an A-type fixture \cite{Chacaltana:2010ks}, call it $\mathsf{T}_{P_\text{L,R}}\{Y_1,Y_2,Y_3\}$, with three regular punctures
\begin{align}\label{eq:fix}
    Y_1 &= [M_\text{L,R}-n_6, M_\text{L,R}-n_6-n_5, \ldots, M_\text{L,R}-n_6-n_5-n_4-n_3-n_2-n_1,1^k]\ ,\\
    Y_2 &= [2M_\text{L,R}+2n_{4'}+n_{3'}+n_{2'}, 2M_\text{L,R}+n_{4'}+n_{3'}+n_{2'}, 2M_\text{L,R}+n_{4'}+n_{3'}]\ ,\\
    Y_3 &= [3M_\text{L,R}+2n_{4'}+2n_{3'}+n_{2'}, 3M_\text{L,R}+2n_{4'}+n_{3'}+n_{2'}]\ ,
\end{align}
which are integer partitions of
\begin{equation}
P_\text{L,R} \equiv 6M_\text{L,R}+k-n_1-2n_2-\ldots-6n_6=6M_\text{L,R}+2n_{2'}+3n_{3'}+4n_{4'}\ .
\end{equation}
The $n_i,n_{i'}$ are the multiplicites of the parts in a Kac label of $k$ as in \eqref{eq:kac}. This fixture can be understood as a modification of the one realizing the rank-$6M$ $E_8$ Minahan--Nemeschansky theory (which has $Y_1=[M^6],Y_2=[(2M)^3],Y_3=[(3M)^2]$ and is indeed of type $A_{6M-1}$). 

The dimension of the HB of this fixture can easily be computed as follows \cite{Chacaltana:2012zy}:
\begin{align}
    2\dim_\mathbb{H} \text{HB}_\text{4D}(\mathsf{T}_{P_\text{L,R}}\{Y_1,Y_2,Y_3\}) = 3 \dim_\mathbb{R} SU(P_\text{L,R})-\rank SU(P_\text{L,R})-\sum_{i=1}^3 \dim_\mathbb{C}Y_i\ ,
\end{align}
where by $\dim_\mathbb{C}Y_i$ we mean the complex dimension of the nilpotent orbit of $\mathfrak{su}(P_\text{L,R})$ defined by the partition $Y_i=[n_1,\ldots,n_p]$ of $P_\text{L,R}$, which is given by
\begin{equation}
    \dim_\mathbb{C}Y_i=P_\text{L,R}^2 -\sum_{j=1}^{p'} s_j^2\ ,
\end{equation}
with $Y_i^\text{t}=[s_1,\ldots,s_{p'}]$ the transpose partition of $Y_i$ (obtained by reflexion along a diagonal).

\subsection{Matching dimensions for \texorpdfstring{$k=2$}{k=2}}

We can now compute all relevant dimensions as explained at the beginning of this section, and perform various nontrivial checks of our proposal. 

We begin with the orbi-instantons of section \ref{sub:k=2study}. For the Kac labels $[1^2],[2],[2']$ of $k=2$ we find:\footnote{\ As a curiosity, we point out that $\dim_\mathbb{H}\text{CB}_\text{3D}^\infty=\dim_\mathbb{H}\text{HB}_\text{6D,$\mu_\text{L,R}$}^\infty$ also equals the dimension of certain strata (or symplectic leaves) in the double affine Grassmannian of $E_8$ specified by the Kac diagram $\mu_\text{L,R}$, which together with $N_\text{L,R}$ and $k$ identifies the chosen orbi-instanton \cite{Fazzi:2023ulb}. Notice that the dimensions appearing in 
\eqref{eq:dim12}-\eqref{eq:dim2}-\eqref{eq:dim2p} match with those in \cite{Mekareeya:2017jgc} only upon sending $N_\text{L,R} \to N_\text{L,R}-N_{\mu_\textsc{l,r}} \equiv N_\text{there}$. This is because of the different definition of the length of the plateau between the two papers.\label{foot:orbidim}}
\begin{equation}\label{eq:dim12}
    \begin{matrix} [1^2] \\ {\scriptstyle M_\text{L,R}=N_\text{L,R}+2} \\ {\scriptstyle N_{\mu_\textsc{l,r}}=l_\text{L,R}=2} \end{matrix} : \begin{cases}\dim_\mathbb{H}\text{CB}_\text{3D} = 4+N_\text{L,R}  \\
    \dim_\mathbb{H}\text{CB}_\text{3D}^\infty = 4+N_\text{L,R}+29 M_\text{L,R} = 62+30N_\text{L,R} \\ %= \dim_\mathbb{H}\text{CB}_\text{3D}+3\cdot 29 \\
    \dim_\mathbb{H}\text{HB}_\text{6D} = 1\cdot 2 + 2\cdot 1 + N_\text{L,R} \cdot 2\cdot 2 - N_\text{L,R} \cdot (2^2-1) = 4+N_\text{L,R}\\
    \dim_\mathbb{H}\text{HB}_\text{6D}^\infty = 4+N_\text{L,R}+29 M_\text{L,R} = 62+30N_\text{L,R} \\
    \dim_\mathbb{H} \text{HB}_\text{4D}(\mathsf{T}_{6(N_\text{L,R}+2)}\{Y_1,Y_2,Y_3\}) = 62+30N_\text{L,R}
    \end{cases}\hspace{-.3cm},
\end{equation}
\begin{equation}\label{eq:dim2}
    \begin{matrix} [2] \\ {\scriptstyle M_\text{L,R}=N_\text{L,R}+1} \\ {\scriptstyle N_{\mu_\textsc{l,r}}=l_\text{L,R}=1} \end{matrix} : \begin{cases}\dim_\mathbb{H}\text{CB}_\text{3D} = 4+N_\text{L,R} \\
    \dim_\mathbb{H}\text{CB}_\text{3D}^\infty = 4+N_\text{L,R} +29 M_\text{L,R}= 33+30N_\text{L,R} \\ %= \dim_\mathbb{H}\text{CB}_\text{3D}+3\cdot 29 \\
    \dim_\mathbb{H}\text{HB}_\text{6D} = 2\cdot 2 + N_\text{L,R} \cdot 2\cdot 2 - N_\text{L,R} \cdot (2^2-1)= 4+N_\text{L,R}\\
    \dim_\mathbb{H}\text{HB}_\text{6D}^\infty = 4+N_\text{L,R}+29 M_\text{L,R} = 33+30N_\text{L,R}\\
    \dim_\mathbb{H} \text{HB}_\text{4D}(\mathsf{T}_{6(N_\text{L,R}+1)}\{Y_1,Y_2,Y_3\}) = 33+30N_\text{L,R}
    \end{cases}\hspace{-.3cm},
\end{equation}
\begin{equation}\label{eq:dim2p}
    \begin{matrix} [2'] \\ {\scriptstyle M_\text{L,R}=N_\text{L,R}} \\ {\scriptstyle N_{\mu_\textsc{l,r}}=l_\text{L,R}=0} \end{matrix} : \begin{cases}\dim_\mathbb{H}\text{CB}_\text{3D} = 16+N_\text{L,R} \\
    \dim_\mathbb{H}\text{CB}_\text{3D}^\infty = 16+N_\text{L,R}+29 M_\text{L,R}=16+30N_\text{L,R} \\ %= \dim_\mathbb{H}\text{CB}_\text{3D}+3\cdot 29 \\
    \dim_\mathbb{H}\text{HB}_\text{6D} = {\scriptstyle 16\cdot 2 \cdot \frac{1}{2} + 2\cdot 2 +  (N_\text{L,R}-1) \cdot 2\cdot 2 - (2^2-1) - (N_\text{L,R}-1) \cdot (2^2-1)} =16+N_\text{L,R}\\
    \dim_\mathbb{H}\text{HB}_\text{6D}^\infty = 17+N_\text{L,R}+ 29M_\text{L,R} =16+30 N_\text{L,R}\\
    \dim_\mathbb{H} \text{HB}_\text{4D}(\mathsf{T}_{6N_\text{L,R}+2}\{Y_1,Y_2,Y_3\}) = 16+30 N_\text{L,R}
    \end{cases}\hspace{-.3cm}.
\end{equation}
We are finally ready to test our proposal \eqref{eq:generalfin}-\eqref{eq:generalinf}: the CB dimension of the 3D magnetic quiver in \eqref{eq:generalinf} has to match the dimension of the HB of the 6D LST at infinite coupling, which may alternatively be found as the dimension of the HB of a \emph{new} class-S theory, call it $\mathsf{S}_{M'}\{Y'_1,Y'_2,Y'_3,Y'_4\}$, obtained by colliding $Y_1^\text{L}$ with $Y_1^\text{R}$ ``along'' their $[1^k]$ part (i.e. gauging a diagonal $SU(k)$ subgroup of the flavor symmetries associated with $Y_1^\text{L}$ and $Y_1^\text{R}$).
Therefore:
\begin{align}
    \dim_\mathbb{H}\text{HB}_\text{6D}^{M_\text{s}} (\text{LST}_{(\mu_\text{L},\mu_\text{R})}) =& \,30N+61k-\left\langle \bm{w}_\text{L},\bm{\rho} \right\rangle-\left\langle \bm{w}_\text{R},\bm{\rho} \right\rangle-1 \\
    =&\dim_\mathbb{H}\text{HB}_{\text{6D},\mu_\text{L}}^\infty + \dim_\mathbb{H}\text{HB}_{\text{6D},\mu_\text{R}}^\infty - \dim_\mathbb{R} SU(k)_\text{diag} \\
    =&\dim_\mathbb{H}\text{CB}^\infty_\text{3D}\text{\eqref{eq:generalinf}} \\    
    =&\dim_\mathbb{H} \text{HB}_\text{4D}(\mathsf{T}_{P_\text{L}}\{Y_1^\text{L},Y_2^\text{L},Y_3^\text{L}\})\, + \nonumber \\ 
    &+ \dim_\mathbb{H} \text{HB}_\text{4D}(\mathsf{T}_{P_\text{R}}\{Y_1^\text{R},Y_2^\text{R},Y_3^\text{R}\})- \dim_\mathbb{R} SU(k)_\text{diag}\\
    =&\dim_\mathbb{H} \text{HB}_\text{4D}(\mathsf{S}_{M'}\{Y'_1,Y'_2,Y'_3,Y'_4\}) \label{eq:classSdim}\ .
\end{align}
It is straightforward to check that this is indeed the case for all possibilities $(\mu_\text{L},\mu_\text{R})$ of $k=2$. Recalling the definition $N=N_\text{L}+N_\text{R}$, we find:
\begin{align}
\label{eq:dimLST1k1k} 
%\begin{array}{@{}>{\displaystyle}l@{}>{\displaystyle{}}l@{}}
    \dim_\mathbb{H}\text{HB}^{M_\text{s}}_\text{6D}(\text{LST}_{([1^2],[1^2])}) &= 30N+61\cdot 2-0-0-1  \\
    &= 62+30N_\text{L}+62+30N_\text{R}-3 = 121+30N \ ,\nonumber \\
    \dim_\mathbb{H}\text{HB}^{M_\text{s}}_\text{6D}(\text{LST}_{([2],[2])})  &= 30N+61\cdot 2-29-29-1 \\
    &=  33+30N_\text{L}+33+30N_\text{R}-3 = 63 +30N\ , \nonumber \\
    \dim_\mathbb{H}\text{HB}^{M_\text{s}}_\text{6D}(\text{LST}_{([2'],[2'])})  &= 30N+61\cdot 2-46-46-1 \\
    &= 16+30N_\text{L}+16+30N_\text{R}-3 = 29 +30N\ , \nonumber \\
    \dim_\mathbb{H}\text{HB}^{M_\text{s}}_\text{6D}(\text{LST}_{([1^2],[2])}) &= 30N+61\cdot 2-0-29-1 \\
    &= 62+30N_\text{L}+33+30N_\text{R}-3 =  92 +30N\ , \nonumber \\
    \dim_\mathbb{H}\text{HB}^{M_\text{s}}_\text{6D}(\text{LST}_{([1^2],[2'])}) &= 30N+61\cdot 2-0-46-1 \\
    &= 62+30N_\text{L}+16+30N_\text{R}-3 = 75  +30N\ , \nonumber \\
    \dim_\mathbb{H}\text{HB}^{M_\text{s}}_\text{6D}(\text{LST}_{([2],[2'])}) &= 30N+61\cdot 2-29-46-1 \\
    &= 33+30N_\text{L}+16+30N_\text{R}-3 = 46 +30N\ . \nonumber
%\end{array}
\end{align}

\section{3D T-dualities}
\label{sec:3Ddual}

In \cite{DelZotto:2022ohj,DelZotto:2022xrh} the authors considered the action of T-duality on the $(e')$ LSTs (essentially obtained by swapping the roles of NS5s and D6's in their Type I' engineering, i.e. via a 9-11 flip in the Ho\v{r}ava--Witten M-theory setup), and proposed a series of $(o')$ duals. In this section we would like to construct the 3D magnetic quivers of the proposed T-duals and make some comments about their relation with those presented in section \ref{sub:rule}.

To construct the $(o')$ T-duals (all coming from the $(o)$ LST on a $\mathbb{C}^2/\mathbb{Z}_{2\tilde{k}}$ orbifold) we must  specify an embedding $\lambda : \mathbb{Z}_{2\tilde{k}} \to Spin(32)/\mathbb{Z}_2$. We restrict our attention to the heterotic string ``without vector structure'' (in the language on \cite{Witten:1997bs}), i.e. the second Stiefel--Whitney class of the compactification vanishes \cite{Berkooz:1996iz}.\footnote{\ For the case in which it does not vanish, see \cite{Aspinwall:1996vc}.} The embedding is concretely determined by the relative position of the 16 D8's with respect to the $\tilde{k}$ physical NS5s along the interval. 

Rather than constructing the $(o')$ T-duals in full generality (i.e. for any choice of $M,k$ on the $(e')$ side), we will focus on a few concrete examples. Take $k= 2\tilde{k}=2$. In this case $\lambda$ is determined by a choice of two integers $w_i$ such that $16=w_1+w_2$ (the numbers of D8's before and after the NS5). It is convenient to parameterize them as
\begin{equation}\label{eq:ws}
    w_1=2p\ , \quad w_2=16-2p\ ,
\end{equation}
and we can restrict our attention to the cases $p=0,\ldots,4$ without loss of generality.\footnote{\ We also neglect the case denoted $w_1=w_2=8^*$ in \cite{DelZotto:2022ohj} for the reasons explained therein (briefly, it is equivalent to $w_1=w_2=8$ upon shifting $\tilde{N}\to \tilde{N}-1$).}  The electric quiver and Type I' engineering are given by, respectively:
\begin{equation}\label{eq:so32typeI'}
    [SO(4p)] \, \overset{\mathfrak{usp}(2\tilde{N})}{1}\, \overset{\mathfrak{usp}(2\tilde{N} +8-2p)}{1}\,  [SO(32-4p)] \ , \quad \begin{tikzpicture}[scale=.75,baseline]
	% NS5
	\node at (0,0) {};
	%\draw[draw,fill=black,circle,label=above:{\footnotesize $k$},minimum size=3.5pt,inner sep=0pt, outer sep=0pt] (3.6,0) circle (0.1cm);
 \node[draw,fill=black,circle,label=above:{\footnotesize 1},minimum size=5pt,inner sep=0pt, outer sep=0pt] (CircleNode) at (3.6,0) {};

	%D6
	
	\draw[solid,black,thick] (0,0)--(3.6,0) node[black,midway,yshift=0.3cm] {\footnotesize $2\tilde{N}$};
	\draw[solid,black,thick] (3.6,0)--(7.2,0) node[black,midway,yshift=0.3cm] {\footnotesize $2\tilde{N}+8-2p$};

	%D8-O8
	\draw[dashed,black,very thick] (0,-.5)--(0,.5) node[black,midway, xshift =0cm, yshift=-1.5cm] {} node[black,midway, xshift =0cm, yshift=1.5cm] {};
	\draw[solid,black,very thick] (0.25,-.5)--(0.25,.5) node[black,midway, xshift =0cm, yshift=+.75cm] {\footnotesize $2p$};
        \draw[solid,black,very thick] (7.2-0.25,-.5)--(7.2-0.25,.5) node[black,midway, xshift =0cm, yshift=+.75cm] {\footnotesize $16-2p$};
        \draw[dashed,black,very thick] (7.2,-.5)--(7.2,.5) node[black,midway, xshift =0cm, yshift=-1.5cm] {} node[black,midway, xshift =0cm, yshift=1.5cm] {};    
\end{tikzpicture}.
\end{equation}
To read off the magnetic quiver, we first perform $8-2p$ Hanany--Witten moves (i.e. we move $8-2p$ D8's from the right to the left and across the NS5, generating $8-2p$ D6's behind them, and leaving only 8 D8's in the right stack), and then we lift the NS5 off of the D6's (stretching D4's between D6's and NS5-D6's):
\begin{equation}\label{fig:o'i'}
    \begin{tikzpicture}[baseline,scale=1]
	% NS5
	\node at (0,0) {};
	%\draw[fill=black,label=right:$\times 5$] (0.5+.2,0.5) circle (0.075cm);
	 \node[draw,fill=black,circle,label=above:{\footnotesize 1},minimum size=5pt,inner sep=0pt, outer sep=0pt] (CircleNode) at (1.8,0.5) {};
	\node at (-0.8,1.25) {\footnotesize $\overbrace{\hspace{1.65cm}}^{8-2p}$};
 %D6
	
        \draw[solid,black,thick] (0.30,-0.4)--(0.55,0);
        \draw[solid,black,thick] (0.30,-0.2)--(0.55,0);
        \draw[solid,black,thick] (0.3,0.2)--(0.55,0);
        \draw[solid,black,thick] (0.3,0.2)--(0,0.2) node[black,midway,yshift=0.2cm] {\scriptsize $1$};;
        \draw[solid,black,thick] (-2.5,-0.4)--(0.30,-0.4) node[black,xshift=-2.25cm,yshift=0.2cm] {\scriptsize $2\tilde{N}$};
        \draw[solid,black,thick] (-1.6,-0.2)--(0.30,-0.2) node[black,xshift=-1.45cm,yshift=0.2cm] {\tiny $2\tilde{N}+1$};
        \draw[solid,black,thick] (0.55,0)--(3,0) node[black,midway,yshift=-0.3cm] {\scriptsize $2(\tilde{N}+4-p)$};
	\draw[solid,black,thick] (3,0)--(3.25,0.2); 
        \draw[solid,black,thick] (3,0)--(3.25,-0.2);
        \draw[solid,black,thick] (3.25,0.2)--(5,0.2) node[black,midway,yshift=0.2cm] {\scriptsize $\tilde{N}+4-p$};
        \draw[solid,black,thick] (3.25,-0.2)--(5,-0.2) node[black,midway,yshift=-0.25cm] {\scriptsize $\tilde{N}+4-p$};

	%D8-O8
	\draw[dashed,black,very thick] (-2.5,-.5)--(-2.5,.5) node[black,midway, xshift =0cm, yshift=-1.5cm] {} node[black,midway, xshift =0cm, yshift=1.5cm] {};
	\draw[solid,black,very thick] (-2.25,-.5)--(-2.25,.5) node[black,midway, xshift =0cm, yshift=+.75cm] {\footnotesize $2p$};
	\draw[solid,black,very thick] (-1.6,-.5)--(-1.6,.5) node[black,midway, xshift =0cm, yshift=+.75cm] {\footnotesize $1$};
        \draw[solid,black,very thick] (-0.7,-.5)--(-0.7,.5) node[black,midway, xshift =0cm, yshift=+.75cm] {\footnotesize $1$};
        \draw[solid,black,very thick] (0,-.5)--(0,.5) node[black,midway, xshift =0cm, yshift=+.75cm] {\footnotesize $1$};
        \path (-0.7,0)--(0,0) node[black,midway,yshift=0cm] {\footnotesize $\cdots$};

        \draw[solid,black,thick] (4.5+0.5,-0)--(4.65+0.5,0) node[black,midway,yshift=-0.2cm] {};
        \draw[solid,black,thick] (4.5+0.5,-0.4)--(4.65+0.5,-0.4) node[black,midway,yshift=-0.2cm] {};
        \draw[solid,black,thick] (4.65+0.5,0.2)--(4.8+0.5,0.2) node[black,midway,yshift=-0.2cm] {};
        \draw[solid,black,thick] (4.65+0.5,-0.2)--(4.8+0.5,-0.2) node[black,midway,yshift=-0.2cm] {};
        \draw[solid,black,thick] (4.8+0.5,0)--(4.95+0.5,0) node[black,midway,yshift=-0.2cm] {};
        \draw[solid,black,thick] (4.8+0.5,-0.4)--(4.95+0.5,-0.4) node[black,midway,yshift=-0.2cm] {};
        \draw[solid,black,thick] (4.95+0.5,0.2)--(5.1+0.5,0.2) node[black,midway,yshift=-0.2cm] {};
        \draw[solid,black,thick] (4.95+0.5,-0.2)--(5.1+0.5,-0.2) node[black,midway,yshift=-0.2cm] {};
        \draw[solid,black,thick] (5.1+0.5,0)--(5.25+0.5,0) node[black,midway,yshift=-0.2cm] {};
        \draw[solid,black,thick] (5.1+0.5,-0.4)--(5.25+0.5,-0.4) node[black,midway,yshift=-0.2cm] {};
        \draw[solid,black,thick] (5.25+0.5,0.2)--(5.4+0.5,0.2) node[black,midway,yshift=-0.2cm] {};
        \draw[solid,black,thick] (5.25+0.5,-0.2)--(5.4+0.5,-0.2) node[black,midway,yshift=-0.2cm] {};
        \draw[solid,black,thick] (5.4+0.5,0)--(5.55+0.5,0) node[black,midway,yshift=-0.2cm] {};
        \draw[solid,black,thick] (5.4+0.5,-0.4)--(5.55+0.5,-0.4) node[black,midway,yshift=-0.2cm] {};
        \draw[solid,black,thick] (5.55+0.5,0)--(5.7+0.5,-0.1) node[black,midway,yshift=-0.2cm] {};
        \draw[solid,black,thick] (5.7+0.5,-0.1)--(5.55+0.5,-0.2) node[black,midway,yshift=-0.2cm] {};

	%D8-O8

        \draw[solid,black,very thick] (4.5+0.5,-.5)--(4.5+0.5,.5) node[black,midway, xshift =0cm, yshift=+.75cm] {\footnotesize $1$};
        \draw[solid,black,very thick] (4.65+0.5,-.5)--(4.65+0.5,.5) node[black,midway, xshift =0cm, yshift=+.75cm] {\footnotesize $1$};
        \draw[solid,black,very thick] (4.8+0.5,-.5)--(4.8+0.5,.5) node[black,midway, xshift =0cm, yshift=+.75cm] {\footnotesize $1$};
        \draw[solid,black,very thick] (4.95+0.5,-.5)--(4.95+0.5,.5) node[black,midway, xshift =0cm, yshift=+.75cm] {\footnotesize $1$};
        \draw[solid,black,very thick] (5.1+0.5,-.5)--(5.1+0.5,.5) node[black,midway, xshift =0cm, yshift=+.75cm] {\footnotesize $1$};
        \draw[solid,black,very thick] (5.25+0.5,-.5)--(5.25+0.5,.5) node[black,midway, xshift =0cm, yshift=+.75cm] {\footnotesize $1$};
        \draw[solid,black,very thick] (5.4+0.5,-.5)--(5.4+0.5,.5) node[black,midway, xshift =0cm, yshift=+.75cm] {\footnotesize $1$};
        \draw[solid,black,very thick] (5.55+0.5,-.5)--(5.55+0.5,.5) node[black,midway, xshift =0cm, yshift=+.75cm] {\footnotesize $1$};

        \draw[dashed,black,very thick] (5.7+0.5,-.5)--(5.7+0.5,.5) node[black,midway, xshift =0cm, yshift=-1.5cm] {} node[black,midway, xshift =0cm, yshift=1.5cm] {};
\end{tikzpicture}\ .
\end{equation}
The $2\tilde{N}$ D6's in the left portion of the setup (those which cross the left O8) must be broken along the $2p$ D8's, following the same pattern as seen on the right. 

Calling $L=2\tilde{N}+8-2p$, we can read off the finite-coupling magnetic quiver; for $p=0$ we have
\begin{equation}\label{eq:p0quiv}
\scriptstyle
     (L-8)/2-\hspace{-.15cm}\overset{\underset{\scriptstyle \vert}{\scriptstyle (L-6)/2}}{(L-6)}\hspace{-.15cm}-(L-5)-(L-4)-(L-3)-(L-2)-(L-1)-\overset{\underset{\scriptstyle \vert}{\scriptstyle 1}}{L}-L-L-L-L-L-\hspace{-.15cm}\overset{\underset{\scriptstyle \vert}{\scriptstyle L/2}}{L}\hspace{-.15cm}-L/2\ ,
\end{equation}
with $\dim_\mathbb{H}\text{CB}_\text{3D}=15L-28=30\tilde{N}+92$. %, \red{and $\dim_\mathbb{H}\text{CB}_\text{3D}^\infty=15L=30(\tilde{N}+4-p)$ upon performing a single small $E_8$ instanton transition NOT $Spin(32)$ transition?}.
For $p=1,2,3$ we have instead
\begin{equation}\label{eq:pneq0quiv}
\scriptstyle
     (L-8+2p)/2\, -\underbrace{\hspace{-.1cm}\overset{\underset{\scriptstyle \vert}{\scriptstyle (L-8+2p)/2}}{\scriptstyle(L-8+2p)}\scriptstyle -\cdots-\scriptstyle(L-8+2p)}_{2p-1}-\underbrace{\scriptstyle(L-7+2p)-\cdots-(L-2)-(L-1)}_{7-2p}-\overset{\underset{\scriptstyle \vert}{\scriptstyle 1}}{L}-L-L-L-L-L-\hspace{-.15cm}\overset{\underset{\scriptstyle \vert}{\scriptstyle L/2}}{L}\hspace{-.15cm}-L/2\ ,
\end{equation}
and for $p=4$ we have
\begin{equation}\label{eq:p4quiv}
     L/2-\hspace{-.15cm}\overset{\underset{\scriptstyle \vert}{\displaystyle L/2}}{L}\hspace{-.15cm}-L-L-L-L-L-\overset{\underset{\scriptstyle \vert}{\displaystyle 1}}{L}-L-L-L-L-L-\hspace{-.15cm}\overset{\underset{\scriptstyle \vert}{\displaystyle L/2}}{L}\hspace{-.15cm}-L/2\ ,
\end{equation}
with
\begin{equation}\label{eq:o'dimCBfin}
    \dim_\mathbb{H}\text{CB}_\text{3D}=15L-28+p(2p-1)=30\tilde{N}+92-31p+2p^2 = \begin{cases}
        30\tilde{N}+63 & p=1 \\
        30\tilde{N}+38 & p=2 \\
        30\tilde{N}+17 & p=3 \\
        30\tilde{N} & p=4
    \end{cases}\ .
\end{equation}
The $p=4$ quiver (which is good in the sense of \cite{Gaiotto:2008ak}) is special, and engineers (through its CB) the moduli space of $L/2=\tilde{N}$ instantons of $SO(32)$ on $\mathbb{C}^2/\mathbb{Z}_2$ \cite{Mekareeya:2015bla}. It has $G_J^\text{IR} = SO(16)\times SO(16) \times SU(2)$. %It also coincides with the magnetic quiver of the class-S theory of type $A_{2\tilde{N}-1}$ with punctures:
%\begin{equation}
 %   \underbrace{[\tilde{N},\tilde{N}]\ , \cdots \ , [\tilde{N},\tilde{N}]}_{4}\ , \ \underbrace{[\tilde{N},\tilde{N}]\ , \cdots \ , [\tilde{N},\tilde{N}]}_{4}\ , \
%\end{equation}

How do we obtain the infinite-coupling version of the above magnetic quivers? We simply need to perform one small $SO(32)$ instanton transition \cite{Witten:1995gx,Intriligator:1997kq,Hanany:1997gh} (i.e. bring the NS5 into one of the O8's), which again turns one tensor into twenty-nice hypers. We propose this is done by adding an affine $D_{16}$ Dynkin-shaped quiver to the magnetic quivers (akin to the more usual $E_8$ case):
\begin{equation}
    D_{16}^{(1)}: 1- \overset{\overset{\displaystyle 1}{\vert}}{2} - 2 -2-2-2-2-2-2-2-2-2-2-\overset{\overset{\displaystyle 1}{\vert}}{2}-1\ .
\end{equation}
This is compatible with the \eqref{eq:p0quiv}-\eqref{eq:pneq0quiv}-\eqref{eq:p4quiv} quiver ``shapes'', and adds 29 quaternionic units to the dimension of the CB (once we subtract the overall decoupled $U(1)$). Adding $D_{16}^{(1)}$ once we obtain:
\begin{equation}
     \scriptstyle
     (L-6)/2-\hspace{-.15cm}\overset{\underset{\scriptstyle \vert}{\scriptstyle (L-4)/2}}{(L-4)}\hspace{-.15cm}-(L-3)-(L-2)-(L-1)-L-(L+1)-(L+2)-(L+2)-(L+2)-(L+2)-(L+2)-(L+2)-\hspace{-.15cm}\overset{\underset{\scriptstyle \vert}{\scriptstyle (L+2)/2}}{(L+2)}\hspace{-.15cm}-(L+2)/2\ , \label{eq:p0infquiv}
     \end{equation}
     for $p=0$;
     \begin{equation}
      \scriptstyle
     (L/2-3+p)\, -\underbrace{\hspace{-.025cm}\overset{\underset{\scriptstyle \vert}{\scriptstyle (L/2-3+p)}}{\scriptstyle(L-6+2p)}\scriptstyle -\cdots-\scriptstyle(L-6+2p)}_{2p-1}-\underbrace{\scriptstyle(L-5+2p)-\cdots-L-(L+1)}_{7-2p}-\underbrace{\scriptstyle (L+2)-\cdots-\hspace{-.15cm}\overset{\underset{\scriptstyle \vert}{\scriptstyle (L+2)/2}}{\scriptstyle (L+2)}}_{7}\hspace{-.025cm}-(L+2)/2\ ,\label{eq:pneq0infquiv}
\end{equation}
for $p=1,2,3$;
\begin{equation}\label{eq:p4infquiv}
   (L+2)/2-\underbrace{\overset{\underset{\scriptstyle \vert}{\displaystyle(L+2)/2}}{ (L+2)}-\cdots-\overset{\underset{\scriptstyle \vert}{\displaystyle(L+2)/2}}{ (L+2)}}_{13}-(L+2)/2
\end{equation}
for $p=4$.

Remembering that $L=2\tilde{N}+8$ or $L=2\tilde{N}+8-2p$ (if $p\neq 0$), we obtain:
\begin{equation}\label{eq:CBinfty}
    \dim_\mathbb{H}\text{CB}_\text{3D}^\infty \begin{array}{c}
    \eqref{eq:p0infquiv}\\
       \eqref{eq:pneq0infquiv}\\
         \eqref{eq:p4infquiv}
    \end{array}\hspace{-.2cm} = 15L+1+p(2p-1)=30\tilde{N}+121-31p+2p^2 = \begin{cases}
        \scriptstyle 30\tilde{N}+121 & \scriptstyle p=0 \\
        \scriptstyle 30\tilde{N}+92 & \scriptstyle p=1 \\
        \scriptstyle 30\tilde{N}+67 & \scriptstyle p=2 \\
        \scriptstyle 30\tilde{N}+46 & \scriptstyle p=3 \\
        \scriptstyle 30\tilde{N}+29 & \scriptstyle p=4
    \end{cases},
\end{equation}
and (minimum) symmetries in the IR
\begin{align}
    &G_J^\text{IR}(p=0) = SO(16)\times SU(8)\times U(1)\ , \\
    &G_J^\text{IR}(p=1,2,3) = SO(16) \times SO(4p)\times SU(8-2p) \times U(1) \ , \\
    &G_J^\text{IR}(p=4) = SO(32) \label{eq:p=4aff}\ .
\end{align}%
%where by $\widehat{SO}(32)$ we mean $D_{16}^{(1)}$ as an algebra.  
For $p=0$ it is reasonable to expect the enhancement
\begin{equation}\label{eq:enh1}
 SO(16) \times SU(8) \times U(1) \to SO(16) \times SO(16) \to SO(32)   
\end{equation}
to match with the Type I' setup in \eqref{eq:so32typeI'}; for $p=1,2,3$
\begin{equation}\label{eq:enh2}
    \begin{split}
 & SO(4p) \times SU(8-2p) \times U(1) \times SO(16) \to \\ & SO(4p) \times SO(16-4p) \times SO(16) \to \\ & SO(4p) \times  SO(32-4p)   \ ,
\end{split}
\end{equation}
by the same token. For $p=4$ the flavor algebra is naively affine $SO(32)$, but one $U(1)$ decouples. We can decide to decouple the $U(1)$ center of one of the $U((L+2)/2)$ groups (turning into $SU((L+2)/2))$, so that we are left with a finite $SO(32)$. The infinite-coupling magnetic quiver however is bad, as we will review below.

At this point, one may correctly wonder whether the magnetic quivers at infinite coupling for the $E_8\times E_8$ and $Spin(32)/\mathbb{Z}_2$ strings, i.e. \eqref{eq:genericmagquivinf} and \eqref{eq:p0infquiv}-\eqref{eq:pneq0infquiv}-\eqref{eq:p4infquiv} respectively, are related in any way (at least for $k=2\tilde{k}=2$). We propose the following picture. Denoting $\text{LST}_{(w_1,w_2)}$ the $\tilde{k}=1$ $(o')$ LSTs engineered by \eqref{eq:so32typeI'}, the T-dualities found in \cite{DelZotto:2022ohj,DelZotto:2022xrh} identify $\text{LST}_{(\mu_\text{L},\mu_\text{R})}$ (at $k=2\tilde{k}=2$) and $\text{LST}_{(w_1,w_2)}$ in the following way:
\begin{subequations}\label{eq:tdual}
\begin{align} 
    &\text{LST}_{([1^2],[1^2])} = \text{LST}_{(0,16)}\ , &&\text{LST}_{([2],[2])} = \text{LST}_{(4,12)}\ , &&\text{LST}_{([2'],[2'])} = \text{LST}_{(8,8)}\ , \\
    &\text{LST}_{([1^2],[2])} = \text{LST}_{(2,14)}\ , &&\text{LST}_{([1^2],[2'])} = \text{LST}_{(4,12)}\ , &&\text{LST}_{([2],[2'])} = \text{LST}_{(6,10)}\ .
\end{align}
\end{subequations}
In all cases but those in the central column (where two different $E_8\times E_8$ LSTs are mapped to the same $Spin(32)/\mathbb{Z}_2$ one) the 3D CBs at infinite coupling have the same dimension upon identifying $N=\tilde{N}$, again as already predicted in \cite{DelZotto:2022ohj,DelZotto:2022xrh}. The 3D HB dimensions on the contrary do not match. Remember however that T-duality between LSTs is an equivalence between \emph{compactified} theories (i.e. between effective descriptions in 5D), so we expect the explicit choice of Wilson lines on the circle to play a crucial role. For instance, the flavor symmetries $F_\text{L}\times F_\text{R}$ and $SO(4p)\times SO(32-4p)$ need to be broken to a common subgroup for the matching to occur. (There are also constraints on the so-called two-group structure constants that have to be satisfied by the T-dualities \cite{DelZotto:2020sop,DelZotto:2022ohj}.) This suggests that the infinite-coupling magnetic quivers for both sides should be \emph{modified} to accommodate this, rather than being considered appropriate descriptions of the compactified LSTs at face value. Once that is done, the two magnetic quivers should become IR dual (upon choosing an appropriate CB vacuum) and it is reasonable to expect that they can also be obtained as magnetic quivers of 5D QFTs representing the compactified LSTs.

The above point can be illustrated rather concretely. Consider e.g. the following T-duality:
\begin{equation}
    \text{LST}_{([2'],[2'])} = \text{LST}_{(8,8)}\ , \quad \dim_\mathbb{H}\text{HB}^{M_\text{s}}_\text{6D} = \dim_\mathbb{H}\text{CB}_\text{3D}^\infty = 30N+29=30\tilde{N}+29\ .
\end{equation}
The electric quivers read\footnote{\ We are making a small deviation from the notation used in \eqref{eq:genelequivLST} and \eqref{eq:usp}. Here $N=0$ means zero full instantons.}
\begin{align}
    &\text{LST}_{([2'],[2'])}: && [SO(16)] \, \overset{\mathfrak{usp}(2)}{1}\, \underbrace{\overset{\mathfrak{su}(2)}{2}\cdots \overset{\mathfrak{su}(2)}{2}}_{N}\overset{\mathfrak{usp}(2)}{1} \,[SO(16)] \ ,\\
   &\text{LST}_{(8,8)}: && [SO(16)] \, \overset{\mathfrak{usp}(2\tilde{N})}{1}\, \overset{\mathfrak{usp}(2\tilde{N})}{1} \,[SO(16)] \ ,
\end{align}
so that the flavor symmetries already match in 6D, and we naively expect that no Wilson line has to be turned on on the circle.\footnote{\ This construction can also be easily generalized to the case of even $k=2\tilde{k}>2$:
\begin{align}
    &[SO(16)] \, \overset{\mathfrak{usp}(2\tilde{k})}{1}\, \underbrace{\overset{\mathfrak{su}(2\tilde{k})}{2}\cdots \overset{\mathfrak{su}(2\tilde{k})}{2}}_{N} \overset{\mathfrak{usp}(2\tilde{k})}{1} \,[SO(16)] \ ,\\
   &[SO(16)] \, \overset{\mathfrak{usp}(2\tilde{N})}{1}\, \underbrace{\overset{\mathfrak{su}(2\tilde{N})}{2}\cdots \overset{\mathfrak{su}(2\tilde{N})}{2}}_{\tilde{k}-1} \overset{\mathfrak{usp}(2\tilde{N})}{1} \,[SO(16)] \ .
\end{align}}
Then the magnetic quivers at infinite coupling read, respectively (remember that for $[2']$ we have $N_{\mu_\textsc{l,r}}=0$):
\begin{equation}\label{eq:Td1}
\begin{split}
    & \scriptstyle  2N_\text{L} -(1+4N_\text{L}) -  \overset{\underset{\scriptstyle \vert}{\scriptstyle  (1+3N_\text{L})}}{\scriptstyle (2+6N_\text{L})}- (2+5N_\text{L})- (2+4N_\text{L})- (2+3N_\text{L})- (2+2N_\text{L})- (2+N_\text{L})-2-\\
    &\scriptstyle-(2+N_\text{R})-(2+2N_\text{R})-(2+3N_\text{R})-(2+4N_\text{R})- (2+5N_\text{R})-\overset{\underset{\scriptstyle \vert}{\scriptstyle  (1+3N_\text{R})}}{\scriptstyle (2+6N_\text{R})}-(1+4N_\text{R}) -2N_\text{R} 
    \end{split}
    \end{equation}
    (where $\cdots - 2-$ in the first line is connected to the second line in the obvious way) and
    \begin{equation}\label{eq:Td2}
    \displaystyle
     (\tilde{N}+1) -  \underbrace{\overset{\underset{\scriptstyle \vert}{\displaystyle (\tilde{N}+1)}}{\displaystyle(2\tilde{N}+2)}- \displaystyle(2\tilde{N}+2) - \cdots - \displaystyle(2\tilde{N}+2) - \overset{\underset{\scriptstyle \vert}{\displaystyle (\tilde{N}+1)}}{\displaystyle(2\tilde{N}+2)}}_{13} - (\tilde{N}+1) \ .
\end{equation}
We know T-duality imposes $N=N_\text{L}+N_\text{R}=\tilde{N}$ in the 6D setups. The first model has $G_J^\text{IR}=SO(16)\times SO(16) \times U(1)^2$, of rank 18, but each of the two $U(1)$'s is known to enhance to $SU(2)$ (because of the isometry of the $\mathbb{C}^2/\mathbb{Z}_2$ orbifold); the second however has $G_J^\text{IR}=SO(32)$ from \eqref{eq:p=4aff}, of rank 16. 

In fact \eqref{eq:Td2} is known to be bad in the sense of \cite{Gaiotto:2008ak}: some of the (dressed) monopole operators have zero or negative R-charge (below the unitarity bound). There is an overall decoupled $U(1)$ which is bad, as it has no flavors, adding a $\mathbb{C}\times \mathbb{C}^\ast \cong \mathbb{R}^3 \times S^1$ ``direction'' to the CB. This was cured in \cite{Cremonesi:2014xha} by adding an ``over-extending'' flavor node to the $D_n^{(1)}$ quiver: $\scriptstyle \tilde{N} -  \underbrace{\overset{\underset{\scriptscriptstyle \vert}{\scriptstyle \tilde{N}}}{\scriptstyle 2\tilde{N}} \scriptstyle -2\tilde{N} - \cdots - 2\tilde{N} - \overset{\underset{\scriptscriptstyle \vert}{\scriptstyle \tilde{N}}}{\scriptstyle 2\tilde{N}}}_{n} - \scriptstyle \tilde{N} - \scriptstyle \boxed{1}$. Now the rightmost $U(\tilde{N})$ is overbalanced, $G_J^\text{IR}=SO(32)\times SU(2)$, and the quiver engineers through its CB the reduced moduli space of $\tilde{N}$ instantons of $SO(32)$ on $\mathbb{C}^2$.\footnote{\ The over-extending procedure is presumably implemented by ``reducing the flavor symmetry one box at a time'' \cite{Gaiotto:2012uq}. We would like to thank S. Cremonesi for discussion on this and related points.} Since in our present situation we cannot add a flavor brane by hand to make \eqref{eq:Td2} over-extended, we simply turn one $U$ group into $SU$, and we interpret the mismatch in the ranks of $G_J^\text{IR}$ as a diagnostic for the ``incompleteness'' of \eqref{eq:Td2} to described the \emph{compactified} LST. The mismatch presumably goes away once we have a proper understanding of the T-dual magnetic quivers from 5D.

For zero instantons (i.e. when $N=\tilde{N}=0$) we can formulate a more precise statement. The two magnetic quivers are \emph{identical},   collapsing to the $D_{16}^{(1)}$ quiver itself. For $N=0$ there is no instanton transition, and indeed looking at the finite-coupling magnetic quivers \eqref{eq:2'-2'}-\eqref{eq:p4quiv} we already recognize the shape and ranks of $D_{16}^{(1)}$. Looking at the Type I' engineerings,
\begin{equation}
   (e'): \begin{tikzpicture}[scale=.75,baseline]
	% NS5
	\node at (0,0) {};
	%\draw[draw,fill=black,circle,label=above:{\footnotesize $k$},minimum size=3.5pt,inner sep=0pt, outer sep=0pt] (3.6,0) circle (0.1cm);
 %\node[draw,fill=black,circle,label=above:{\footnotesize 1},minimum size=5pt,inner sep=0pt, outer sep=0pt] (CircleNode) at (3.6,0) {};

	%D6
	
	\draw[solid,black,thick] (0,0)--(4.2,0) node[black,midway,yshift=0.2cm] {\footnotesize $2$};

	%D8-O8
	\draw[dashed,black,very thick] (0,-.5)--(0,.5) node[black,midway, xshift =0cm, yshift=-1.5cm] {} node[black,midway, xshift =0cm, yshift=1.5cm] {};
	\draw[solid,black,very thick] (0.25,-.5)--(0.25,.5) node[black,midway, xshift =0cm, yshift=+.75cm] {\footnotesize $8$};
        \draw[solid,black,very thick] (4,-.5)--(4,.5) node[black,midway, xshift =0cm, yshift=+.75cm] {\footnotesize $8$};
        \draw[dashed,black,very thick] (4.25,-.5)--(4.25,.5) node[black,midway, xshift =0cm, yshift=-1.5cm] {} node[black,midway, xshift =0cm, yshift=1.5cm] {};    
\end{tikzpicture}
\hspace{2cm}
(o'):
\begin{tikzpicture}[scale=.75,baseline]
	% NS5
	\node at (0,0) {};
	%\draw[draw,fill=black,circle,label=above:{\footnotesize $k$},minimum size=3.5pt,inner sep=0pt, outer sep=0pt] (3.6,0) circle (0.1cm);
 \node[draw,fill=black,circle,label=above:{\footnotesize 1},minimum size=5pt,inner sep=0pt, outer sep=0pt] (CircleNode) at (2.125,0) {};

	%D6
	
	%\draw[solid,black,thick] (0,0)--(3.6,0) node[black,midway,yshift=0.2cm] {\footnotesize $2\tilde{N}$};
	%\draw[solid,black,thick] (3.6,0)--(7.2,0) node[black,midway,yshift=0.2cm] {\footnotesize $2\tilde{N}+8-2p$};

	%D8-O8
	\draw[dashed,black,very thick] (0,-.5)--(0,.5) node[black,midway, xshift =0cm, yshift=-1.5cm] {} node[black,midway, xshift =0cm, yshift=1.5cm] {};
	\draw[solid,black,very thick] (0.25,-.5)--(0.25,.5) node[black,midway, xshift =0cm, yshift=+.75cm] {\footnotesize $8$};
        \draw[solid,black,very thick] (4,-.5)--(4,.5) node[black,midway, xshift =0cm, yshift=+.75cm] {\footnotesize $8$};
        \draw[dashed,black,very thick] (4.25,-.5)--(4.25,.5) node[black,midway, xshift =0cm, yshift=-1.5cm] {} node[black,midway, xshift =0cm, yshift=1.5cm] {};    
\end{tikzpicture}\ ,
\end{equation}
we see that T-duality of the LSTs (i.e. a 9-11 flip in M-theory) exchanges $k=2\tilde{k}=2$ D6's with $\tilde{k}=1$ NS5 \cite{DelZotto:2022ohj}, as expected. For the $(e')$ string $D_{16}^{(1)}$ arises from the breaking pattern of the D6s onto the $8+8$ D8s just as in \eqref{eq:2'-2'}. Then, because of T-duality, we learn that the $(o')$ string must have the same light magnetic degrees of freedom.

A quick glance at the mismatching dimensions between equations \eqref{eq:dimLST1k1k} (and following) and \eqref{eq:CBinfty} seems to suggest the following explanation. In one case we clearly see that $75 - 29 = 46$, hence we are one instanton transition away from the ``right'' dimension ($N$ and $\tilde{N}$ need to be shifted by one unit). In the other case the bad quiver is off by $4$ quaternionic units from the good one on the dual side. This also seems to indicate we are missing some degrees of freedom on the $(o')$ side. We do not have a good explanation of this remark at this stage, and, as we have stressed above, we believe the source of the explanation will lie in the 5D  matching of HBs across T-dual theories. We plan to come back to this question in future work \cite{inprogrororo}.

\medskip

We close this section with an intriguing observation. $D_{16}$ and $E_8^2$ are the only two even unimodular (i.e. self-dual) lattices in dimension 16 \cite{Conway2010-vh}, and are known to be related by T-duality of the two 9D heterotic strings \cite{Ginsparg:1986bx,Mohaupt:1992jf}, where these lattices play a central role in the construction of the (chiral bosonic) worldsheet CFT at $c_\text{L}=16,c_\text{R}=0$ \cite{Gross:1984dd} (see e.g. \cite{BoyleSmith:2023xkd,Rayhaun:2023pgc} for a modern perspective),\footnote{\ There are only two such CFTs \cite{dong2002holomorphic}, and they are precisely the worldsheet CFTs of the two 10D spacetime-supersymmetric heterotic strings.} and in the umbral moonshine (see e.g. \cite{Kachru:2016nty} for an introduction to this subject). For 3D heterotic strings with sixteen supercharges, i.e. compactified on a $T^7$, \cite{Kachru:2016ttg} showed that at special points the heterotic moduli space splits into a sum of two lattices, the $E_8$ root lattice and any of the twenty-four 24-dimensional Niemeier lattices.\footnote{\ We would like to thank B. C. Rayhaun for discussion on this point.} For 3D heterotic strings with eight supercharges, i.e. on $T^3\times \text{K3}$, it is possible that the moduli space at special points splits into two 16-dimensional lattices, each of which being a copy of either $D_{16}$ or $E_8^2$. Therefore it is not inconceivable to think that T-duality of the two heterotic strings on $T^3\times \text{K3}$ (see e.g. \cite{Berkooz:1996iz}) along an $S^1\subset T^3$ gives rise to the equivalence between the 3D magnetic quivers \eqref{eq:Td1}-\eqref{eq:Td2} when $N=\tilde{N}=0$.\footnote{\ For a suggestive relation between 2D fermionic chiral CFTs built out of the $D_{16}$ or $E_8^2$ lattices that uses fermionization, see \cite{BoyleSmith:2023xkd}.} In fact, the maximal gauge symmetry enhancement for toroidal compactifications of the heterotic strings can be obtained by means of an ``extended Dynkin diagram'' \cite{Fraiman:2018ebo,Font:2020rsk,Fraiman:2021soq,Fraiman:2021hma} (see also \cite{Fraiman:2023cpa} for the non-supersymmetric version of this statement).\footnote{\ This diagram has appeared for the first time in \cite{Ginsparg:1986bx,Mohaupt:1992jf,Cachazo:2000ey} in the heterotic context, and in \cite{Goddard:1983at,Vinberg_1972} in others.} It is amusing to notice that for the enhancement to $E_8\times E_8$ this diagram is nothing but \eqref{eq:genericmagquivinf}, which has a physical realization as a 3D QFT in our work. It also has a physical realization as the intersection graph of two-cycles of a real K3 on which one has compactified F-theory \cite{Cachazo:2000ey}, such a configuration being again dual to a brane setup in Type I', or to the 9D heterotic string with a certain $SO(32)$ Wilson line on the circle.

%%%%%%%%%%%%%%%%%%%%%%%%%%%%%%%%%
\section{Conclusions}
\label{sec:conc}
%%%%%%%%%%%%%%%%%%%%%%%%%%%%%%%%%

In this section we would like to discuss the implications of our findings in a broader context.

First of all, given the validity of \eqref{eq:classSdim}, it would be interesting to construct explicitly $\mathsf{S}_{M'}\{Y'_1,Y'_2,Y'_3,Y'_4\}$. This class-S theory is given by two fixtures connected by a tube (i.e. an $\mathcal{N}=2$ vector multiplet gauging an $SU(k)$ flavor symmetry), so it must be a sphere with four punctures $Y_1',\ldots,Y_4'$. % given by $Y_2^\text{L,R},Y_3^\text{L,R}$.
The collision of $Y_1^\text{L}$ and $Y_2^\text{R}$ along $[1^k]$ can be computed via the ``OPE of punctures'' technique introduced in \cite{Chacaltana:2012ch}. Once we have constructed $\mathsf{S}_{M'}\{Y'_1,Y'_2,Y'_3,Y'_4\}$, it should be easy to take the mirror of its circle compactification \`a la \cite{Benini:2010uu} and confirm that this is precisely our star-shaped, four-arm 3D quiver in \eqref{eq:generalinf}. For consistency, given that the latter's central node is $U(k)$, it should be possible to also understand $\mathsf{S}_{M'}\{Y'_1,Y'_2,Y'_3,Y'_4\}$ as a class-S theory of type $A_{k-1}$ (in some duality frame).

The second task is to understand the \emph{geometry}, not just the dimension, of the CB of \eqref{eq:generalinf},\footnote{\ The quantum corrected CB is notoriously hard to compute in nonabelian (quiver) gauge theories because of perturbative (at one loop) and nonperturbative corrections. Classically, it is given by $(\mathbb{R}^3 \times S^1)^{r_\text{V}}/\text{Weyl}(G)$ if the (product) gauge group $G$ has rank $r_\text{V}$, and the UV symmetry acting on it is given by $U(1)^{r_\text{V}}$. See \cite{Bullimore:2015lsa} for a proposal on how to compute quantum corrections in general.} i.e. the geometry of the HB of the LST at ``infinite coupling'' which realizes the heterotic hypermultiplet moduli space, extending nontrivially the $N=0$ results by Witten \cite{Witten:1999fq} and Sen \cite{Sen:1997js}.\footnote{\ In our language, they only considered the case $(\mu_\text{L},\mu_\text{R})=([1^k],[1^k])$.} For $k=2$ Witten found that the moduli space is smooth, and is the Atiyah--Hitchin manifold of $\dim_\mathbb{H}=1$, i.e. the simplest hyperk\"ahler space. For higher $k$ Sen found that the space is (the smoothing of) a multi-Taub--NUT one, with topology $(\mathbb{R}^3\times S^1)^k/S_k$ ($S_k$ being the symmetric group of $k$ letters, whose standard action coincides with that of $\text{Weyl}(SU(k))$).\footnote{\ Satisfactorily, for both theories the techniques of \cite{Bullimore:2015lsa} yield the same quantum corrected (i.e. smooth) result.} Also in presence of small instantons (i.e. when $N\neq 0$), string theory suggests \cite{Hanany:1999ui} that the heterotic hypermultiplet moduli space should again be smooth. (The smoothness statement is translated into D-brane charge conservation in the Type I' engineering.) Thanks to the mapping of the problem to the LST setup, we can put forth the following picture. It has been proposed \cite{Fazzi:2022hal,Fazzi:2022yca,Fazzi:2023ulb} that the HB at infinite coupling of an orbi-instanton is a stratum of the so-called affine Grassmannian of $E_8$ (more precisely, of the \emph{double} affine Grassmannian of $E_8$ \cite{Braverman:2007dvq,Finkelberg:2017nbc} once one accounts for small $E_8$ instanton transitions). Strata and slices are classified \cite{malin-ostrik-vybornov}, and the connection to CBs of 3D $\mathcal{N}=4$ theories has already been made in multiple papers (see e.g. \cite{Nakajima:2015txa,Braverman:2016wma,Braverman:2016pwk,Bullimore:2015lsa,Bourget:2021siw} and references therein). It remains to be understood how to ``glue'' two such strata, coming from the right and left orbi-instanton needed to construct the wanted LST at infinite coupling. Because 3D $\mathcal{N}=4$ CBs, or 6D $(1,0)$ HBs, are hyperk\"ahler cones (and thus may be $c_1=0$ examples of symplectic singularities \cite{beauville}) it is natural to apply the holomorphic symplectic quotient construction of \cite{Moore:2011ee} (w.r.t. the diagonal action of the $[SU(k)]$ that we are gauging, i.e. the central $\mathfrak{su}(k)$ algebra at finite coupling in the LST). The outcome should be a new hyperk\"ahler space which coincides with the HB of the class-S theory $\mathsf{S}_{M'}\{Y'_1,Y'_2,Y'_3,Y'_4\}$.\footnote{\ The construction generalizes the better known hyperk\"ahler quotient of \cite{Hitchin:1986ea}, and requires to define 2D TQFTs associated with 4D class-S theories. For us, the two class-S theories are the fixtures with punctures as in \eqref{eq:fix}. Then one applies the ``$\eta_{SU(k)_\mathbb{C}}$-functor'' of \cite{Moore:2011ee} to construct the 2D TQFT associated with the sphere with 4 punctures obtained by gluing $Y_1^\text{L}$ with $Y_1^\text{R}$ along $[1^k]$ with a tube. See also \cite{Dancer:2020wll}.} Checking smoothness is likewise nontrivial.

A third direction would be to investigate further into the action of T-dualities on 3D quivers, and determine whether they always (or, if not, under which conditions) induce 3D dualities between magnetic quivers associated with compactified LSTs. It is also known that the LSTs enjoy higher-form symmetries \cite{DelZotto:2022ohj}, and it would be interesting to determine their avatar in 3D (for instance, as a choice of global structure for the groups appearing in the magnetic quiver).\footnote{\ We would like to thank N. Mekareeya for discussions on this point. Subtle effects are known to arise for compactifications of 6D theories even on tori \cite{Gukov:2020btk}. See e.g. \cite{Carta:2023bqn} for concrete examples.}

Finally, we note that the $(2,0)$ LSTs (which are also classified by ADE groups \cite{Witten:1995zh,Seiberg:1997zk,Losev:1997hx}) compactified on a cylinder with punctures (associated with partitions of any semisimple, simply laced Lie algebra, i.e. of ADE type) have been studied in \cite{Aganagic:2015cta,Haouzi:2016ohr,Haouzi:2016yyg}. They admit a description as 4D quivers depending on the punctures (reminiscent of class-S constructions), and as 2D $(2,0)$ SCFTs of type ADE if we take a further compactification on $T^2$ followed by the field theory limit $M_\text{s} \to \infty$ (which is reminiscent of the AGT correspondence \cite{Alday:2009aq}). The Coulomb moduli of this 2D SCFT are the same as those of a 3D $\mathcal{N}=4$ SCFT, whose CB is given by a slice in an affine Grassmannian. It would be very interesting to investigate whether a similar analysis carries over to the $(1,0)$ LSTs (in particular $(e')$), and if so whether there is any connection with our conjecture on the holomorphic symplectic quotient along $SU(k)$ of two slices of the affine Grassmannian of $E_8$ realizing the HB or the LSTs.

An obvious extension of this work would be to repeat the whole construction for $\mathbb{C}^2/\mathbb{D}_k$ and $\mathbb{C}^2/\Gamma_E$ orbifolds of the heterotic string. The D and E-type orbi-instantons lack a simple ``Kac label classification'', but can nonetheless be constructed and given an F-theory electric quiver \cite{DelZotto:2014hpa,Frey:2018vpw}. In type D the associated magnetic quivers can be constructed extending the rules in \cite{Cabrera:2019dob,Cabrera:2019izd,Sperling:2021fcf}. It should then be feasible to propose an analog of \eqref{eq:generalinf}. (In type E, the orbi-instanton electric quivers can once again be constructed \cite{Frey:2018vpw}, but there is no known construction of the associated magnetic quivers.) It would also be interesting to construct renormalization group flows $\text{LST}_{(\mu_\text{L},\mu_\text{R})}\to \text{LST}_{(\mu'_\text{L},\mu'_\text{R})}$ between LSTs defined by different Kac labels at fixed $k$. This possibility was already mentioned in \cite{DelZotto:2022ohj}, and is obvious from the perspective of the orbi-instanton constituents, for which it has been thoroughly investigated in \cite{Frey:2018vpw,Fazzi:2022yca,Fazzi:2023ulb,Giacomelli:2022drw}. We plan to come back to this in the future.

%%%%%%%%%%%%%%%%%%%%%%%%%%%%%%%%%%%
\section*{Acknowledgments}
%%%%%%%%%%%%%%%%%%%%%%%%%%%%%%%%%%%

We would like to thank Sergio Benvenuti, Stefano Cremonesi, Simone Giacomelli, Paul Levy, Muyang Liu, Lorenzo Mansi, Noppadol Mekareeya, Paul-Konstantin Oehlmann, Sav Sethi and Yuji Tachikawa for interesting discussion and useful correspondence, and Craig Lawrie for interesting discussion and for informing us of an upcoming publication on related subjects \cite{Lawrie:2023uiu}.\footnote{\ Citation added in v3.} MF would like to thank the University of Milano--Bicocca, the Pollica Physics Center, the University of Lancaster for their kind hospitality and support during various stages of this work, and the ``Symplectic Singularities and Supersymmetric QFT'' conference held at UPJV in Amiens for providing a stimulating environment. SG thanks the University of Milano--Bicocca for hospitality during various stages of this work. MF and SG gratefully acknowledge support from the Simons Center for Geometry and Physics, Stony Brook University (2023 Simons Physics Summer Workshop) at which some of the research for this paper was performed. The work of MDZ and MF has received funding from the European Research Council (ERC) under the European Union's Horizon 2020 research and innovation program (grant agreement No. 851931). MDZ also acknowledges support from the Simons Foundation Grant \#888984 (Simons Collaboration on Global Categorical Symmetries). The work of MF is also supported in part by the Knut and Alice Wallenberg Foundation under grant KAW 2021.0170, the Swedish Research Council grant VR 2018-04438, the Olle Engkvists Stiftelse grant No. 2180108. The work of SG was conducted with funding awarded by the Swedish Research Council grant VR 2022-06157.

%%%%%%%%%%%%%%%%%%%%%%%%%%%%%%%%%%%
%%%%%%%%%%%%%%%%%%%%%%%%%%%%%%%%%%%

\bibliography{main}
\bibliographystyle{at}

\end{document}